\documentclass[traditabstract]{aa}

\usepackage{graphicx}
\usepackage{txfonts}

\usepackage{epsfig}
\usepackage{natbib}

\def\rsun{R$_\odot$}

\newcommand{\beq}{\begin{equation}}
\newcommand{\eeq}{\end{equation}}

\begin{document}

\title{Hinode EIS line widths in the quiet corona up to 1.5~\rsun}
\author{G. Del Zanna\inst{1}  
\and
G.R. Gupta\inst{1,2}
\and H.E. Mason\inst{1}
}
\institute{DAMTP, Centre for Mathematical Sciences,  
 University of Cambridge,  Wilberforce Road, Cambridge CB3 0WA UK
\and 
Udaipur Solar Observatory, Physical Research Laboratory, Badi Road, Udaipur
313 001, India
}

 \date{Received  ; accepted }

 \abstract{We present an analysis of several Hinode EIS observations of coronal
 line widths in the quiet Sun, up to 1.5~\rsun\ radial distances. No significant 
variations are found, which indicates no damping of Alfv\'en waves in the 
quiescent corona. 
 However, the uncertainties in estimating the instrumental width 
mean that a firm conclusion cannot be reached. 
We present a discussion of various EIS instrumental issues and suggest that 
the strongest lines, from  \ion{Fe}{xii} at 193.5 and 195.1~\AA, 
have anomalous instrumental widths. We also show how line widths in EIS are 
uncertain when the signal is low, and that the instrumental variation along
the slit is also uncertain. 
  We also found an 
anomalous decrease (up to 40\%) in the intensities of these lines in many 
off-limb and active region observations, and suggest that this
is due to opacity effects. We find that 
the most reliable measurements are obtained from the weaker lines.
 }

\maketitle

\section{Introduction }

It has long been known that  coronal 
lines  always exhibit non-thermal  broadenings
\citep[see the recent review in ][]{delzanna_mason:2018}, 
i.e. widths in excess of the expected thermal widths,
on the basis of the estimates of the ion temperatures.
Earlier Skylab and SMM observations of coronal forbidden lines indicated 
excess widths of 15--20 km\,s$^{-1}$
\citep[see, e.g.][]{doschek_feldman:1977_ntv,mason:1990}.
This  excess broadening is often interpreted as a signature of waves that are propagating
in the solar atmosphere and could contribute to the coronal heating
\citep[see, e.g.][]{hollweg:1978,hollweg:1984,parker:1988,van_ballegooijen_etal:2011}.
{
Variations (both increases and 
decreases) in the excess broadening  with height above the limb  
have been reported.  
The  decreases in the excess broadening seen  in coronal holes and quiet Sun 
areas have been interpreted as caused by the damping of Alfv\'en waves in the corona.

We only consider here  quiet Sun off-limb observations. There are surprisingly few 
observations with varying radial distance reported in the literature.
The measurements are notoriously challenging, as the signal rapidly decreases off-limb,
and there is always a large uncertainty in measuring ion temperatures.
Also, estimating instrumental broadenings is often a challenge in the EUV/UV.
\cite{doschek_feldman:1977_ntv} measured the widths of several 
coronal iron UV forbidden lines
above the limb and found  excess widths of about 20 km/s.
\cite{hassler_etal:1990}  found increases in the excess widths
of \ion{Mg}{x} lines, observed with a rocket spectrum. 
In contrast, SoHO CDS off-limb observations of the quiet corona 
 showed  a decrease  in the line widths of \ion{Mg}{x} \citep{harrison_etal:2002}.
However, an east-west  variation of the CDS instrumental width 
was later found, thus removing  evidence of variations \citep{wilhelm_etal:2005}.

{
Most of the SoHO SUMER off-limb observations of quiet regions 
have shown little variation in the excess line widths up to about 1.3~\rsun\
\citep[see ][]{doyle_etal:1998,doschek_feldman:2000,landi_feldman:2003,wilhelm_etal:2004_widths}.
\cite{seely_etal:1997} found a small decrease in the  excess line widths from about 20
km/s near the limb to about 10 km/s up to about  1.2~\rsun, assuming 
uniform ion temperatures for all elements.
 \cite{landi_feldman:2003} performed a very detailed analysis, measured the 
ionisation temperatures, and found nearly constant
excess line widths up to 1.3~\rsun. The values were higher than those 
found by \cite{seely_etal:1997} though, about 30 km/s.
 \cite{landi_etal:2006} found instead small increases, of up to 10 km/s,
 in coronal  lines observed by SUMER  out to 1.6~\rsun.

In contrast, there are two reported Hinode EIS off-limb observations of the quiet Sun 
where a decrease in the excess widths were found: 
\cite{hahn_savin:2014} measured widths along  expected locations
of magnetic field lines, while \cite{gupta:2017} 
analysed a  `quiet' off-limb  area close to an  active region
(which was the focus of an  analysis by \citealt{odwyer_etal:11}).

As SUMER had a much better spectral resolution than CDS and EIS,
we can summarise the previous observations by stating that there is no 
significant evidence for variations in the excess widths of 
coronal lines  in the quiet Sun, out to about 1.3~\rsun. 
Our aim is to analyse further Hinode EIS observations and extend these results
to further distances.

}

As pointing Hinode  outside the solar limb is normally 
not allowed and the EIS slits are aligned in the N-S direction, 
most off-limb quiet Sun EIS observations have been  restricted to
about 1.2~\rsun. Here, we present  results of a special EIS campaign 
where the EIS slit reached  1.5~\rsun, and which we have used to 
assess  line width variations up to such distances. 
{
The paper is organised as follows: Section~2 presents a summary of 
the data analysis and a few  instrumental issues relevant for this paper.
Section~3 presents a sample of 
results from this  special EIS campaign. As we did not find 
strong evidence for a decrease in the line widths 
and found  an anomalous behaviour in the 
strongest lines, we have re-analysed in Section~4  
the \cite{gupta:2017} observations. We carried out an analysis of the possible reasons, summarised in 
Section~5 and detailed in an Appendix. 
Section~6 presents our conclusions.
}

\begin{figure*}[!htbp]
\centerline{\includegraphics[width=11.cm,angle=90]{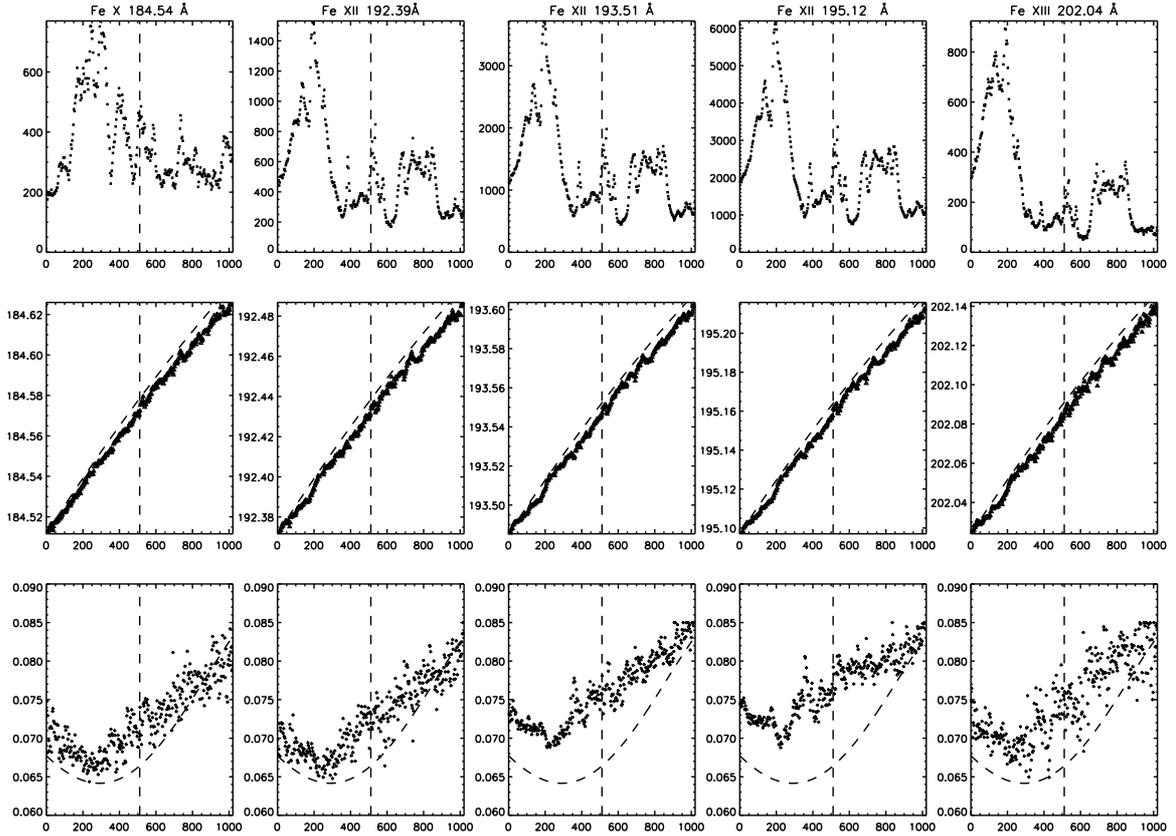}}
\caption{The top row shows the peak DN in a selection of SW lines, as a function of the 
position along the slit, for a 160s exposure with the 2\arcsec\ slit on the  quiet 
Sun, recorded on 2006-10-28. The dashed vertical line indicates the mid position
of the 1024 slit. 
The middle row shows  the line centroid positions; the dashed line indicates the 
estimated variation as available in the EIS software notes. 
The bottom row shows the FWHM in the lines, while the dashed line indicates the 
estimated variation as available in the EIS software notes.}
\label{fig:tilt_2_20061028}
\end{figure*}

\section{EIS data analysis and instrumental issues}

The EIS data have been processed with custom-written 
software  \citep[see, e.g. ][]{delzanna_etal:2011_aia}.
The analysis follows closely the procedures developed within the EIS 
software. The hot, warm pixels, and those affected by cosmic rays
or particle events are flagged as `missing'. The missing
pixels are then replaced with interpolated values with a 
series of interpolations along different  directions.
The  exposures are then visually inspected. As the dark frames 
(bias) are very variable and unknown, a base minimum value is removed
from each exposure (unlike the standard software where an average
of the bottom pixels is removed - which causes negative DN in the 
spectra). 
A large quantity of dust on the EIS CCD affects the 193.5~\AA\ line,
so that area was blanked out during the analysis.
To obtain an uncertainty in each pixel, we convert the 
observed DN into detected photon events, assuming Poisson noise
and adding the read-out noise.

In principle, saturation of an EIS pixel should occur 
at a value of 16383 in data numbers (DN). Indeed the standard software assumes a 
saturation threshold of 16000 DN. 
{ 
However, we have  found some evidence for a  small non-linear 
behaviour in the line intensities when peak counts are above 
8000--10000 DN, as described in the Appendix.

The narrow 1\arcsec\ and 2\arcsec\ EIS slits are 1024  pixels 
(1 pixel is equal to about 1\arcsec) long,
which are equivalent to about 1024\arcsec. However, when exposures are telemetred to the 
ground, only a portion of the slits (up to 512 pixels) is downloaded.
Regular observations extract the central regions of the slits.
A region of the Sun is scanned by `rastering', i.e. moving in the E-W 
direction the image of the slits onto the detectors.

The images of the two narrow slits (1 and 2\arcsec) onto the detector present several 
instrumental issues which are different for the two slits. 
The main ones are a slant of the spectra compared to the detector pixels
\citep{delzanna_ishikawa:09}, and a tilt and width of the spectral lines which 
 vary with position along the slit.
We have corrected the EIS full spectra for the slant and 
tilt by rotation, following the prescriptions  in 
\cite{delzanna_ishikawa:09}.
}
 We adopted the same procedure developed by W. Thompson for the  SoHO CDS, which suffered similar 
instrumental issues as EIS. We first rotate the spectra. 
We note that once the spectra are rotated, a shift along the slit 
of about 18 EIS pixels has always been present. 
The spectra in the two EIS  channels (short: SW and long: LW) are then 
corrected for the differences in the solar X and Y.
In earlier observations, a shift of about 2\arcsec\ in solar X 
was also present, meaning that the SW and LW spectra never observed 
simultaneously the same region on the Sun.


The {\em cfit} programs, developed by S.V.Haugan for the SoHO CDS,
were used, with several modifications, to fit Gaussian profiles
and  a  linear background  to the spectra in DN. 
As the EIS profiles are dominated by the instrumental width,
they are always close to Gaussian, for the unblended strong lines
which we consider here. We have also checked that the total 
signal obtained by summing the counts over the line profiles is
close (within at most a few percent) to the value obtained from 
the Gaussian parameters 
{ (see the Appendix).

The radiometric calibration is only applied
later to the integrated intensities. We used the 
\cite{delzanna:13_eis_calib} calibration, although some results
obtained with the ground calibration \citep{lang_etal:06} are also shown.
}
We have also processed some of the observations presented here 
with the standard EIS software, and found very similar results.

\subsection{Instrumental variations of the EIS line widths and anomalous Fe XII widths}

\begin{figure}[!htbp]
\centerline{\includegraphics[width=7.5cm,angle=0]{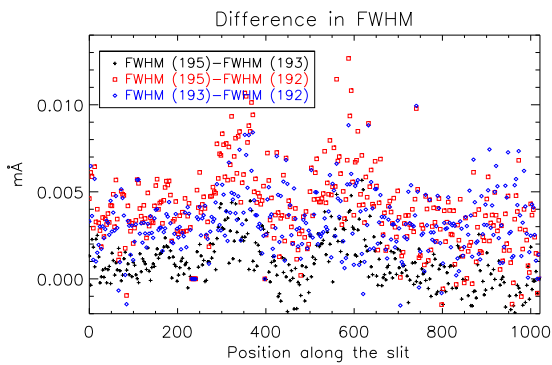}}
\caption{The differences in the FWHM of the 
three main \ion{Fe}{xii}  lines, as a function of the 
position along the slit, for the same observation in Fig.~\ref{fig:tilt_2_20061028}.
}
\label{fig:20061028_fwhm_ratios}
\end{figure}

{

Observed line widths FWHM  typically contain an instrumental 
width, a thermal width which depends on the ion temperature, and an excess width $\xi$.
Assuming thermal broadening:  

\begin{equation}
FWHM^2 =  w_{\rm I}^2 \, +  \, 4 \, \ln 2 \, \left ({\lambda_0 \over c} \right )^2 \, 
\left ({2 k T_{\rm i} \over M} + \xi^2 \right )  \;\;,
\end{equation}
where $w_{\rm I}$ is the instrumental FWHM,  $k$ is Boltzmann's constant, $M$ is the mass 
of the ion,  and $T_{\rm i}$ the ion temperature.

As the widths of the EIS lines are dominated by the instrumental 
width, a careful assessment of instrumental issues should be carried out.
A thorough discussion of the EIS instrumental width and its 
variation in time, along the two narrow slits and as a function of  mirror position is 
beyond the scope of the present paper. However, here and in the Appendix 
we present some key examples which are relevant for this paper.

\cite{hara_etal:2008} noted in their Appendix that the instrumental widths
of the EIS lines in the SW channel must be significantly larger than 
the values measured on the ground. 
In most literature, the instrumental widths 
as described in the EIS software note 4  \citep{young_eis_width}) are used. They were obtained from 
observations in Nov-Dec 2009 and do not have a wavelength (or time) dependence.  
The main assumption was that the smallest observed widths 
would provide a measure of the instrumental width. 
The same observations were used to measure the tilt \citep{young_eis_sn4}.

We have performed our own analyses and found some variations and differences with the 
values recommended in the EIS software notes.
To illustrate the main effects relevant to this paper, we show in Fig.~\ref{fig:tilt_2_20061028}
the peak intensities, position and full-width at half-maximum (FWHM) of the main SW lines from 
a 2\arcsec\ slit observation on the quiet Sun obtained on 
2006-10-28 at 11:17 UT (bottom  512 pixels of the 1024-pixel slit)
and 11:23 UT (top 512 pixels of the slit) with 5 exposures.
We selected the longest (160s) exposures and 
rebinned the spectra by a factor of 3 along the slit to improve the signal.
The dashed lines in the middle and bottom rows are the recommended 
values for the variation in the position and the 
instrumental FWHM estimated by P.R. Young from the Nov-Dec 2009 observations.
It is clear that some departures from the recommended 
values are present.  In particular, the slit width seems to be overestimated at places.

The first main point of Fig.~\ref{fig:tilt_2_20061028} is to show that the bottom of the 1024\arcsec\
slit has a much lower  width than the central/top part,
where the width increases. 
The initial decrease and then increase in the bottom 
of the slit is not  obvious in other datasets (cf. the Appendix) and is within the 
scatter of the measurements, which is typically around 3 m\AA. 
The fact that the bottom is the best part of the slit  is the main 
reason why the special off-limb sequence discussed below was designed to 
obtain data from this region of the CCD.

The second main point of the Figure is to show that the 
widths of the three \ion{Fe}{xii} 192.4, 193.5, and 195.1~\AA\ lines are 
consistently different  at all locations along the slit.
 The widths of the 193.5 and 195.1~\AA\ lines
 are similar, while the width of the 192.4~\AA\ line is smaller
by about 4--5 m\AA, as shown in Fig.~\ref{fig:20061028_fwhm_ratios}. 
We have seen similar differences in the observations we have analysed,
 as shown in the Appendix. 
Lines at shorter and longer wavelengths show widths similar to the weaker
\ion{Fe}{xii} 192.4~\AA\ line, as for the \ion{Fe}{x} and \ion{Fe}{xiii} lines shown in 
Fig.~\ref{fig:tilt_2_20061028}.

We cannot think of any physical reason that can explain the differences
in the widths of the \ion{Fe}{xii}   lines.
The EIS wavelength range is rich in unidentified coronal lines,
as discussed in \cite{delzanna:12_atlas}. However, these are typically
much weaker (a few percent) than the strong \ion{Fe}{xii} lines.
Also, recent laboratory measurements with an  Electron Beam Ion Trap (EBIT) 
do not suggest that  lines from other ions 
or elements  are blending the  \ion{Fe}{xii} lines, 
see \cite{traebert_etal:2014_193}.
We would therefore expect the widths of these three \ion{Fe}{xii}  lines to be the same.
{%
Indeed there is an independent evidence for this. To our knowledge, 
the best measurements of the widths of these lines were obtained in second order
with the SERTS-95 sounding rocket \citep{brosius_etal:98b}. The spectrometer
had  an instrumental 
FWHM resolution of about 30 m\AA\ in second order. 
In their quiet Sun averaged spectra the 193.5 and 195.1~\AA\ lines 
actually had lower observed widths than the 192.4~\AA\ line, however
the signal was low, and measurements had a large uncertainty. 
In the active region spectra, the observed FWHM
for the 192.4, 193.5 and 195.1~\AA\ lines were $51.2 \pm 3$,$49.2 \pm 3$,$49.3 \pm 3$
respectively, i.e. were the same within uncertainties. 

On a side note, the SERTS-95  spectra were radiometrically calibrated by \cite{brosius_etal:98a}
assuming optically thin conditions and the CHIANTI atomic data 
available at the time, so these data cannot be used directly to check the 
intensity ratios of these lines.
We conclude that the larger widths in the two EIS  \ion{Fe}{xii} lines 
(193.5, and 195.1~\AA) are instrumental.
}

Finally, on the basis of the east-west  variation in the instrumental line widths  found in 
SoHO CDS, we have  carried out several tests on observations of the quiet Sun
with the   2\arcsec\ slit to see if any east-west instrumental 
variation is also  present for Hinode EIS.
We analysed both the central part of the slit,  and in particular
 the  bottom of the 2\arcsec\ slit, 
as the main scientific results presented here were obtained with  
that part of the slit. We  did not find any obvious variations,
as shown in the Appendix.

}

\section{The off-limb QS observations out to 1.5~\rsun}

{
One of us (GDZ)  designed an engineering
EIS 'study' (gdz\_off\_limb1\_60) to use the bottom half of the long (1024\arcsec) EIS slit,
so the lower part of the EIS field of view could  reach  1.5~\rsun.
}
A week-long campaign  (Hinode HOP 7) was  coordinated to obtain simultaneous 
 SOHO/Hinode/TRACE/STEREO observations during the SOHO-Ulysses quadrature in May 2007. 
EIS observations were obtained during May 7--10, as outlined in \cite{delzanna_etal:2009ASPC}. 
The EIS  2\arcsec\ slit was moved with 8\arcsec\ jumps, to cover about 500\arcsec\
in the E-W direction with  60s exposures.
As far as we are aware, these are the only EIS observations of the quiet Sun up to 
such large distances above the limb. 

The Sun  was very quiet during the campaign, with the exception of an active region (AR), which was
at the west limb on May 7.
 The AR  produced a filament eruption, where large 
non-thermal widths of more than 100 km/s were observed \citep{delzanna_etal:2009ASPC}. 
On May 10, the main part of the active region was behind the limb,
as shown in  Fig.~\ref{fig:10-may_1}. Some emission associated with the active region 
is still visible in the north portion of the EIS field of view,
but the lower part was quiet.
An off-limb  region about 5$^o$ wide, along the radial sector indicated in 
 Fig.~\ref{fig:10-may_1}  was chosen to obtain averaged spectra, to  improve 
the signal in the far off-limb region.

\begin{figure}[!htbp]
\centerline{\includegraphics[width=9.2cm,angle=0]{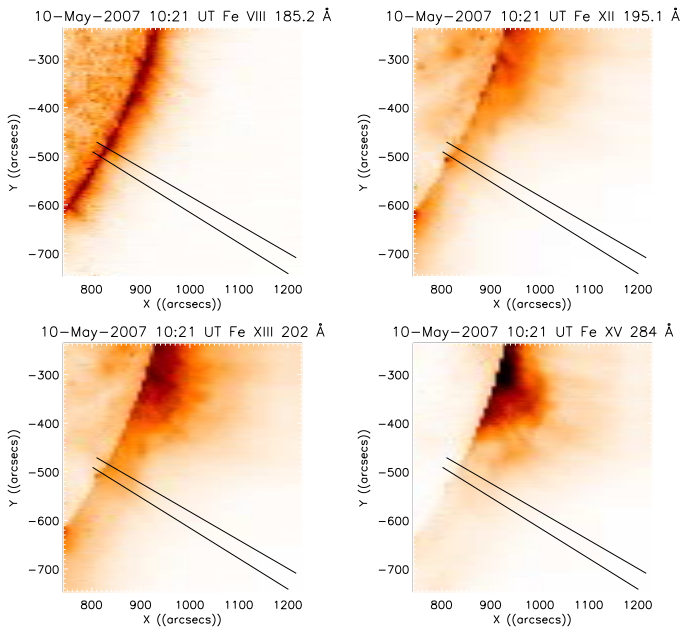}}
\caption{Monochromatic images (negative) in a few spectral lines,
obtained on 2007-05-10 with Hinode EIS. 
The radial sector used to obtain averaged spectra in the quiet corona 
as a function of radial distance is indicated.
}
\label{fig:10-may_1}
\end{figure}

{
A detailed analysis of  temperatures and densities 
obtained from this region has been presented in \cite{delzanna_etal:2018_cosie}.
A remarkable feature was the nearly constant ionisation temperature (around 1.4 MK)
obtained from ratios of  \ion{Fe}{x},  \ion{Fe}{xi},  \ion{Fe}{xii} lines,
extending to greater heights the SUMER results of  \cite{landi_feldman:2003}.
The densities were close to typical quiet Sun values, and scattered light 
in the coronal lines was found to be negligible.
}

The signal in the  \ion{Fe}{xii} 195.1~\AA\ was sufficient to measure its width 
 for each pixel, at least out to  about 1.2~\rsun, 
as shown in Fig.~\ref{fig:10-may_2}. There is an 
indication of a variation in the width, but with a large scatter, largest
at greater distances, where the signal is low, below about 200 DN in the peak of the line
(cf. bottom plot in Fig.~\ref{fig:10-may_2}).
{
This is expected, as the width of weak line profiles tends to increase.
In the Appendix, we show several examples of near-simultaneous observations
with different exposure times, where it is clear that a large uncertainty 
(5 m\AA\ or more) is associated with the widths of lines with peak 
counts below 200 DN, as seen here. 
Note that  we have not constrained the upper limit of the 
 width of the line during the fitting process.
}

\begin{figure}[!htbp]
\centerline{\includegraphics[width=5.cm,angle=0]{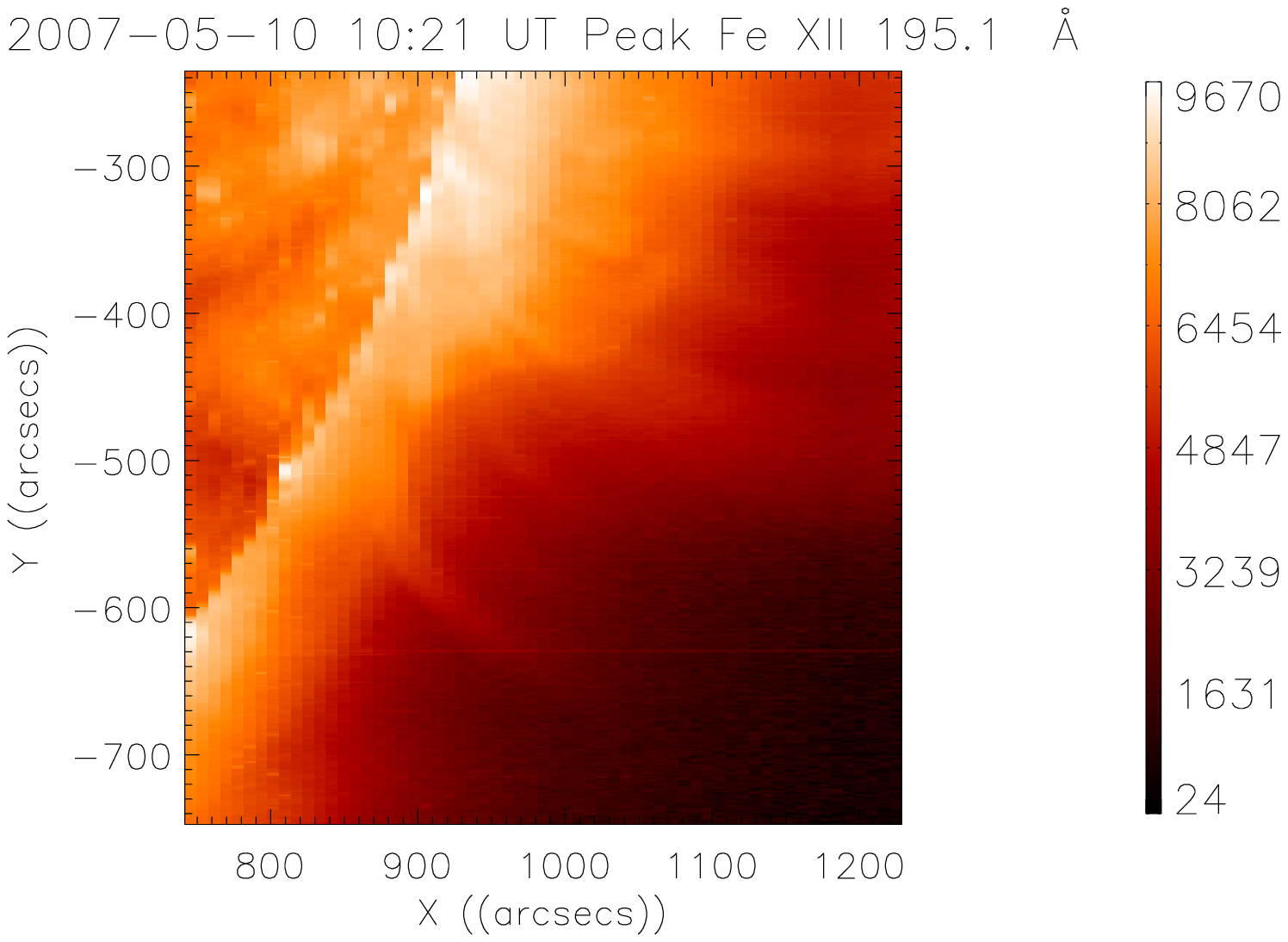}
\includegraphics[width=5.cm,angle=0]{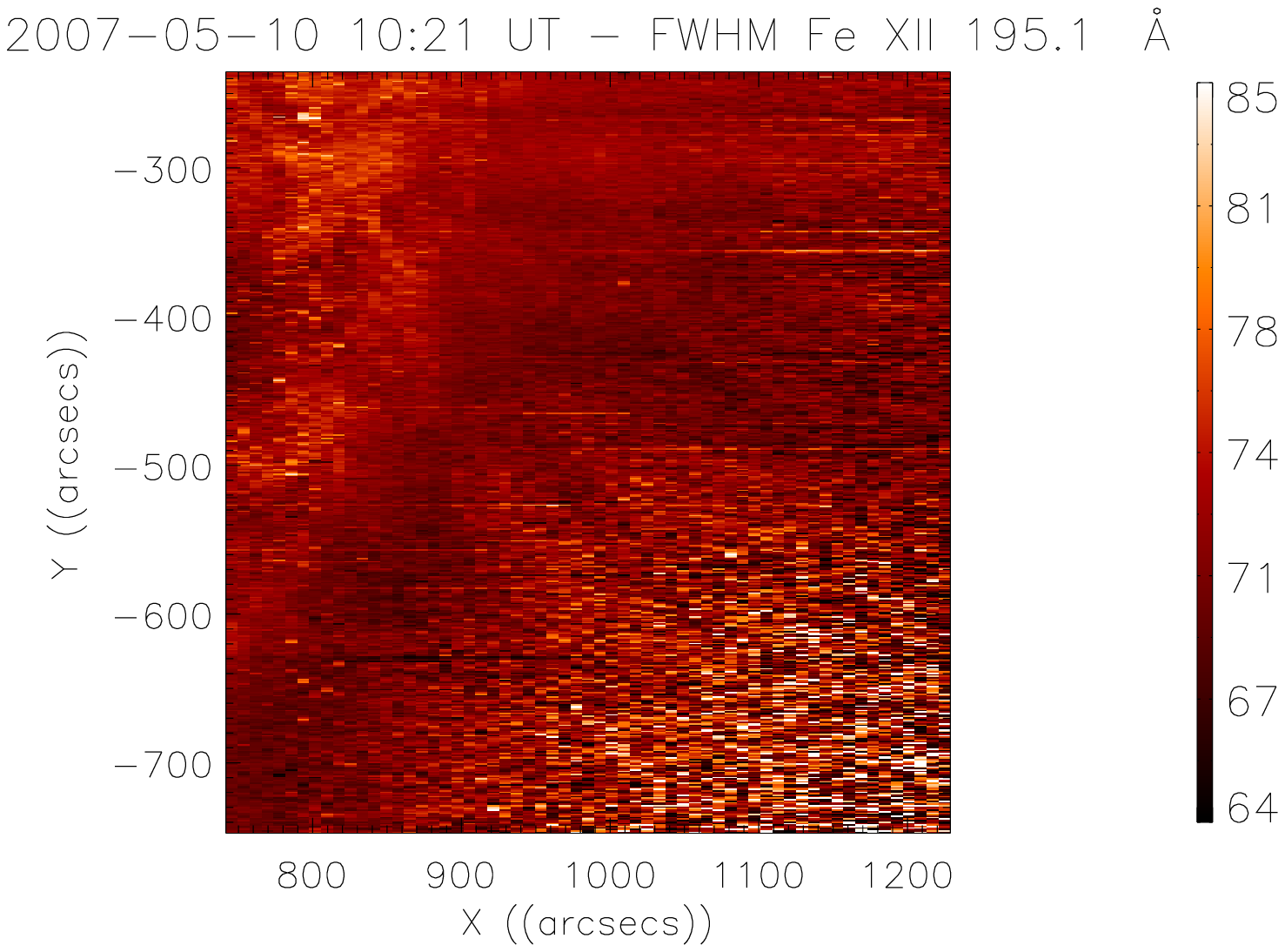}
}
\centerline{\includegraphics[width=6.0cm,angle=0]{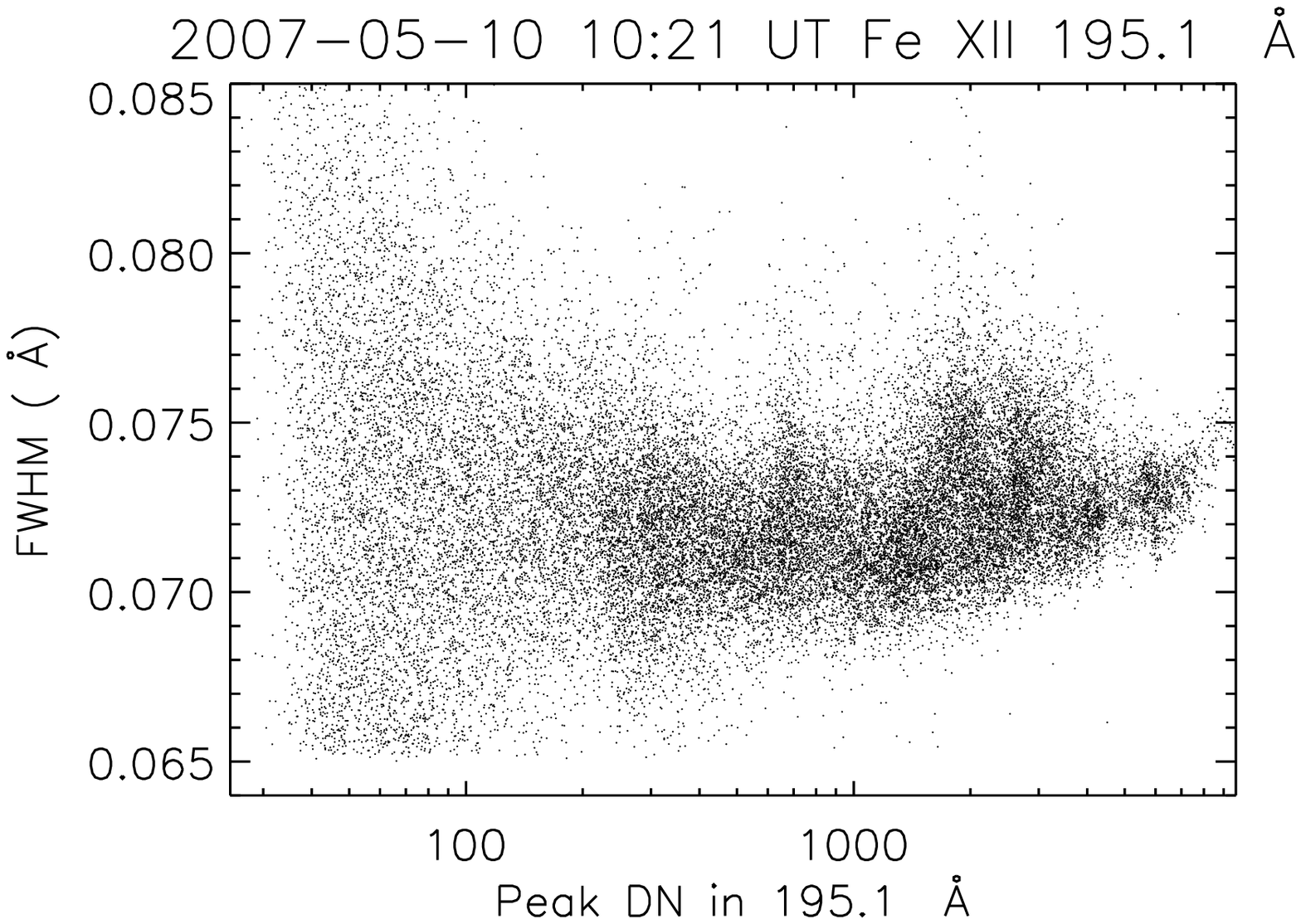}}
\caption{Selected results from the Hinode EIS off-limb observation on 2007-05-10 at 10:21 UT.
Top: images of the  peak intensity (DN) and  FWHM (m\AA) 
in the \ion{Fe}{xii} 195.1~\AA\ line. Bottom:
scatter plot of the FWHM (\AA) as a function of peak intensity.
}
\label{fig:10-may_2}
\end{figure}

\begin{figure}[!htbp]
\centerline{\includegraphics[width=6.5cm,angle=90]{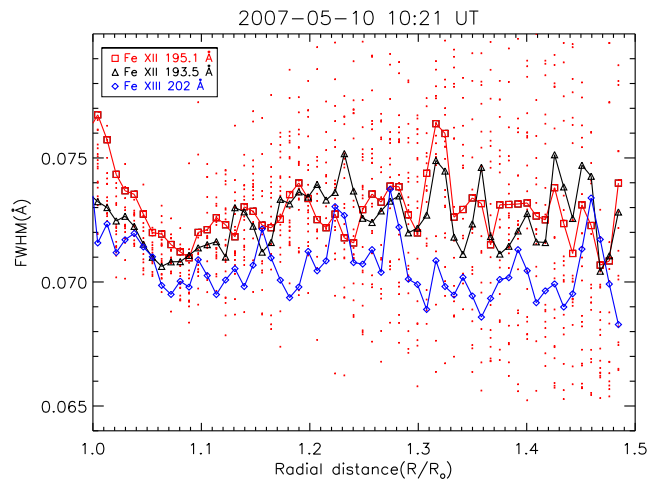}}
\caption{FWHM of the  \ion{Fe}{xii} 195.1, 193.5~\AA\ and  \ion{Fe}{xiii} 202~\AA\ lines,
obtained from averaged spectra along the radial QS sector shown in 
Fig.~\ref{fig:10-may_1}.
The points are  the pixel-by-pixel FWHM of the  195.1~\AA\ line.
}
\label{fig:10-may_3}
\end{figure}

 Fig.~\ref{fig:10-may_3} shows the variation in the widths 
of the  \ion{Fe}{xii} 195.1, 193.5~\AA\ and  \ion{Fe}{xiii} 202~\AA\ lines as  a 
function of radial distance, obtained from the radial QS sector shown in 
Fig.~\ref{fig:10-may_1}. 
{
The  \ion{Fe}{xii} 193.5~\AA\ and 
 \ion{Fe}{xiii} 202~\AA\ lines do not show any significant variations in their
widths. 
Note the instrumental effect, e.g. the fact that the widths of the two strong
\ion{Fe}{xii} lines are larger than that of the \ion{Fe}{xiii} line. 
The weaker \ion{Fe}{xii} 192.4~\AA\ line is not shown in the plot as this line
was not included in the EIS study. 
Another feature is the increased width of the \ion{Fe}{xii} 195.1~\AA\ line
near the limb.  Below we present similar results, which we suggest are 
related to opacity effects.

\begin{figure}[!htbp]
\centerline{\includegraphics[width=7.5cm,angle=0]{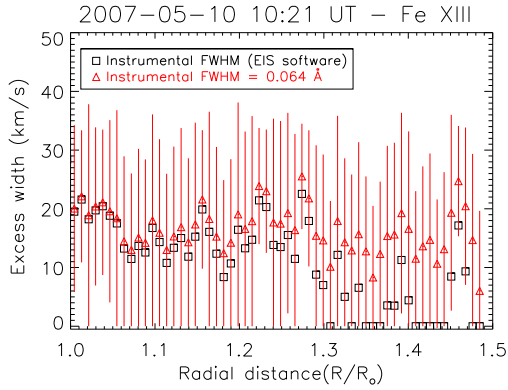}}
\caption{Excess widths from the \ion{Fe}{xiii} 202~\AA\ line as  a 
function of radial distance, obtained from the radial QS sector shown in 
Fig.~\ref{fig:10-may_1}, assuming two different instrumental widths.
}
\label{fig:10-may_ntw}
\end{figure}

The points in  Fig.~\ref{fig:10-may_3} are the single values obtained for the 
 \ion{Fe}{xii} 195.1~\AA\ line from each EIS pixel used to obtain the averaged spectra,
and give an indication of the large uncertainty associated with such measurements
(5 m~\AA\ or more in the weaker off-limb areas). 
The  variations in the widths obtained from the averaged spectra 
are within 3 m~\AA, i.e. well within the large scatter 
in each pixel values and also within the uncertainties in the  small 
variation of the instrumental width in the bottom of the CCD, shown above and further
discussed  in the Appendix.

{%

To illustrate the challenge in obtaining the excess widths, we plot in 
Fig.~\ref{fig:10-may_ntw} the values obtained from  the 
\ion{Fe}{xiii} 202~\AA\ line, subtracting a thermal width 
(assuming $T=1.4$ MK as measured from line ratios) and two different instrumental widths.
A constant value of 0.064~\AA\ (triangles) 
provides excess widths around 15--20 km/s, close to some of 
the  Skylab and SUMER results \citep[see, e.g.][]{seely_etal:1997}.
The error bars on the points are obtained by varying the instrumental 
width by only $\pm$2 m\AA, which is well within the scatter of values 
we have seen in several observations, and adding 
a $\pm$2 m\AA\ uncertainty in the observed widths.  
The boxes are the values obtained with the instrumental width variation
 as obtained by P.R. Young (discussed previously) and available
within the EIS software. The latter instrumental widths suggest
a small decrease at heights greater than  1.3~\rsun,
although in several places they  produce negative excess widths 
(whenever the observed width was smaller than the sum of the 
thermal and instrumental widths, we have set the values of 
the excess width to zeros in the plot).

We have analysed the observations of the week-long campaign, and found similar
results, as shown in the Appendix for the 8th of May 
(see Figs.~\ref{fig:8-may_1},\ref{fig:8-may_2}).
In this case, a marginal increase in the excess widths above 1.3~\rsun\ is present.


As most of the 2007 observations were affected somewhat by the presence 
of the active region, we have recently obtained new observations, in July 2018,
when the Sun was very quiet. The pointing was similar to the 2007 observations,
in the south-west.
As the EIS instrument has degraded and the solar signal is lower, these 
observations have lower quality than the earlier ones. 
 In this case we also found line widths to be nearly constant, as shown in the Appendix.

Therefore, we conclude that, in agreement with the SUMER 
results, there is no significant evidence for a variation of the excess
widths by more than 10 km/s out to 1.3~\rsun. The uncertainties in the 
observed and instrumental widths preclude firm conclusions in the 
1.3--1.5~\rsun\ region.

}

\section{A case study: the 2007-12-17 off-limb observation}

\begin{figure}[!htbp]
\centerline{\includegraphics[width=4.8cm,angle=0]{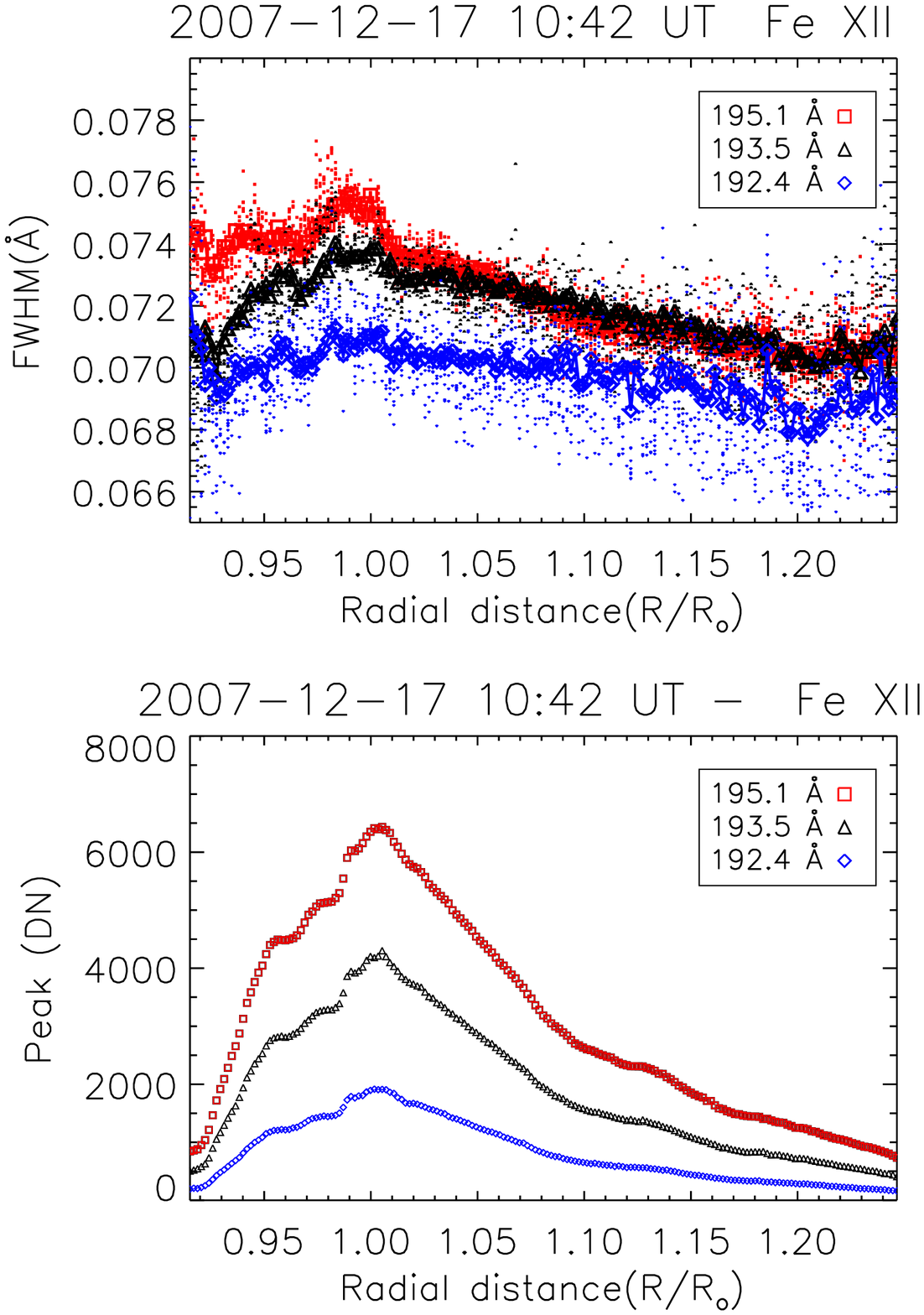}
\includegraphics[width=4.8cm,angle=0]{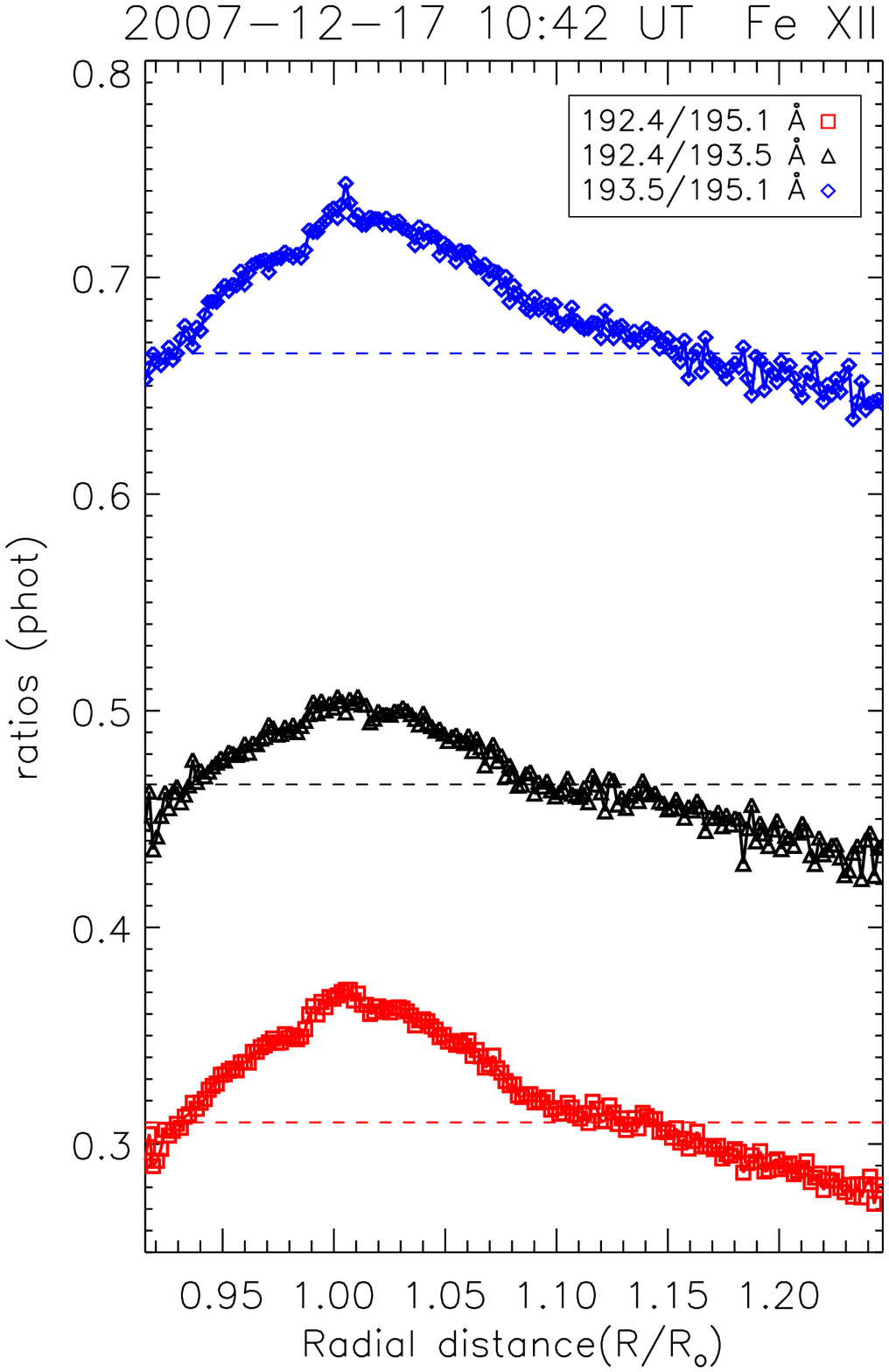}}
\caption{Left: off-limb variations of the FWHM in the three  \ion{Fe}{xii} lines,
obtained from a `quiet' region  30\arcsec\ wide observed on 2007-12-17 
\citep[see ][]{gupta:2017}; 
variations of the peak line intensities (DN).
Right:  variations of the 
calibrated intensity line ratios, with the expected values shown as dashed lines.
}
\label{fig:2007-12-17a}
\end{figure}

\begin{figure}[!htbp]
\centerline{\includegraphics[width=5.0cm,angle=0]{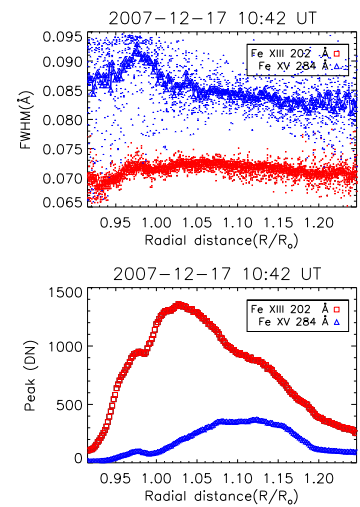}}
\caption{Off-limb variations of the FWHM and peak line intensities 
for the strongest \ion{Fe}{xiii} and \ion{Fe}{xv} lines along the same 
 `quiet' region  30\arcsec\ wide as observed on 2007-12-17 
and as on  Fig.~\ref{fig:2007-12-17a}.  
}
\label{fig:2007-12-17b}
\end{figure}

{
We have re-analysed the same `quiet Sun' off-limb region analysed by 
\cite{gupta:2017}, consisting of an E-W region out to 1.25~\rsun\ and 
30\arcsec\ wide in the N-S direction.
{This was an excellent observation with many spectral lines and 
good signal. 
As we have no indications (see also the Appendix) that 
the instrumental width varies along the E-W direction, its estimated value
was subtracted, along with the thermal width. The resulting
excess widths measured by \cite{gupta:2017} showed several inconsistenciess,
so we wondered if the instrumental and anomaous effects 
we have found could be seen here.

}

The  FWHM of the three strongest \ion{Fe}{xii} lines are shown in Fig.~\ref{fig:2007-12-17a}.
 Both the single pixel values and the  values obtained from the 
averaged spectra are shown. The FWHM of the weaker  192~\AA\ line has a 
small decrease  (2 m~\AA) with radial distance.
The nearly constant line widths in the weaker  \ion{Fe}{xii} line are 
in agreement with the widths of other lines: the strongest \ion{Fe}{xiii} line, at 202~\AA\ 
also shows a nearly constant FWHM with radial distance, see
 Fig.~\ref{fig:2007-12-17b} (top). 
In the LW channel, the \ion{Fe}{xv} line at 284~\AA\ 
also indicates  a nearly constant  line width, although with a larger scatter.

Returning to Fig.~\ref{fig:2007-12-17a}, 
we can clearly see two problems. One is that the instrumental widths
of the stronger  \ion{Fe}{xii} lines are larger, as we have seen. 
The second is that 
 the  widths of 193 and 195~\AA\ lines  increase where the peak intensities are large,
near the solar limb. The width of the stronger 195~\AA\ line  increases 
even more than the  193~\AA\ line.

As already mentioned, we would expect the widths of these three lines 
to be the same. Furthermore, the intensity 
ratios of these lines, also shown in Fig.~\ref{fig:2007-12-17a},
vary with radial distance, contrary to 
their expected values, indicated with dashed lines. 
After an analysis of several observations, discussed below and in the 
Appendix, we found this anomalous behaviour to be typical and 
suggest  that these variations are related to opacity effects
in the stronger lines as discussed below.

}

\section{The anomalous \ion{Fe}{xii} intensity ratios}

{
We have seen that in certain places there appears to be some 
correlation between anomalous widths of the strongest 
 \ion{Fe}{xii} 193.5 and 195.1~\AA\ lines, and 
anomalous intensities, when compared to the weaker 192.4~\AA\ line.
 As these lines are  the strongest features 
in the EIS spectra, being close to the  the peak of the EIS  effective area,
they are commonly used in the literature for a wide range of 
measurements, including line widths, electron densities and temperatures. 
We therefore explored these anomalies by analysing several 
datasets. 
We present here an example case, as a full discussion is 
beyond the scope of the present paper. 
More details are presented in the Appendix.

Fig.~\ref{fig:2013-02-14a} summarises the main results obtained from a 
quiet Sun east limb observation with an  ATLAS\_60 study, recorded 
on 2013-02-14 at 09:49 UT  with 60s exposures. 
There were no active regions nearby, but 
the Sun was relatively active during this period, so the 
strongest 195~\AA\ line was almost saturated near the limb, reaching a 
peak  of 10000 DN, as shown in  Fig.~\ref{fig:2013-02-14a}, top  plot.  
The 192 vs. 195~\AA\ intensity ratio is close to the 
expected value in the dimmest regions (on-disk and far off-limb), 
but shows an anomalous 
behaviour in the brightest  regions close to the limb.
The 192 vs. 195~\AA\ intensity ratio varies by as much as  35\%.
Note that the intensity variations are also related to an increase
in the  width of the 195~\AA\ line. 
A similar behaviour is seen in the 192 vs. 193~\AA\ ratio (not shown).
}

The ratios of the \ion{Fe}{xii} 192.4, 193.5, and 195.1~\AA\ lines 
are not sensitive to density or  temperature.
\cite{delzanna_mason:05_fe_12} identified in laboratory 
plates obtained by B.C. Fawcett a weak density-sensitive line
blending the 195.1~\AA\ line. This transition is less than 1\%
the intensity of the strong line at typical quiet Sun densities, 
so can be ignored. At active region densities, this line can reach 
a 10\% level, and the line profile becomes asymmetric. 

Comparison with the well-calibrated  \cite{malinovsky_heroux:73} intensities
shows exact agreement (within less than 1\%) in the ratios of these \ion{Fe}{xii}
 lines with theory \citep{delzanna_etal:12_fe_12}, as discussed in  \cite{delzanna:13_eis_calib}.
One would therefore expect that these three lines should have 
a constant intensity ratio, with the 
exception of active region observations where the intensity of the 195.1~\AA\ line should increase. 

The 192 vs. 195~\AA\ and 192 vs. 193~\AA\ intensity 
ratios are in excellent agreement with the expected values 
in the dimmest regions for the earlier EIS observations. 
This is because the  \cite{delzanna:13_eis_calib}
calibration was based on earlier quiet Sun observations.
With the ground calibration, the ratios are consistently lower by  up to 15\%.
On a side note, in later observations the ratios have 
often shown decreases (by up to 30\%), 
which would indicate a wavelength-dependent change in the sensitivity
over time.

\begin{figure*}[!htbp]
\centerline{\includegraphics[width=5.5cm,angle=0]{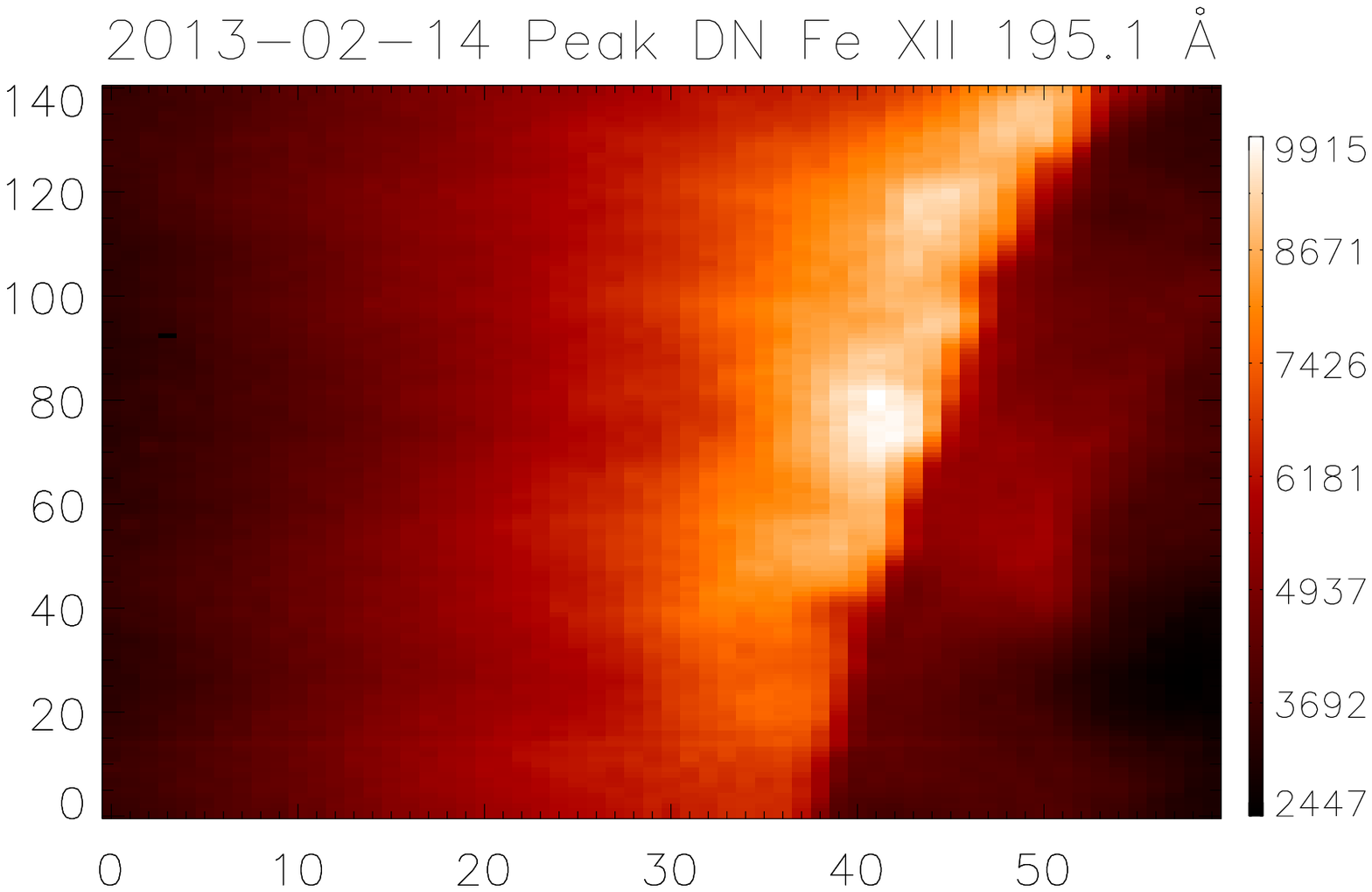}
\includegraphics[width=5.5cm,angle=0]{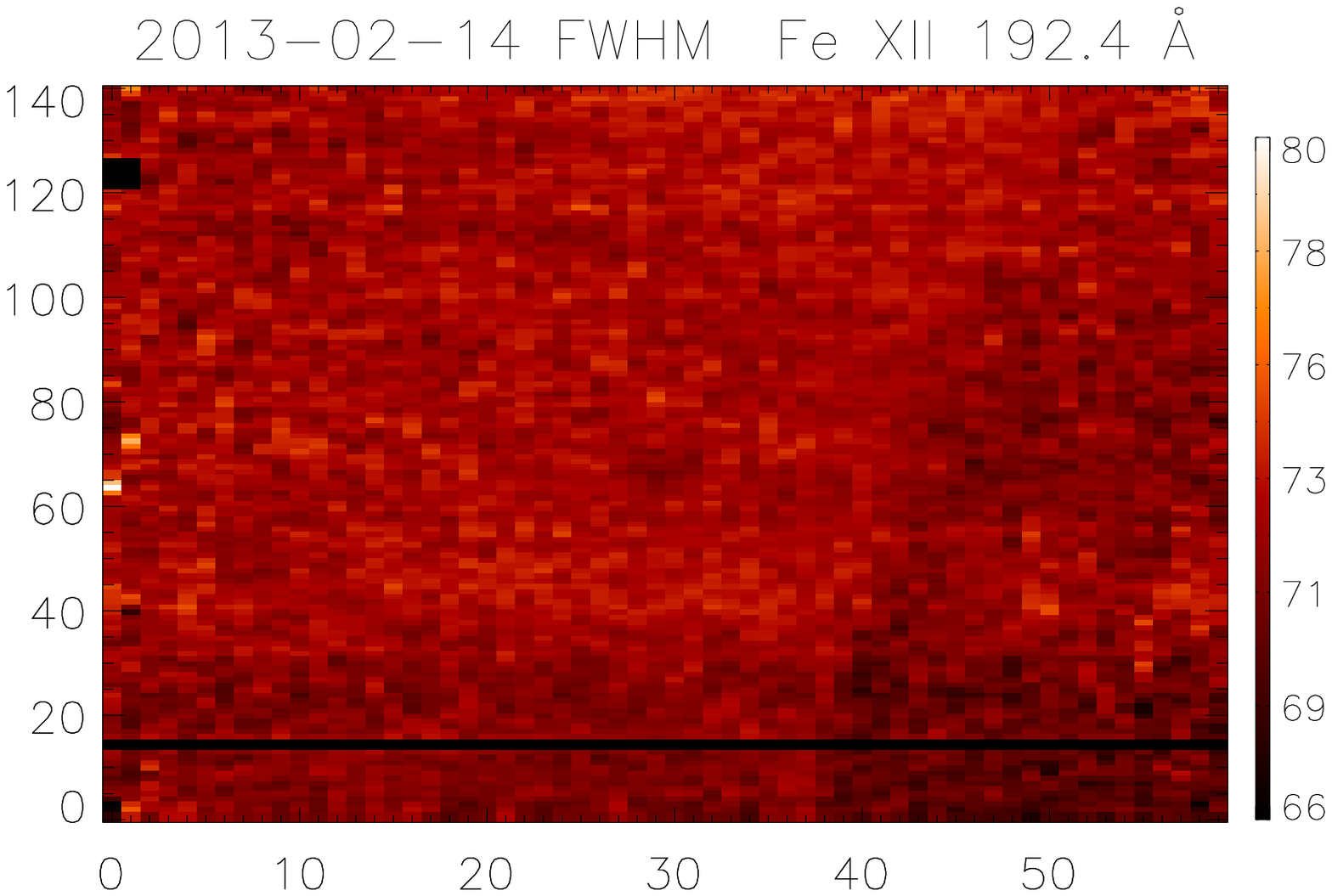}
\includegraphics[width=5.5cm,angle=0]{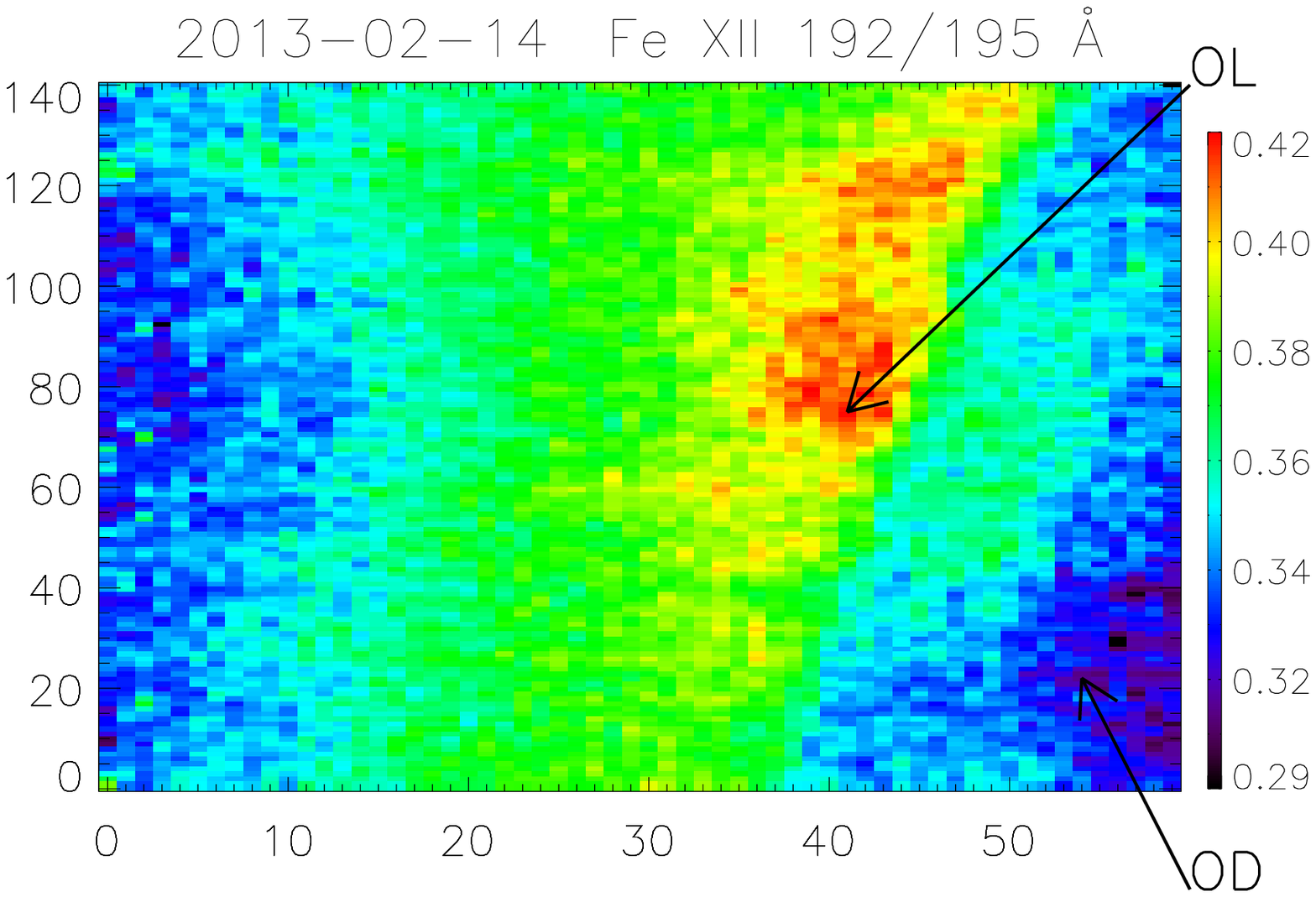}}
\centerline{\includegraphics[width=5.5cm,angle=0]{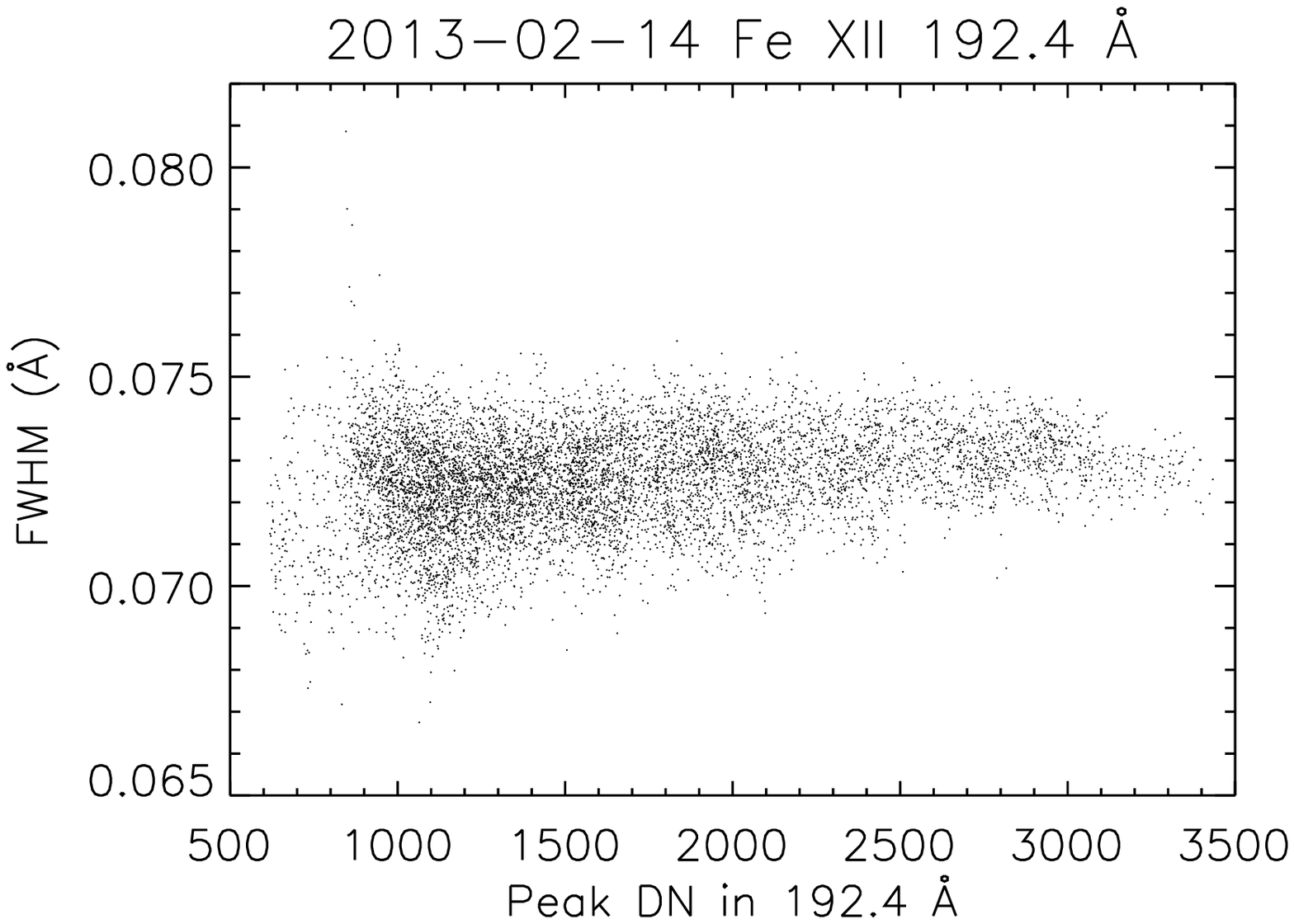}
\includegraphics[width=5.5cm,angle=0]{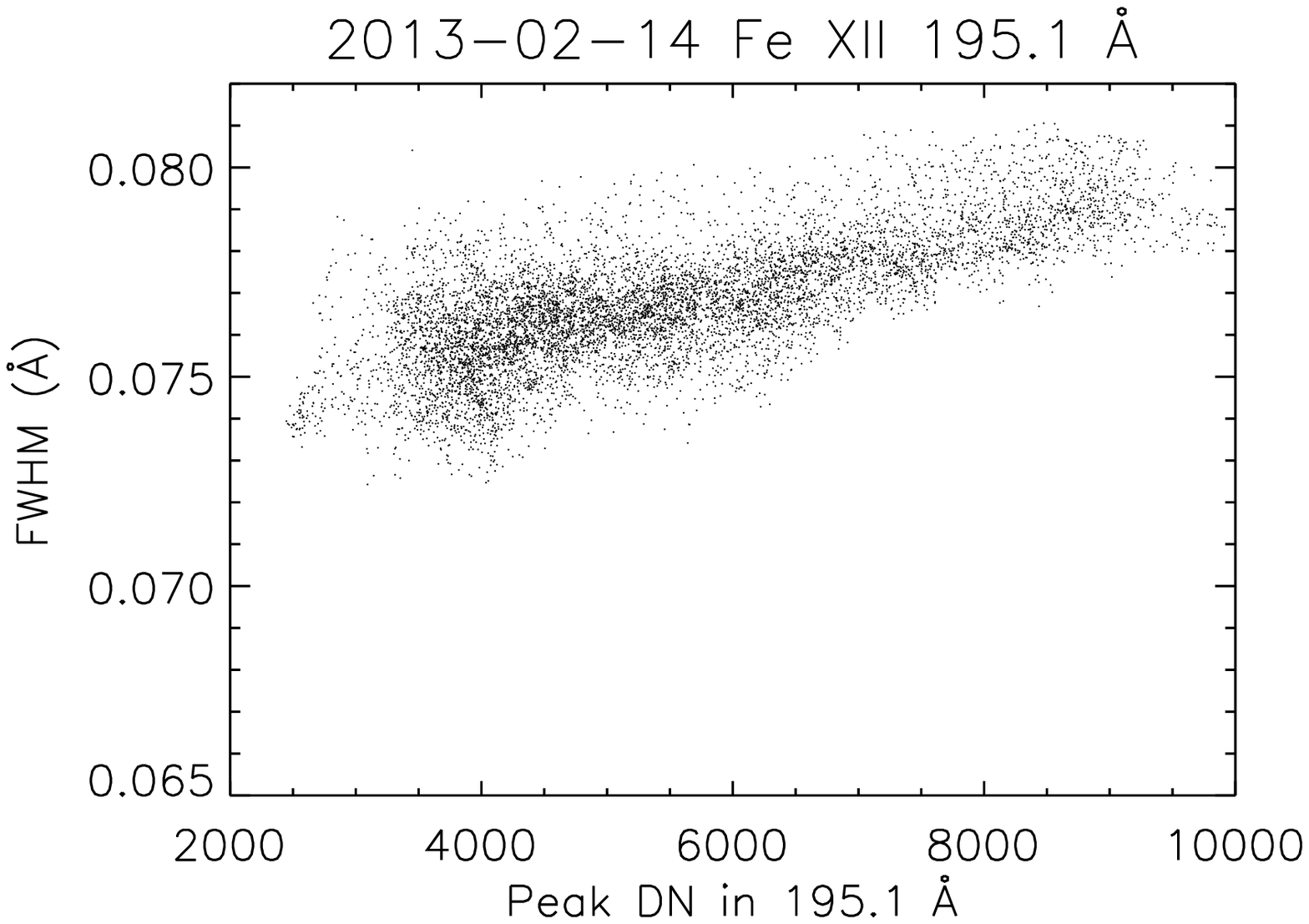}
\includegraphics[width=5.5cm,angle=0]{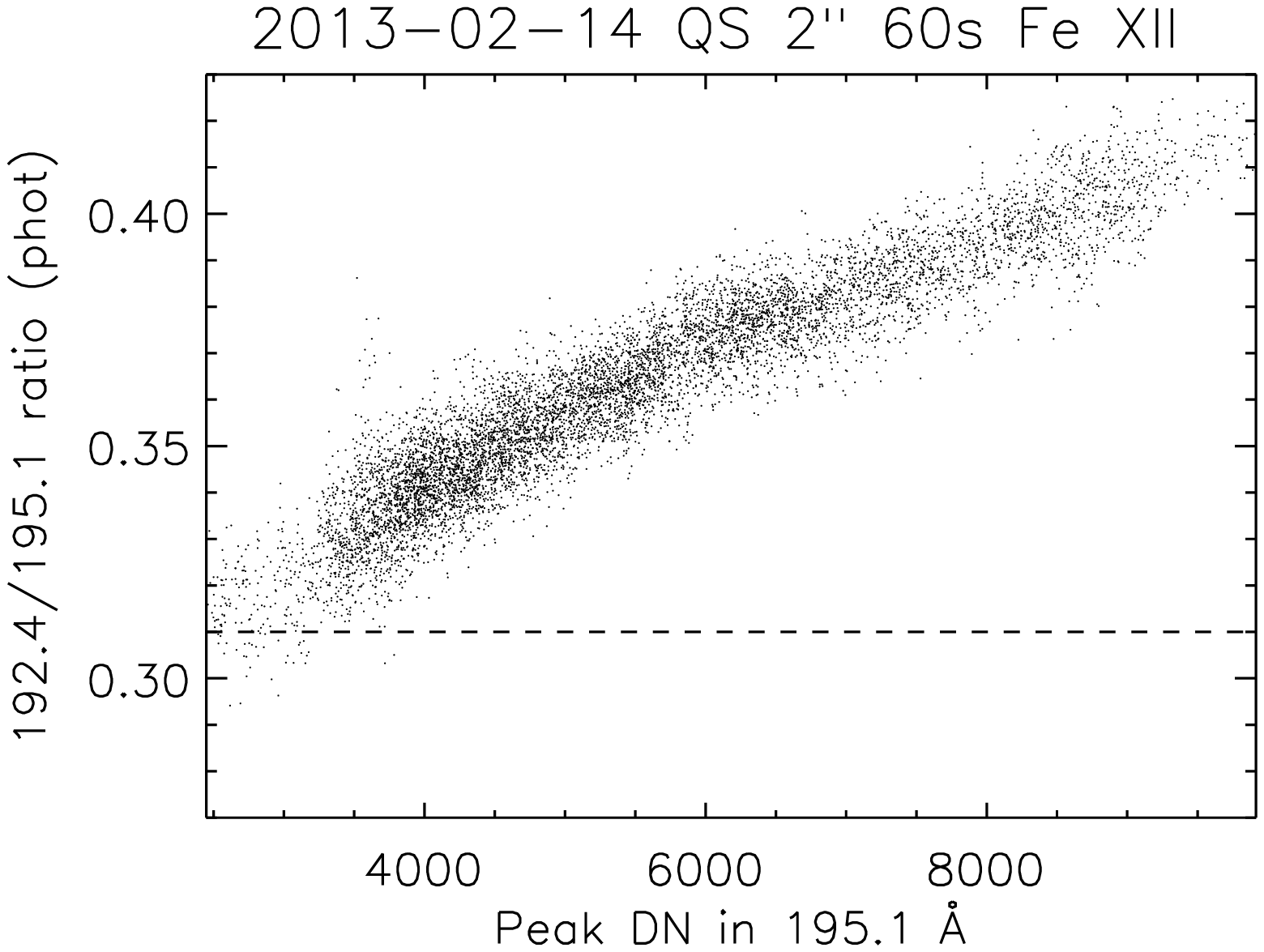}
}
\caption{Quiet Sun East limb observation on 2013-02-14 at 09:49 UT.
The top images show the 
peak intensity of the 195~\AA\ line (DN),  the observed FWHM (\AA) of the 192~\AA\ line,
and the intensity ratio of the 192 and 195~\AA\ lines.
The coordinates are the pixel positions. The arrows indicate two regions,
off-limb (OL) and on-disk (OD) chosen for further studies. 
The lower plots show 
scatter plots of the FWHM in the 192 and 195~\AA\ lines, as a function of their 
peak intensity, and the intensity ratio of the 192 vs. 195~\AA\  lines.
The dashed line indicate the expected value.
}
\label{fig:2013-02-14a}
\end{figure*}

{
The departures of the line ratios show a clear correlation with the 
 intensities of the lines. As shown with some examples in the Appendix, 
we have found such departures in both 1 and 2\arcsec\ slit spectra, in on-disk and off-limb
spectra. 
These departures  are normally larger (of the order of 30--40\%) 
 in  observations  where the peak intensities
are above 2000 DN, either in quiet Sun or in active region observations, 
as shown in the Appendix. 

These anomalous ratios have gone un-noticed in the literature,
but we have found them in earlier (from 2006) and later observations. 
The strongest departures occur where the lines are bright, independent
of the slit used, the pointing (e.g. on-disk, at the limb or off-limb),
and the source region (e.g. quiet Sun or active region).

\cite{young_etal:2009} reported an analysis of EIS \ion{Fe}{xii} 
line ratios to measure electron densities, but also showed 
in an Appendix the same \ion{Fe}{xii} intensity ratios discussed here.
They noted a variation of the ratios along the slit and suggested 
that perhaps this was due to a change in the instrument sensitivity along the 
slit. We reanalysed their 2007 May 6 observation and found that 
also in this case the 192 vs. 195~\AA\ intensity ratio varies 
with the intensity of the lines by about 30\%. Details are presented in the Appendix.

The anomalous behaviour in the intensity ratios is often associated with 
a broadening of the strongest lines, which in principle could be indicative either
of an instrumental non-linear effect in the line cores or opacity. 
However, this anomalous behaviour does not seem to be related 
to  a possible non-linearity in the EIS detector. 
We analysed  several  SYNOP1 studies, where two consecutive
exposures, of 30s and 90s,  are taken with the  1\arcsec\ slit. 
In  most cases  we found that the anomalous ratios were the same in both exposures. 
If the reason for the variation 
in the ratio were due mainly to a non-linearity in the detector, the ratios
should be different, as the  195~\AA\ line is three times stronger
in the 90s exposure.

We can also rule out any blending in the strongest lines as it  
would increase their signal, and not decrease it. 
One effect which would naturally cause these anomalous ratios 
is opacity. 
Optical depth effects in coronal lines have rarely 
been reported, although they are known to be present in 
low-temperature lines formed in high-density regions or close to the 
solar limb. 
}

\begin{figure}[!htbp]
\centerline{\includegraphics[width=6.cm,angle=90]{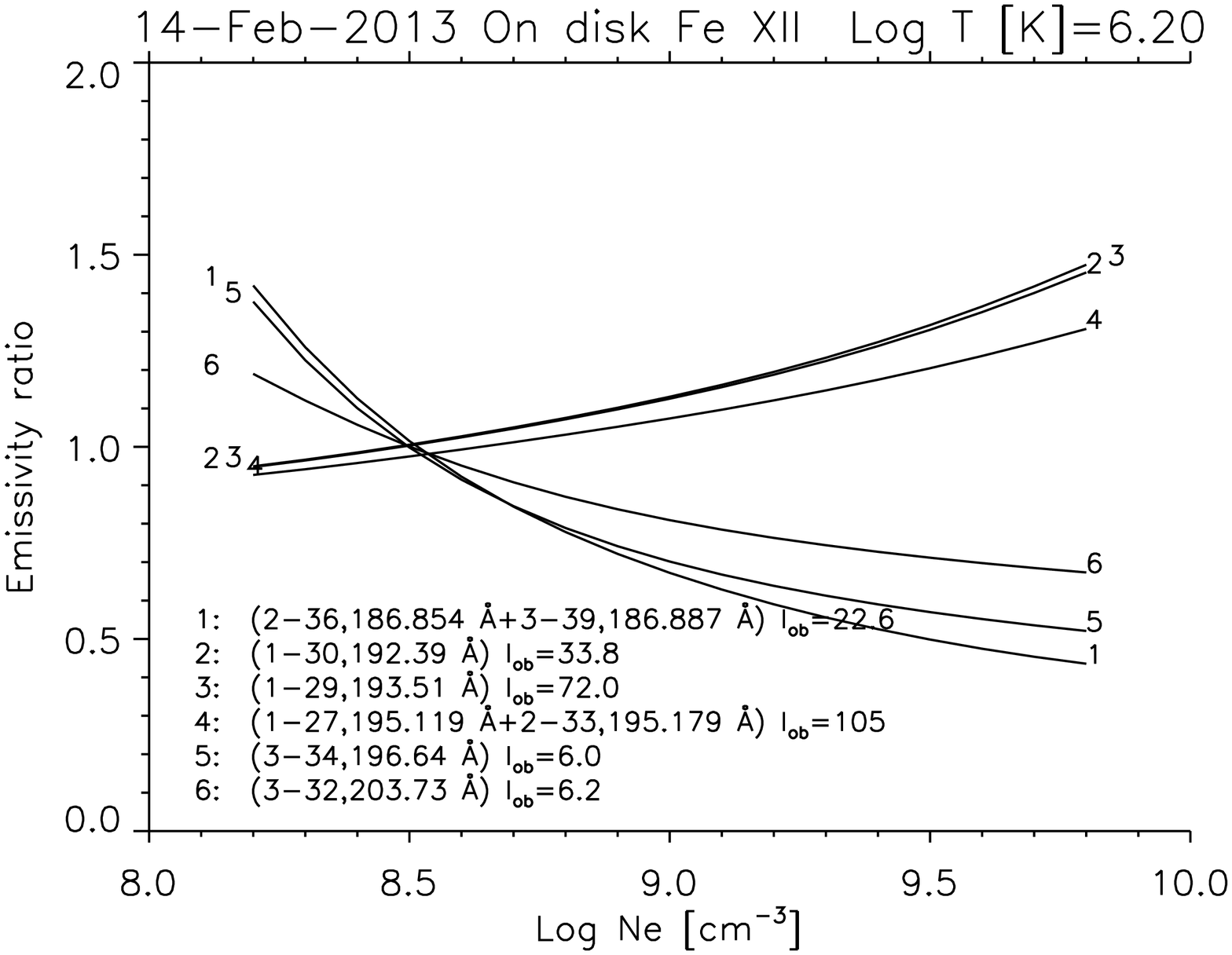}}
\centerline{\includegraphics[width=6.cm,angle=90]{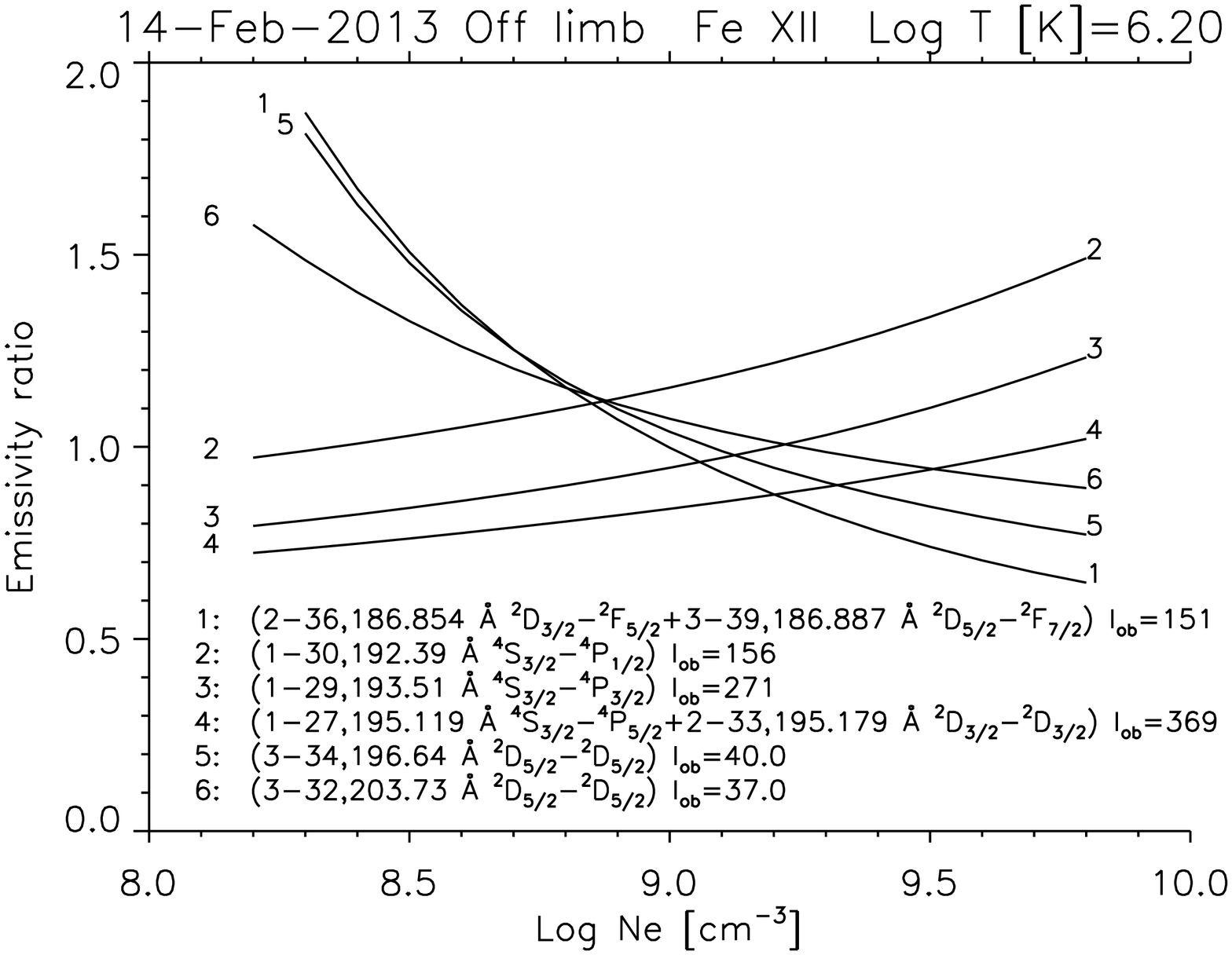}}
\caption{Emissivity ratio plots for the main \ion{Fe}{xii} lines in the 
on-disk (top) and off-limb (bottom) regions indicated in Fig.~\ref{fig:2013-02-14a}.
The calibrated intensities (in photons cm$^{-2}$ s$^{-1}$ arcsec$^{-2}$) 
are shown in the plots ($I_{\rm ob}$).
}
\label{fig:2013-02-14_er_fe_12}
\end{figure}

{
We have selected two regions in the EIS spectra, indicated with 
'OL' and 'OD' in   Fig.~\ref{fig:2013-02-14a}. We averaged spectra in these two regions,
and measured the line intensities.
Using the  \cite{delzanna:13_eis_calib} calibration, we find excellent agreement
(within a few percent)
among the intensities of the \ion{Fe}{xii} lines in the on-disk case, as shown in 
Fig.~\ref{fig:2013-02-14_er_fe_12} (top).
The figure shows the emissivity ratio plots (cf \citealt{delzanna_mason:2018}), wherby 
the intensities of the lines are divided by their emissivities. 
If the plasma along the line of sight is isodensity, the curves 
should provide a crossing.  The crossing indicated for the on-disk quiet Sun an
electron density of 10$^{8.5}$ cm$^{-3}$. 
For the off-limb case, the weaker lines indicate a far higher density
of  10$^{8.8}$ cm$^{-3}$, as one would expect as this is a much brighter region.
The anomaly in the 193.5 and 195.1~\AA\ lines is obvious. 

We then performed a DEM analysis of the off-limb region using 
CHIANTI v.8 \citep{dere_etal:97,delzanna_chianti_v8} and a 
selection of lines, see Fig.~\ref{fig:2013-02-14_ol_hr_dem}.
The peak emission is close to the temperature of formation of \ion{Fe}{xii}
in ionisation equilibrium, and there is very little emission at higher 
temperatures. 
The Sulphur and Iron lines indicate photospheric abundances 
\citep{asplund_etal:2009}. 

Integrating this DEM distribution, we obtain a total column emission measure EM.
 Assuming a uniform distribution of the 
densities along the line of sight (filling factor unity), and the measured density 
(10$^{8.5}$ cm$^{-3}$), we obtain a path length of 2 $\times$10$^{10}$ cm.
As the absolute calibration of the EIS instrument appears correct 
within say 20\%  \citep{delzanna:13_eis_calib}, the main uncertainty in this estimate 
is the absolute value of the Iron abundance. If its value was e.g. higher
by a factor of 2, the EM would be a factor of 2 lower, and the path length 
would be 10$^{10}$ cm.
}

\begin{figure}[!htbp]
\centerline{\includegraphics[width=7.cm,angle=90]{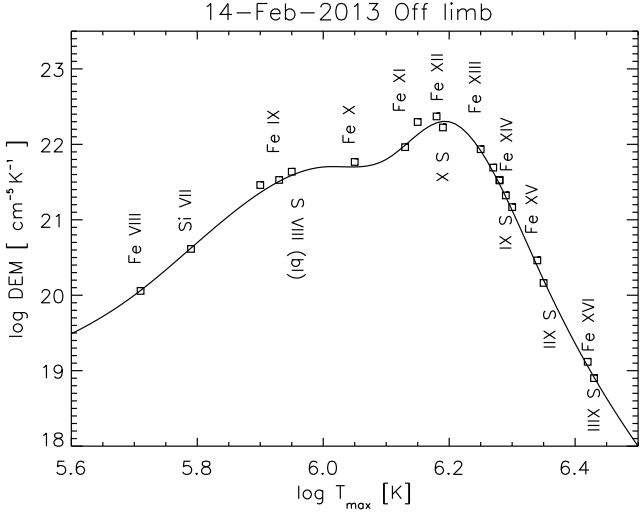}}
\caption{DEM distribution for the off-limb region 
indicated in Fig.~\ref{fig:2013-02-14a}. The points indicate the ratio of the predicted
vs. observed radiance, multiplied by the DEM value at
the temperature of maximum emissivity in each line.
}
\label{fig:2013-02-14_ol_hr_dem}
\end{figure}

The optical thickness at line centre for a spectral line is
\citep[cf.][]{delzanna_mason:2018}:
\begin{equation}
\tau_{\nu_0} = \int  k_{\nu_0}  N_l  \Delta S \;, 
\end{equation}
where $k_{\nu_0}$ is the absorption coefficient at line center (at frequency $\nu_0$),
$N_l$ is the number density of the lower level, and $\Delta S$ the path length.
If photons are scattered within a Doppler profile, i.e. with a Gaussian 
function in frequency, with a half-width $\Delta \nu_D$, the 
optical thickness at line centre is
\begin{equation}
\tau_{0} =  \frac{\pi^{3/2} \, {\rm e}^2} {m_{\rm e}\, c \, \Delta \nu_D} \; f_{lu} \, N_l \, \Delta S \;,
\end{equation}
where $f_{lu}$ is the absorption oscillator strength,  $c$ is the speed of light, 
$m_{\rm e}$ the electron mass.
In terms of wavelengths,  the Doppler width of a  line, equivalent to the 
 half-width $\Delta \nu_D$, is $\Delta \lambda_D = \frac{\Delta \lambda_{FWHM}}{\sqrt{4\, ln 2} }$,
where  $\Delta \lambda_{FWHM}$ is the measured FWHM of the line profile in wavelength. 
So the optical thickness at line centre can be written as
\begin{equation}
\tau_{0} = 8.3  \, 10^{-21} \, f_{lu}  \frac{\lambda^2}{\Delta \lambda_{FWHM}} \; N_l \, \Delta S 
\end{equation}
with $\lambda$ and $\Delta \lambda_{FWHM}$ expressed  in \AA.
For the 195~\AA\ line, $f_{lu}= 2.97/4$, neglecting the weaker line blending the 
main line. 

The  population of the lower level can be written as 
\begin{equation}
N_l = {N_l \over N({\rm Fe XII}) } \, {N({\rm Fe XII}) \over N({\rm Fe})} \,  Ab({\rm Fe})  \,
\frac{N_\mathrm{H}}{N_\mathrm{e}}  \, N_\mathrm{e} \;,
\end{equation}
where $N_l / N({\rm Fe XII})$ = 0.9 is the relative population of the ground state at 
quiet Sun densities,  
${N({\rm Fe XII}) / N({\rm Fe})}$ = 0.25 is the peak relative population of the ion,
$ Ab({\rm Fe}) = 3.16 \, 10^{-5}$ is the Fe photospheric abundance, 
$N_\mathrm{H} /N_\mathrm{e} = 0.83$, and $N_\mathrm{e}$ is the averaged 
electron number density.
We therefore have $\tau_{0} = 6.9 \, 10^{-20} \,  N_\mathrm{e} \; \Delta S $  [cm$^{-2}$] 
assuming $\Delta \lambda_{FWHM}=$0.02~\AA. 

{
With the measured 
averaged density of $7\, 10^{8}$ cm$^{-3}$, and assuming the path length obtained from the 
emission measure, we obtain $\tau_{0} = 0.97$, i.e. a significant  optical thickness at line centre
for the stronger line.
Note that the 192.4~\AA\ line has an oscillator strength a third of the 
195.1~\AA\ line ($f_{lu}= 0.98/4$), so has a smaller opacity at line centre, 0.32.
}

{

To estimate how an opacity at line centre of 1 for the 195.1~\AA\ line 
could  affect the 195/192~\AA\ intensity ratio, 
we assume that the source function $ S_\nu(\tau_\nu)$ does not vary along the line of sight, 
and take  averaged values of all the quantities. 
Of course $ S_\nu(\tau_\nu)$ would depend at each location on 
the local angle-averaged intensity $J_\nu$, even if the plasma is uniformly 
distributed in density. 
Note that this simple approach is similar to the use of averaged escape factors, while
a proper analysis would require a model of the distribution of the density.

With this assumption, the peak intensity of each line is 
\beq
I_\nu =  S_\nu \, \left(1- e^{-\tau_0} \right) \;, 
\eeq
while the the line source function $S_\nu$ is:
\beq
S_{\nu} = {2\,h\, \nu^3 \over c^2} \; \left( {g_u N_l \over g_l N_u} -1 \right)^{-1} \; ,
\eeq
with standard notation. If we calculate, assuming optically thin conditions, the 
relative populations of the two upper levels producing the  195, 192~\AA\ lines, we find that 
$S_{195} / S_{192} = 1.04$. The $I_{195} / I_{192}$ ratio of the peak intensities of the two lines 
for negligible opacitiy is 3.14. For $\tau_0 (195) =1$ we have 
$I_{195} / I_{192} = 2.34$, i.e. we obtain a 25\% reduction in the intensity ratio,
which is similar to the observed values.

Simlar arguments apply to active region observations, as briefly discussed in the 
Appendix.

}

\section{Conclusions}

The  quiet Sun off-limb observations we have analysed  do not 
indicate any significant variation in the widths of the coronal lines, up to 
1.5~\rsun. Some variations are present, but they are within 
the uncertainties of these very difficult measurements, so we cannot 
confirm that there is any  damping of Alfv\'en waves in the lower quiescent corona.

Our results, which extend out to 1.5~\rsun, confirm  those obtained from  SoHO CDS and 
SUMER  in quiet Sun areas where no significant changes in excess line widths were found.
We look forward to the future measurements in the forbidden lines 
which will be obtained by  DKIST. They will  have a much greater accuracy  
in line width measurements, and will provide a range of diagnostics 
\citep[see e.g. ][]{delzanna_deluca:2018}.

We have carried out an analysis of several EIS datasets and found 
that the widths of the two strongest \ion{Fe}{xii} lines 
are consistently larger than expected, and we suggest that this is an 
instrumental issue. We point out that these lines have been widely 
used in the literature to measure excess widths. 

{
We found that the ratios of the  \ion{Fe}{xii} lines are often anomalous,
and suggest that the two strongest lines are affected by opacity in 
most off-limb and active region observations. 
}
Opacity effects are very common in transition region lines, but have 
rarely been reported in coronal lines. There were several papers discussing the 
possibility of scattering of the strongest \ion{Fe}{xvii} line at 
15~\AA\ in  X-ray observations of active regions \citep[see, e.g.][]{schmelz_etal:1997}.
However, most of the discrepancies were due to atomic data and/or  
blending \citep[see][]{delzanna:2011_fe_17}.
\cite{kastner_bhatia:2001} report an analysis of opacity in the resonance 
 \ion{Fe}{xv} line at 284.16~\AA, and suggested, using line ratios, that the line
is affected by opacity in some active region observations. 

{
The fact that  \ion{Fe}{xii} lines appear to be affected by opacity 
in many observations has several implications.
On a positive note,
the ability to measure the optical thickness, in conjunction with 
an independent measurement of the electron density via line ratios
(using weaker lines or other ions), gives
us the possibility to estimate the path lengths of the sources.
We will discuss this issue in a separate paper.  

On a negative note, these opacity effects and anomalous behaviour of the two 
strongest  \ion{Fe}{xii} lines most likely affect previously
published results.
We have also found evidence that other strong lines from other ions 
are also affected by opacity, so previous results 
 based on widths and intensities (e.g. densities, emission measures)
of the strongest lines, especially in 
active regions and flares should be revisited. 
}


\begin{acknowledgements}
GDZ and HEM  acknowledge support  by  STFC (UK) via a 
consolidated grant to the solar/atomic physics group at DAMTP, University of
Cambridge. \\

GRG  acknowledges support from the UK Commonwealth Scholarship Commission via
a Rutherford Fellowship. \\

We warmly thank the effort of all the EIS team members involved in the 
planning of  EIS  observations with  the engineering study.\\

 Hinode is a Japanese mission developed and launched by ISAS/JAXA, 
with NAOJ as domestic partner and
NASA and STFC (UK) as international partners. It is operated by these agencies in
co-operation with ESA and NSC (Norway). 
 CHIANTI is a collaborative project involving the University of Cambridge (UK),
George Mason University, and the University of Michigan (USA). 

\end{acknowledgements}


\bibliographystyle{aa}

\bibliography{../bib}  

\appendix

\section{Single 1\arcsec\ slit observations (SYNOP1) with different exposure times}

\begin{figure*}[!ht]
\centerline{\includegraphics[width=17.0cm,angle=0]{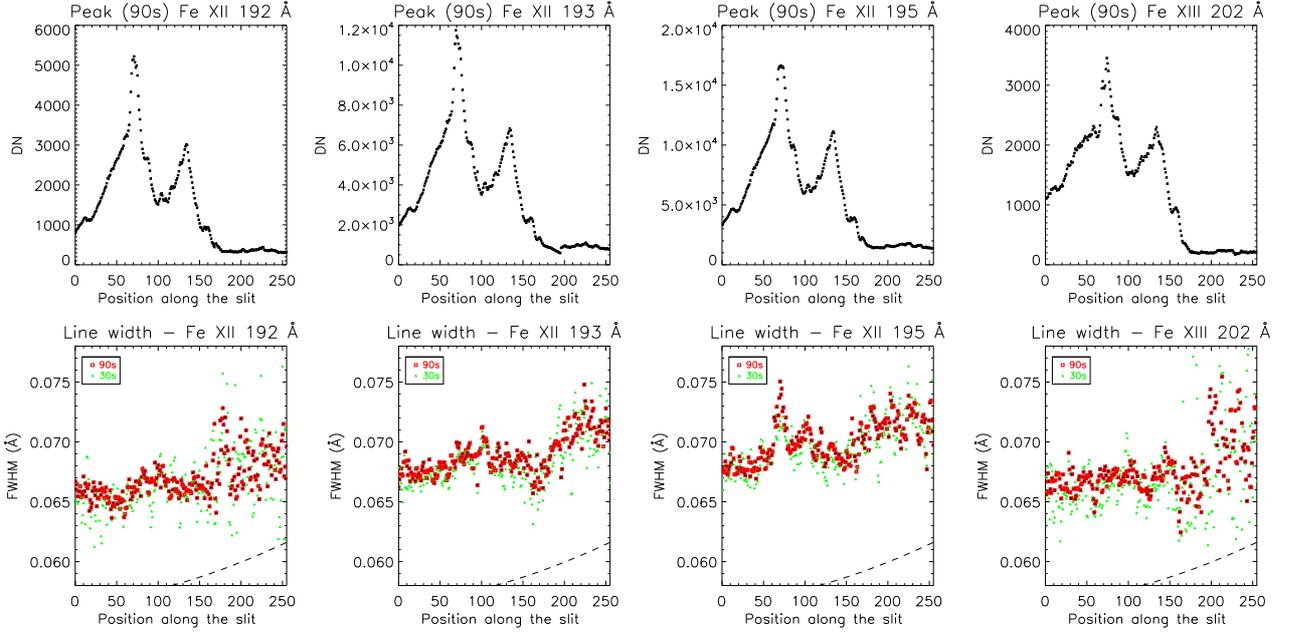}}
\caption{SYNOP1 single-slit observations on 2007-01-06 at 15:47 UT
across an active region.
Top row:  Peak DN  along the slit, from the 90s exposures.
Bottom row: line widths  (FWHM) from the  30 and 90s exposures. The dashed line 
indicate the instrumental FWHM as estimated within the EIS software.
}
\label{fig:synop1a}
\end{figure*}

 \begin{figure}[!htbp]
\centerline{\includegraphics[width=6.5cm,angle=0]{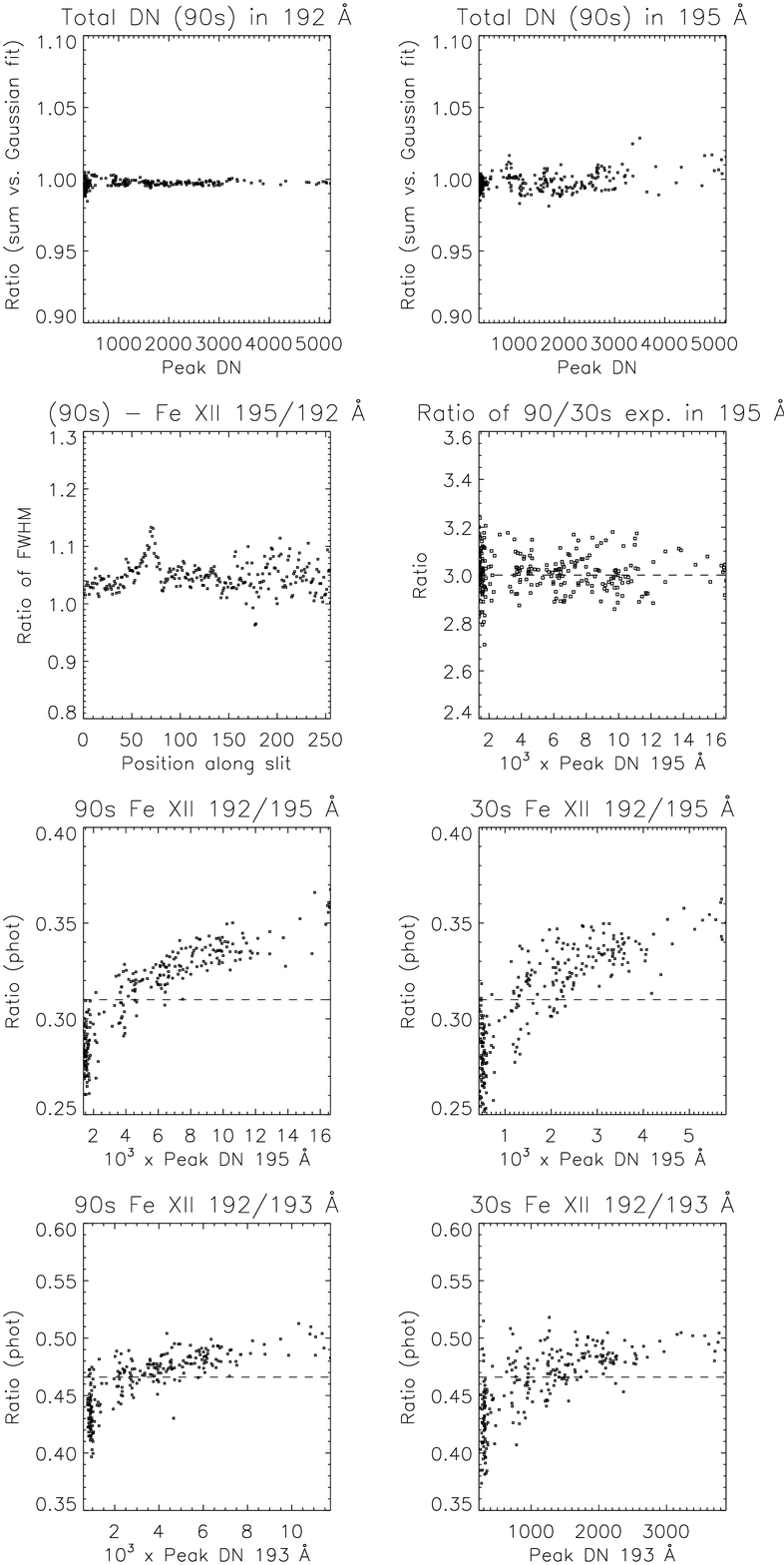}}
\caption{SYNOP1 single-slit observations on 2007-01-06 at 15:47 UT
across an active region.
From top to bottom: ratios of total counts in the 192 and 195~\AA\ lines,
as obtained by summing the pixel values and the Gaussian fit;
ratio of the FWHM in the 195 vs. the 192~\AA\ lines; 
intensity ratio in the 195~\AA\ line, between the 90s and 30s exposures;
 192/195~\AA\ and 192/193~\AA\ line  intensity ratios from the 90s 
and 30s exposures (in photons, dashed lines indicates expected values).
}
\label{fig:synop1b}
\end{figure}

{
We analysed several SYNOP1 studies
(taken on 2006 Dec 23,25,26,29, 2007 Jan  5,6,7,24, 2007 Feb 1, 2007 Jun 8, 9, 
2007 Aug 7, 2008 Jan 7). Most observations were carried out at Sun centre to monitor the degradation 
of the instrument. 
In the SYNOP1 studies,   two consecutive
exposures, of 30s and 90s,  are taken with the  1\arcsec\ slit. }
Having different exposures is useful to assess the uncertainties in 
measuring the line widths and if the anomalous Fe XII ratios could be associated with 
a non-linear behaviour of the instrument. 
 Fig.~\ref{fig:synop1a} shows as one example the intensities and 
line widths in a selection of lines in  one observation across an AR. 
Only the strongest 195~\AA\  line is saturated in the 90s exposure, in one location.
This line shows some increased widths close to 
the saturation area in the 90s exposures, compared to the 30s
one. 
The 195~\AA\ line is always broader than the 192~\AA\ line, as is the 193 one.
The nearby \ion{Fe}{xiii} 202~\AA\ line usually has the same 
width as the 192.4~\AA\ line, as shown in  Fig.~\ref{fig:synop1a} and as
we have seen previously.

The total intensity of the 195~\AA\  line in the 90s exposure is nearly three times the signal in the 30s
exposure (cf Fig.~\ref{fig:synop1b}), meaning that little variability occurred between the two exposures.
This is important, as we would then expect that also line ratios and widths
to be the same in the two exposures.

EIS profiles are typically Gaussian. Indeed the first two plots in  Fig.~\ref{fig:synop1b} 
show that the intensities obtained from fitting a Gaussian are within a few percent
those obtained by summing the intensities across the whole line profile 
(+/- one FWHM). We found similar results in all observations. 

We note that the standard way to estimate uncertainties on the 
parameters of a Gaussian typically produces very small uncertainties
in the widths. 
 The second row of Fig.~\ref{fig:synop1a} 
shows that there is a consistency of the widths in the two exposures, 
until when lines have peak values lower than about 200 DN.
In this case line widths become very uncertain, with a typical scatter of 5 m\AA.

The  \ion{Fe}{xii} intensity ratios 
become anomalous, as they vary with the intensity of the lines. 
The 192 vs. 195~\AA\ ratio varies by almost 40\%. 
Blending cannot explain this  behaviour: 
any blending in the two stronger lines would make the ratios vary in the opposite sense.
The 195~\AA\ line should increase its intensity in the active region, as the 
density-sensitive line in the self blend should become measurable. 
Instead, the intensity of the 195~\AA\ line, compared to that of the 
192~\AA\ line, decreases.
The important point is that the anomalous ratios are similar in both the
30s and 90s exposures. If the reason for the variation 
in the ratio were due mainly to a non-linearity in the detector, the ratios
should be different, as the  195~\AA\ line is three times stronger
in the 90s exposure. 
We have seen this behaviour in all observations analysed.

Fig.~\ref{fig:synop1_1} shows another SYNOP1  example, indicating very little variability in 
the  \ion{Fe}{xii} intensity ratios. However, the instrumental effect of the 
larger widths in the two stronger  \ion{Fe}{xii} lines is always present.

\begin{figure*}[!htbp]
\centerline{\includegraphics[width=17.5cm,angle=0]{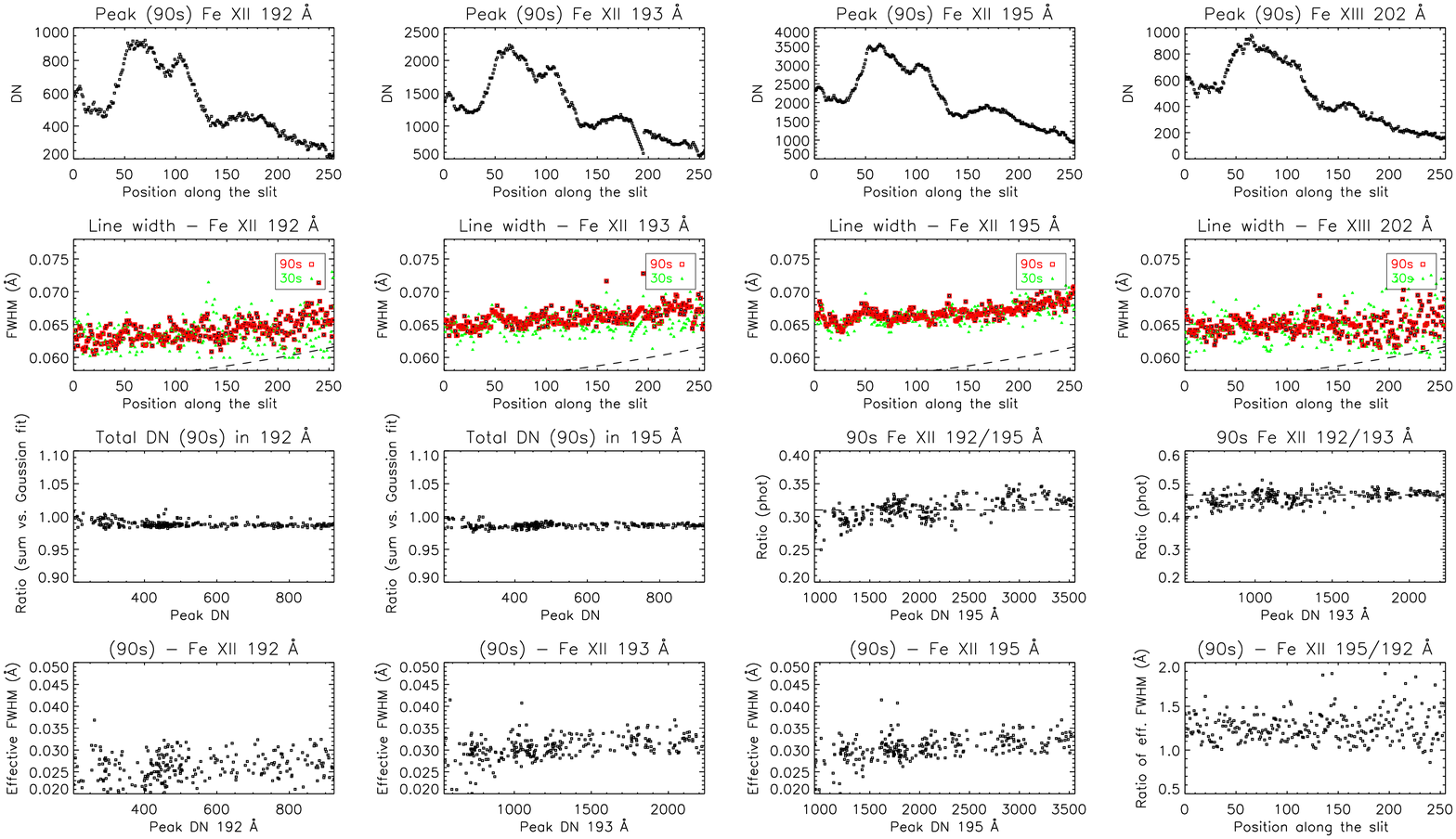}}
\caption{Quiet Sun SYNOP1 observations on 2007-06-09 at 06:12 UT. 
Top row:  Peak DN  along the slit, from the 90s exposures.
Second row: line widths  (FWHM) from the  30 and 90s exposures.
Note the larger scatter in the 30s exposures, for low DN, below 200.
 Dashed lines: instrumental FWHM as in the EIS software.
Third row: ratios of total counts in the 192 and 195~\AA\ lines,
as obtained by summing the pixel values and the Gaussian fit;
 192/195~\AA\ and 192/193~\AA\ line  intensity ratios from the 90s exposures (in photons, 
dashed lines indicates expected values);
Last row: effective FWHM in the three \ion{Fe}{xii} lines, and the ratio of the 
195 vs. the 192 widths.
} 
\label{fig:synop1_1}
\end{figure*}

\section{A possible EIS saturation example}

\begin{figure}[!htbp]
\centerline{\includegraphics[width=7.5cm,angle=0]{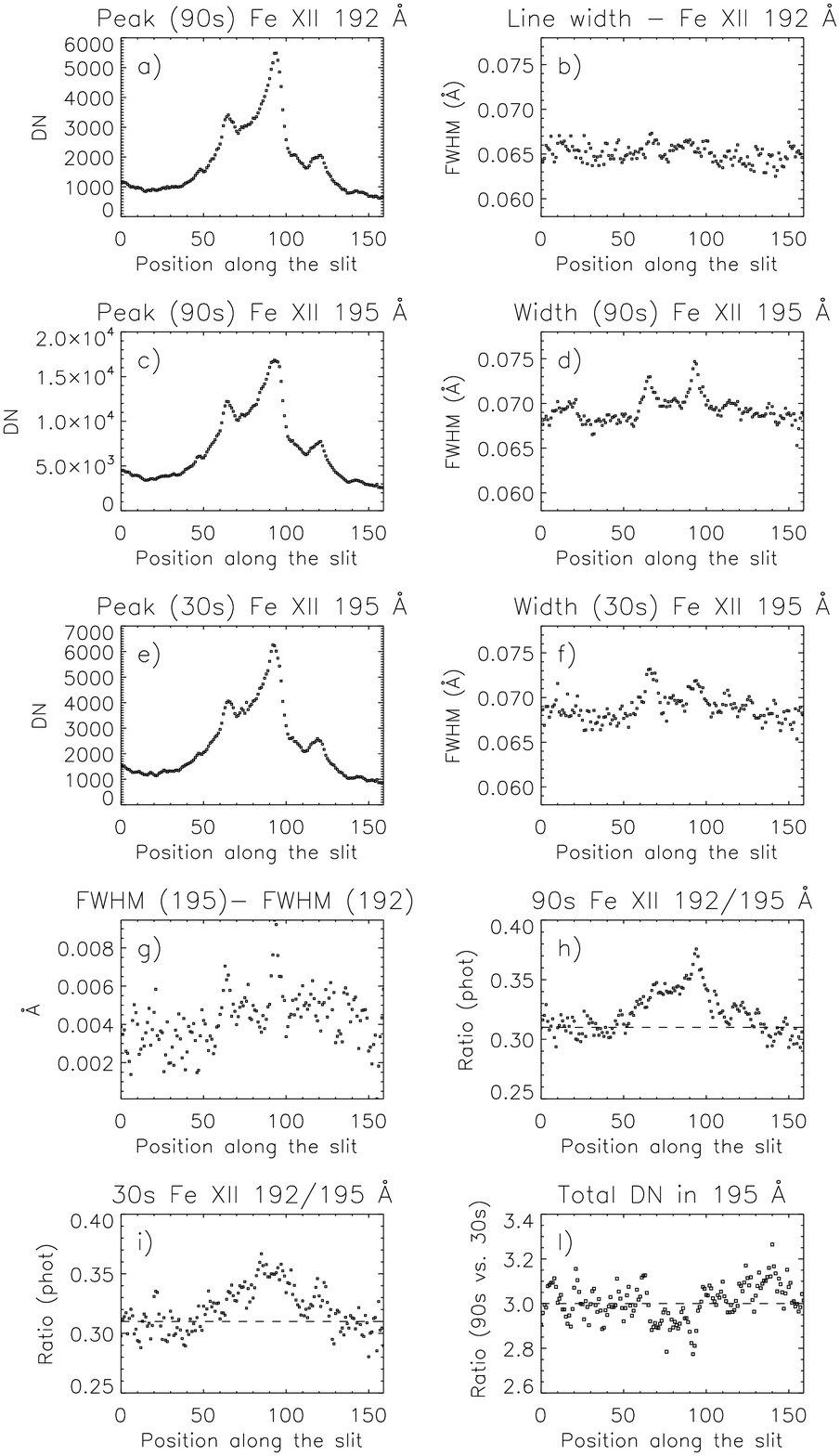}}
\caption{Active region SYNOP1 observations on 2006-12-29 at 11:15 UT. 
 Note that the 195~\AA\ line is saturated in one 
location, in the 90s exposure. Also note the significant variation in the 
 \ion{Fe}{xii} intensity ratios.
}
\label{fig:synop1_2}
\end{figure}

{

 Fig.~\ref{fig:synop1_2} shows one example  where a saturation 
problem appears  to be  present. A portion of the EIS  slit crossed an active region, around 
pixel coordinates 60--100, where counts peak. 
Panels a), b) and c) show the peak counts in the 
  192 and 195~\AA\ lines with the 90s exposures and the 
195~\AA\ line with the 30s exposure. 

The 192~\AA\ line does not show any 
significant variation in the width (panel b). On the other hand, the 
195~\AA\ line shows two strong increases in the width (panels d,f). 
These are the anomalous increases which are common in active region observations.

The second increase in the width around pixel 90 is more pronounced in the 90s exposure
(panel d), where the line reaches the EIS saturation limit,
compared to the increase in the 30s exposure.
This is a different issue and is related to a saturation problem.

The overall difference in the widths of the two lines is around 4 m~\AA\ 
(panel g) as we have seen in other cases. However, the difference increases
up to 9 m~\AA\ where the peak counts are (pixel 90).

Considering first the 30s exposures, where both lines are far from 
the EIS saturation, we see in panel i) that the ratio of the total 
intensities of the 192 and 195~\AA\ lines is increased near the peak
emission, from a value of about 0.3 to 0.35. Considering the 90s exposures,
the ratio is higher near the peak (panel h), an indication of a saturation
problem. Indeed, considering the total intensity of the 195~\AA\ line with the 
90s exposure, one would expect the ratios with the values obtained from the 30s
exposure to be 3, see panel l). However, the ratio consistently drops 
to lower values near the peak emission. This can only 
be an instrumental problem, unless the intensities changed
only in those locations between the two exposures. 
 The variation is not large though.

}

\section{Our 2008-11-19 campaign.}

\begin{figure*}[!htbp]
\centerline{\includegraphics[width=12.5cm,angle=90]{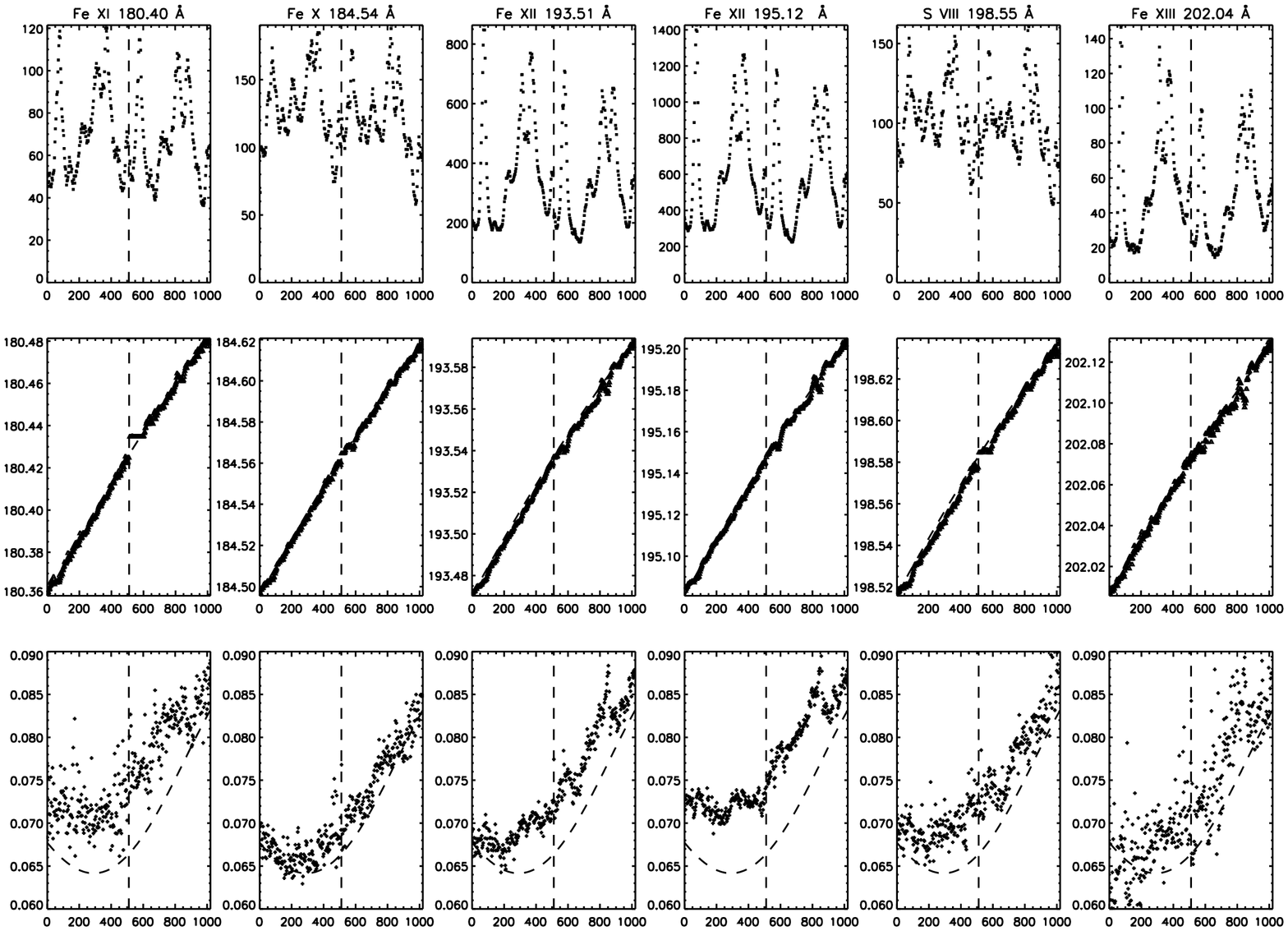}}
\caption{The top row shows the peak DN in a selection of SW lines, as a function of the 
position along the slit, for an average of 15  exposures
of 60s,  with the 2\arcsec\ slit on the  quiet 
Sun, recorded on 2008-11-19 with the EIS study GDZ\_QS1\_NS\_60x512. 
The dashed vertical line indicates the mid position
of the 1024 slit.  The bottom 512\arcsec\ were observed by a raster starting at 
11:38 UT, while the top 512\arcsec\ were observed by a raster starting at
12:13 UT. Note the small temporal variation in the peaks of the lines. 
The middle row shows  the line centroid positions; the dashed line indicates the 
estimated variation as available in the EIS software notes. 
The bottom row shows the FWHM in the lines, while the dashed line indicates the 
estimated variation as available in the EIS software notes.}
\label{fig:tilt_summary_20081119}
\end{figure*}

{
In Nov  2008, we ran a campaign to study the line width variations 
with the  2\arcsec\ slit, scanning a small region of the Sun (60\arcsec\ wide)  with 60s exposure times.
We processed all observations, but found only one set to be useful.
}

Fig.~\ref{fig:tilt_summary_20081119} shows the variations of the 
line positions and FWHM as observed on 2008-11-19 with the EIS study GDZ\_QS1\_NS\_60x512.
The same solar region was observed with the bottom 512 and the top 
512\arcsec\ of the slit, successively. This is clear by looking at the variations of the 
intensities in the lines in the two parts of the CCD, divided by the vertical dashed line in 
Fig.~\ref{fig:tilt_summary_20081119}.
This is important, and shows very little  solar variability. We therefore expect the 
solar line widths to be  the same. 
A small brightening was avoided, and 15 exposures 
were averaged. To further improve the signal, the spectra were rebinned
by a factor of 3 along the slit. 
The  variation of the line positions along the slit agrees 
very well with the values estimated by P.R. Young (shown as dashed lines),
 but the line widths still show some departures. It is worth noting very little 
change in the widths in the bottom part of the slit.

\clearpage

\section{On-disk quiet Sun}

We searched for EIS rasters 
with long exposures and with the full spectral range and the 1\arcsec\ slit
on the quiet Sun and found the first useful one was a study
HPW001\_FULLCCD\_RAST\_128x128\_90S\_SLIT1, i.e. with 90s exposures and 128 
slit positions.
In spite of the long exposures, the strongest 195~\AA\ line only reached a maximum 
of  5400 DN at line centre in a brightening near the limb, see Fig.~\ref{fig:2006-12-23}.
Considering the dimmest regions, there is no obvious trend in the 
 widths of the lines. However, the three lines have a consistently 
different width and the intensity ratios become  anomalous above about 1500 DN in
the brightest region, near the solar limb.
The  \ion{Fe}{xii} intensity ratios are close to their expected values on-disk.

\begin{figure*}[!htbp]
\centerline{\includegraphics[width=5.5cm,angle=0]{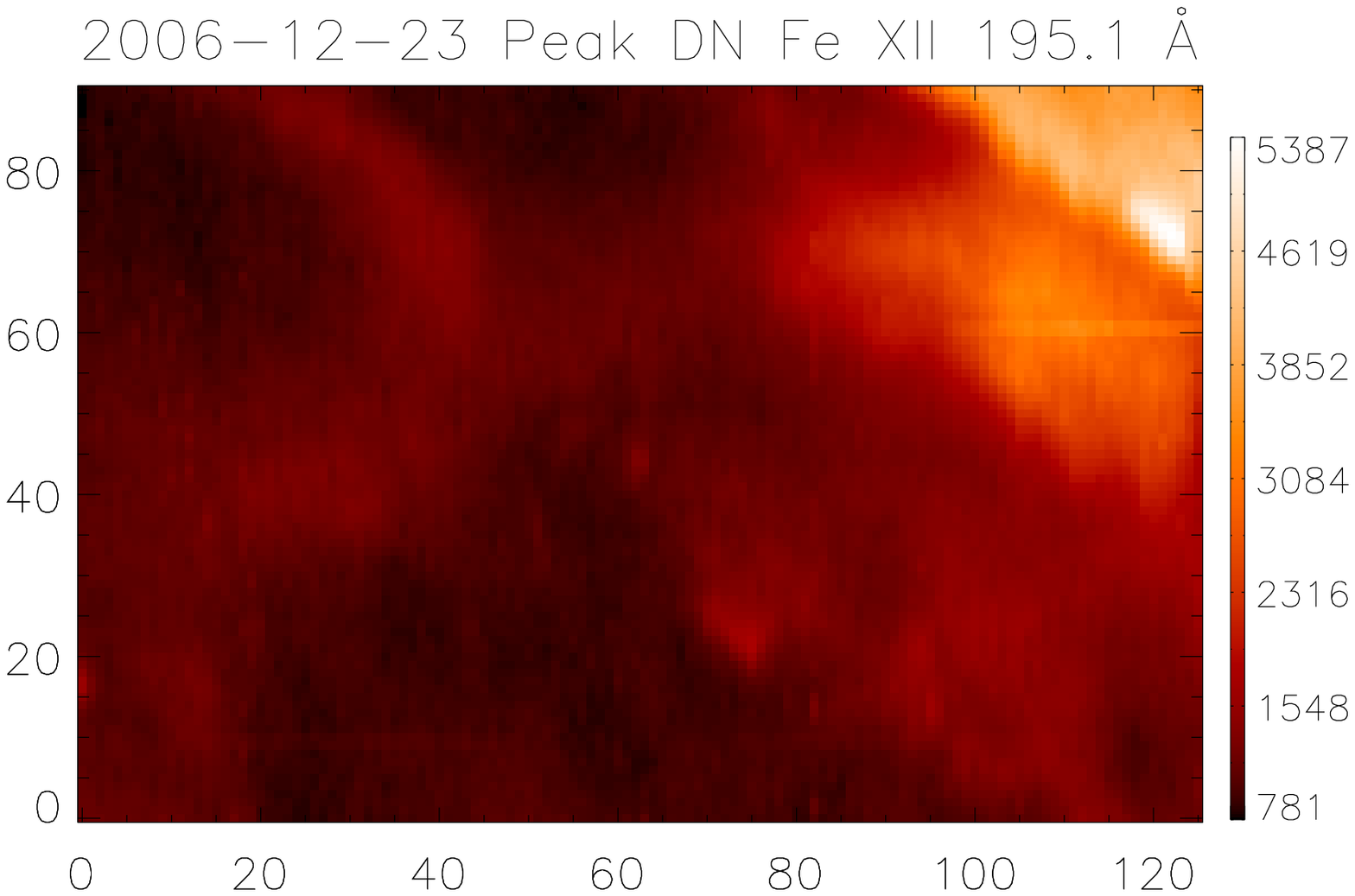}
\includegraphics[width=5.5cm,angle=0]{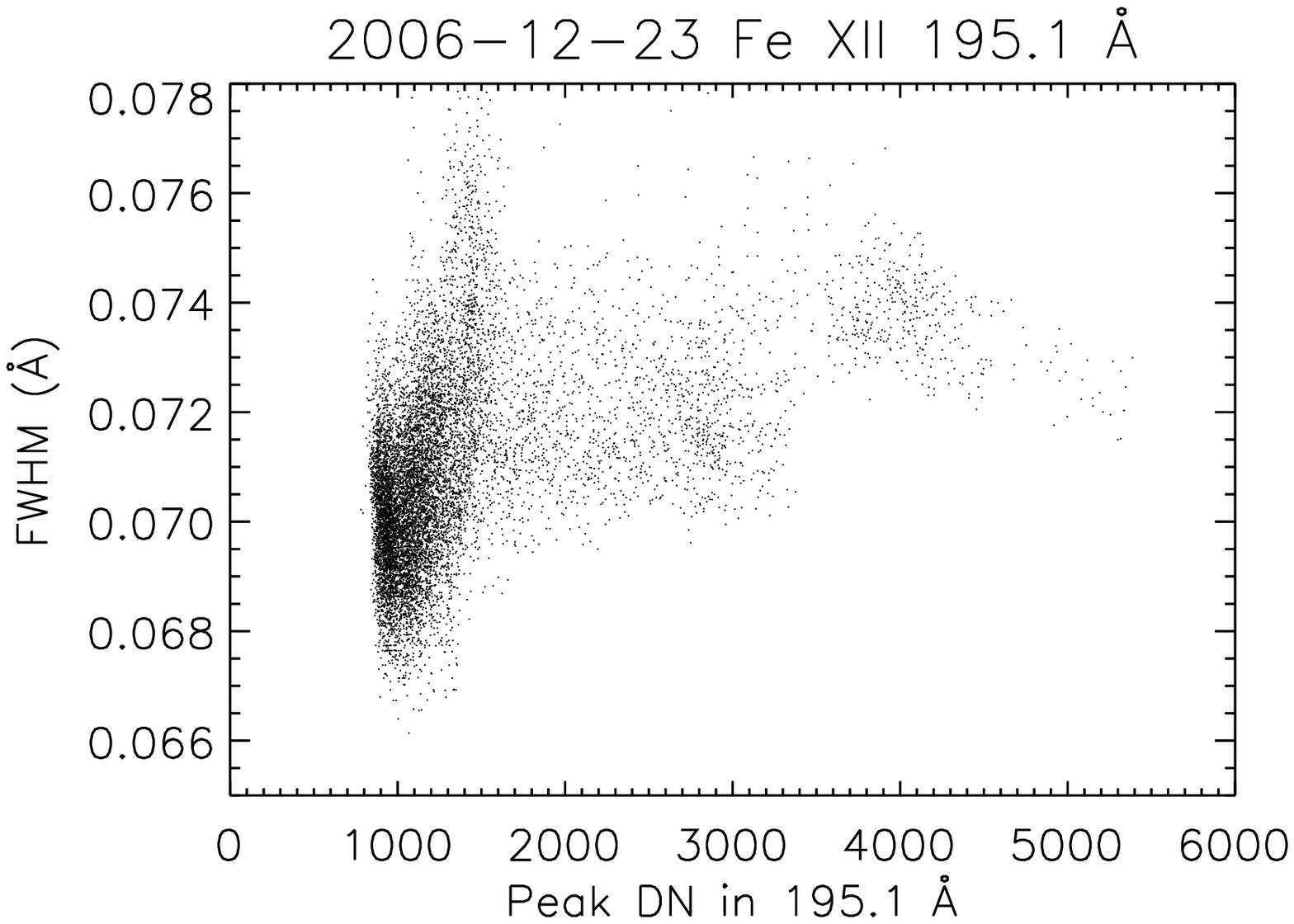}
\includegraphics[width=5.5cm,angle=0]{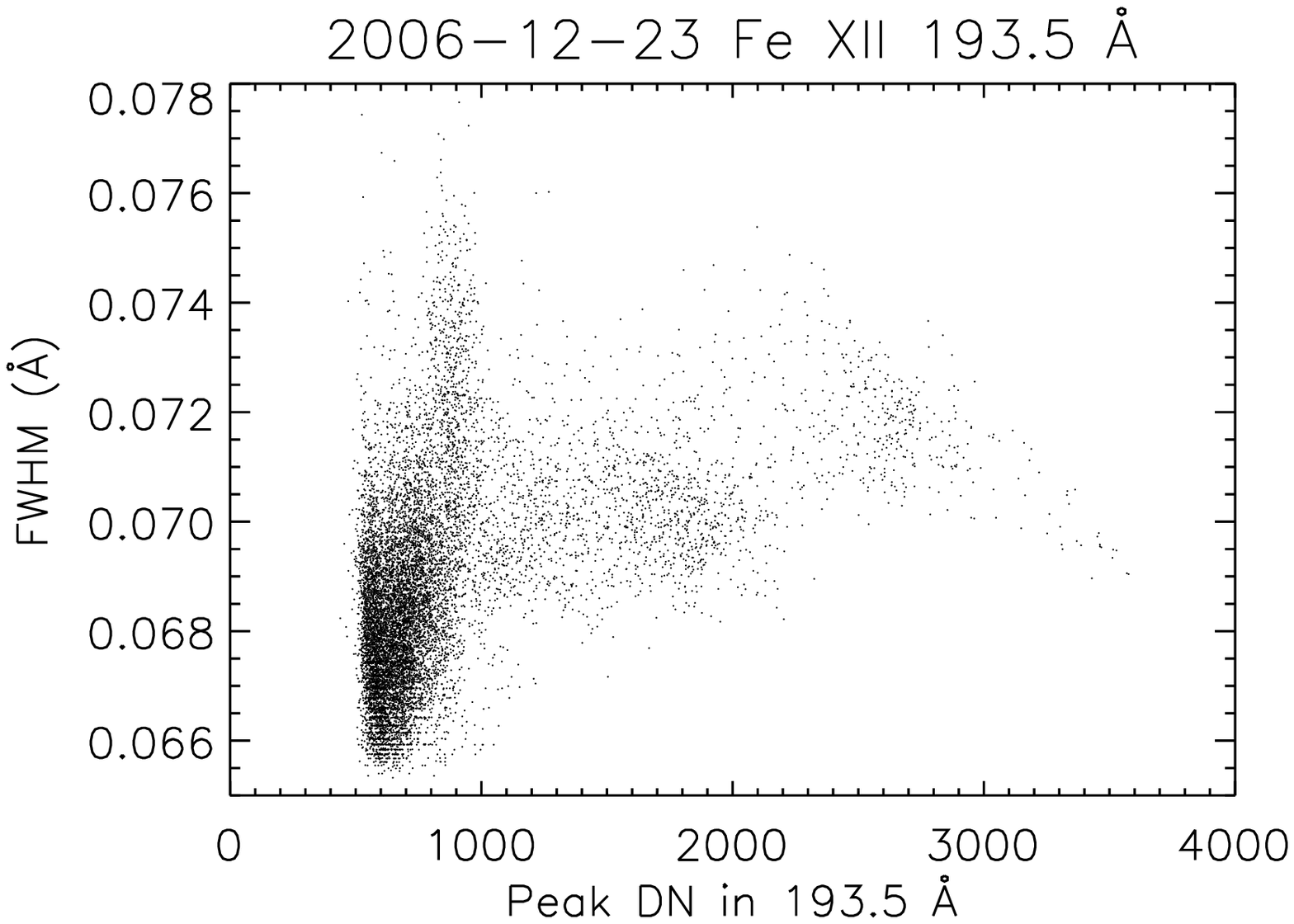}
}
\centerline{\includegraphics[width=5.5cm,angle=0]{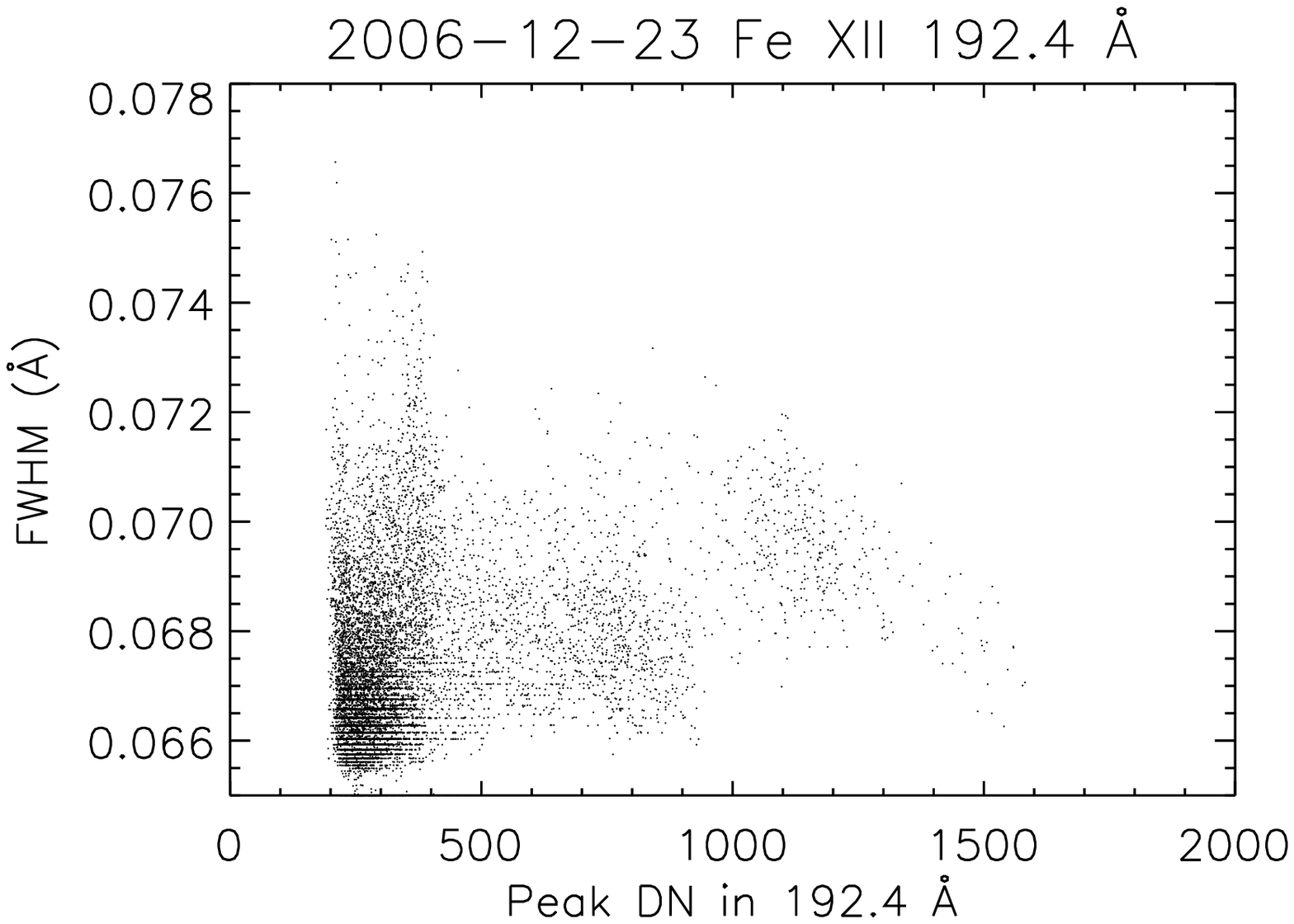}
\includegraphics[width=5.5cm,angle=0]{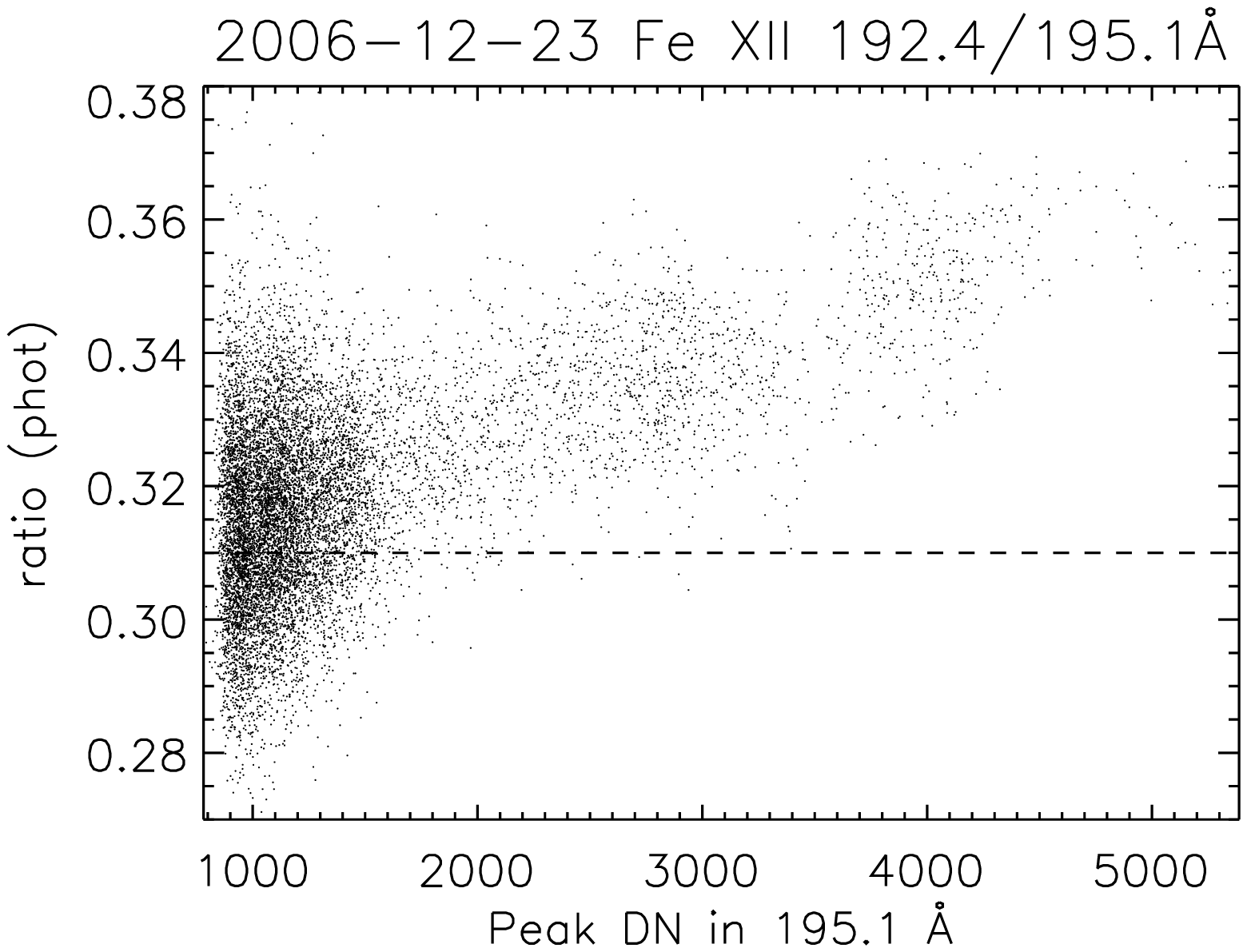}
\includegraphics[width=5.5cm,angle=0]{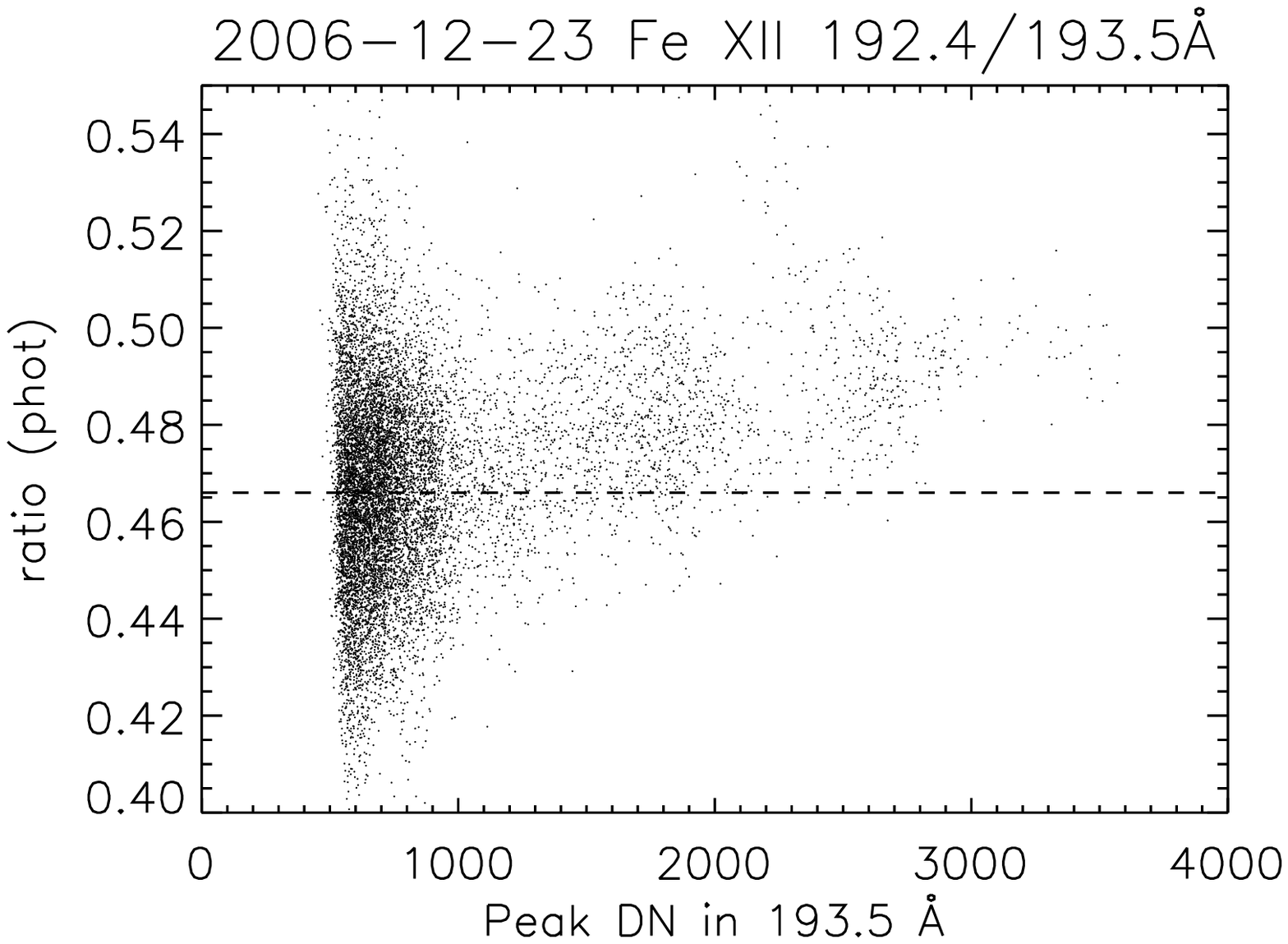}}
\caption{One of the earliest full-spectral EIS observations of the quiet Sun,
with the 1\arcsec\ slit on 2006-12-23. From left to right and top to bottom: 
the peak DN in the \ion{Fe}{xii} 195.1~\AA\ line; the FWHM of the three main 
\ion{Fe}{xii} lines, as a function of their peak intensities (DN);
the intensity ratios of the lines, as a function of the peak intensities
of the strongest line. The dashed lines indicate the expected values.
}
\label{fig:2006-12-23}
\end{figure*}


\begin{figure*}[!htbp]
\centerline{\includegraphics[width=5.5cm,angle=0]{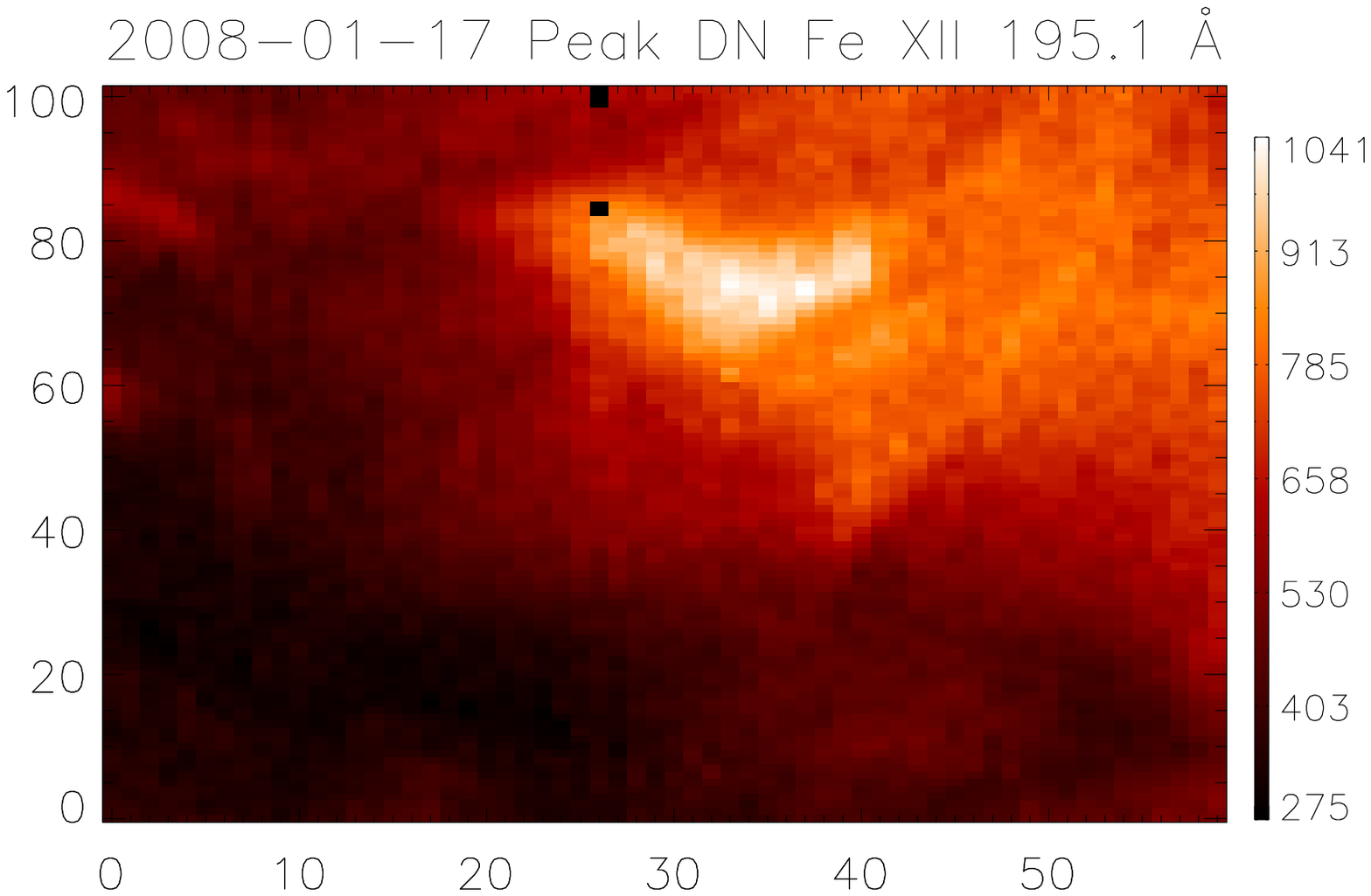}
\includegraphics[width=5.5cm,angle=0]{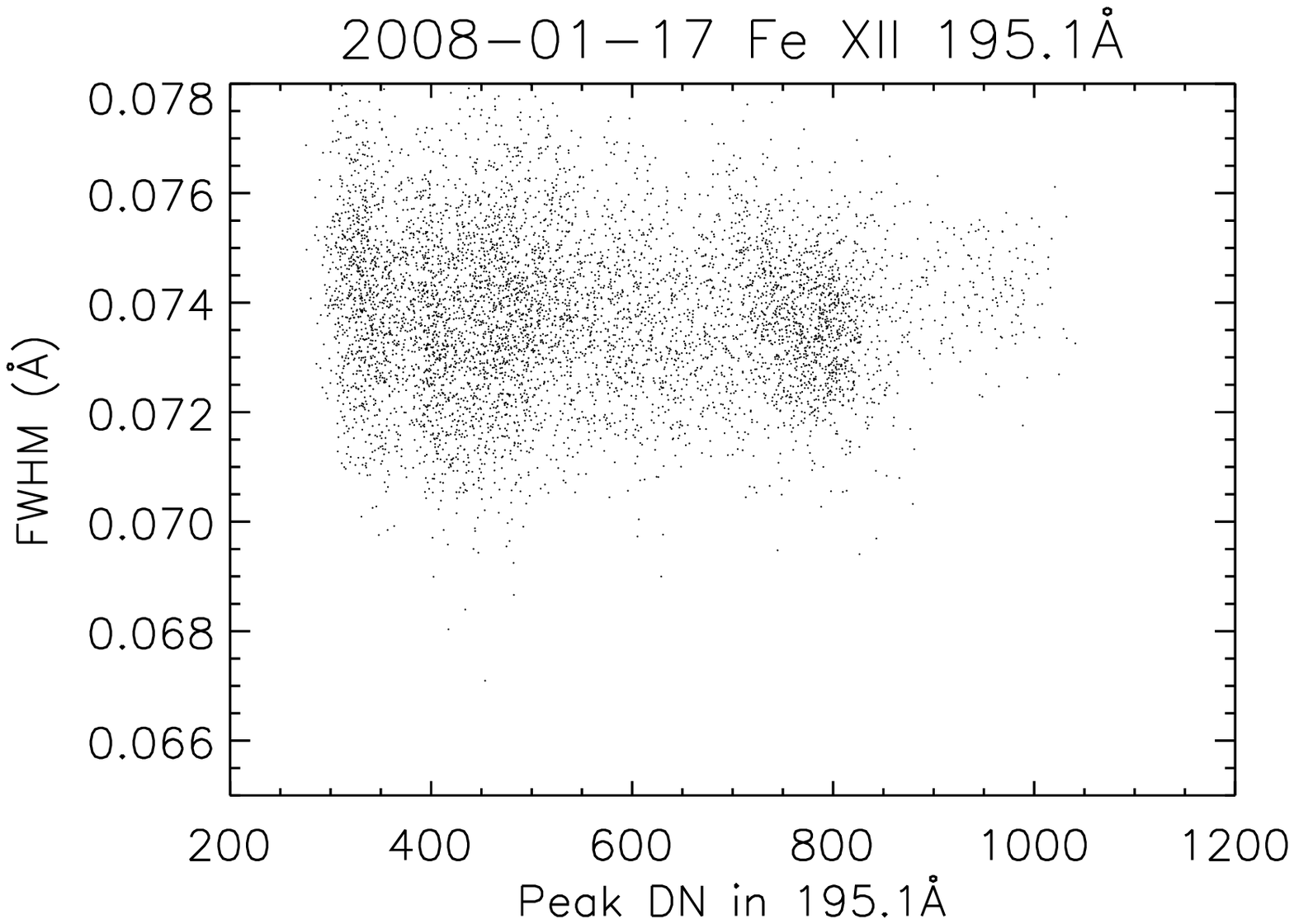}
\includegraphics[width=5.5cm,angle=0]{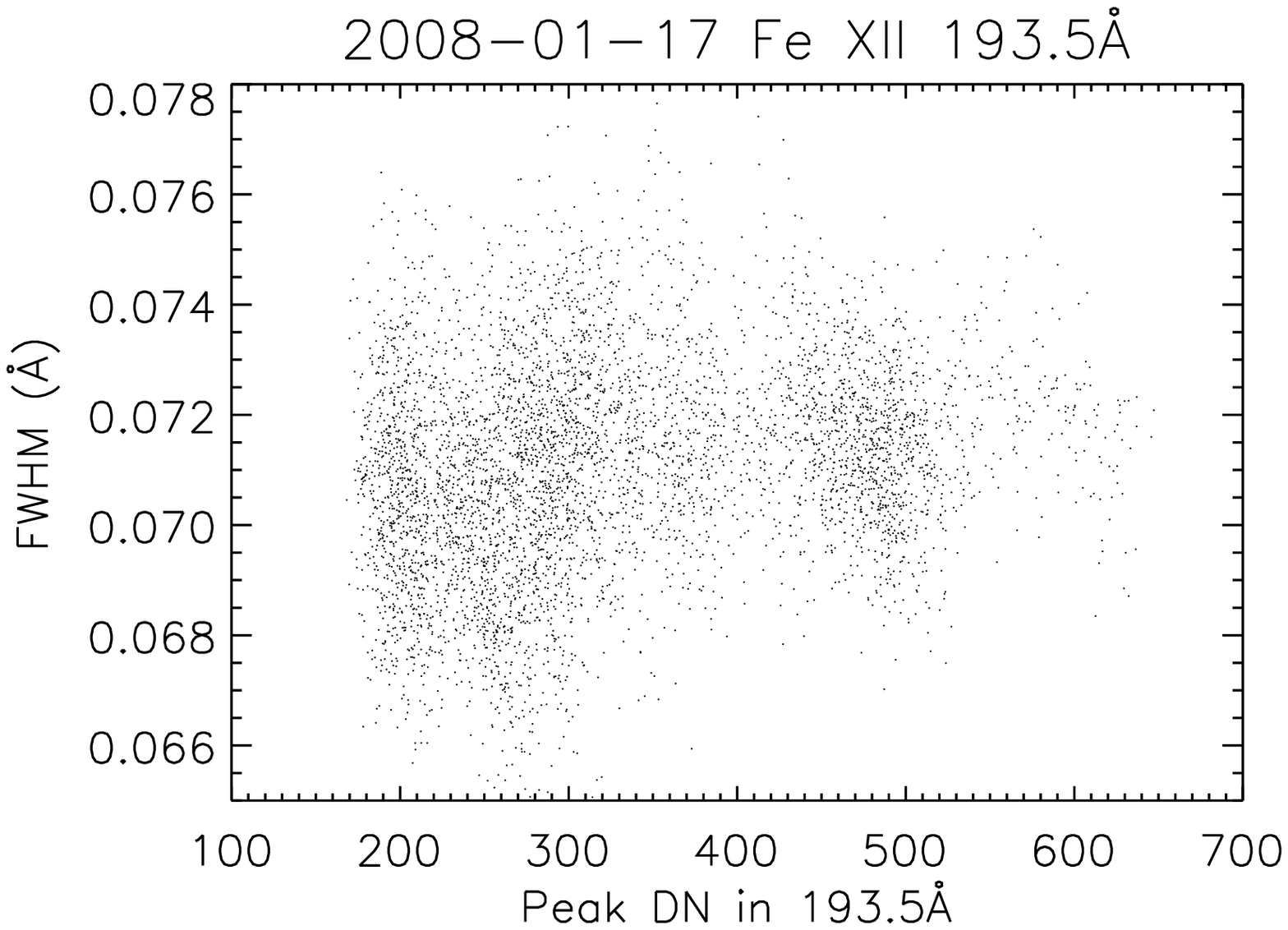}
}
\centerline{\includegraphics[width=5.5cm,angle=0]{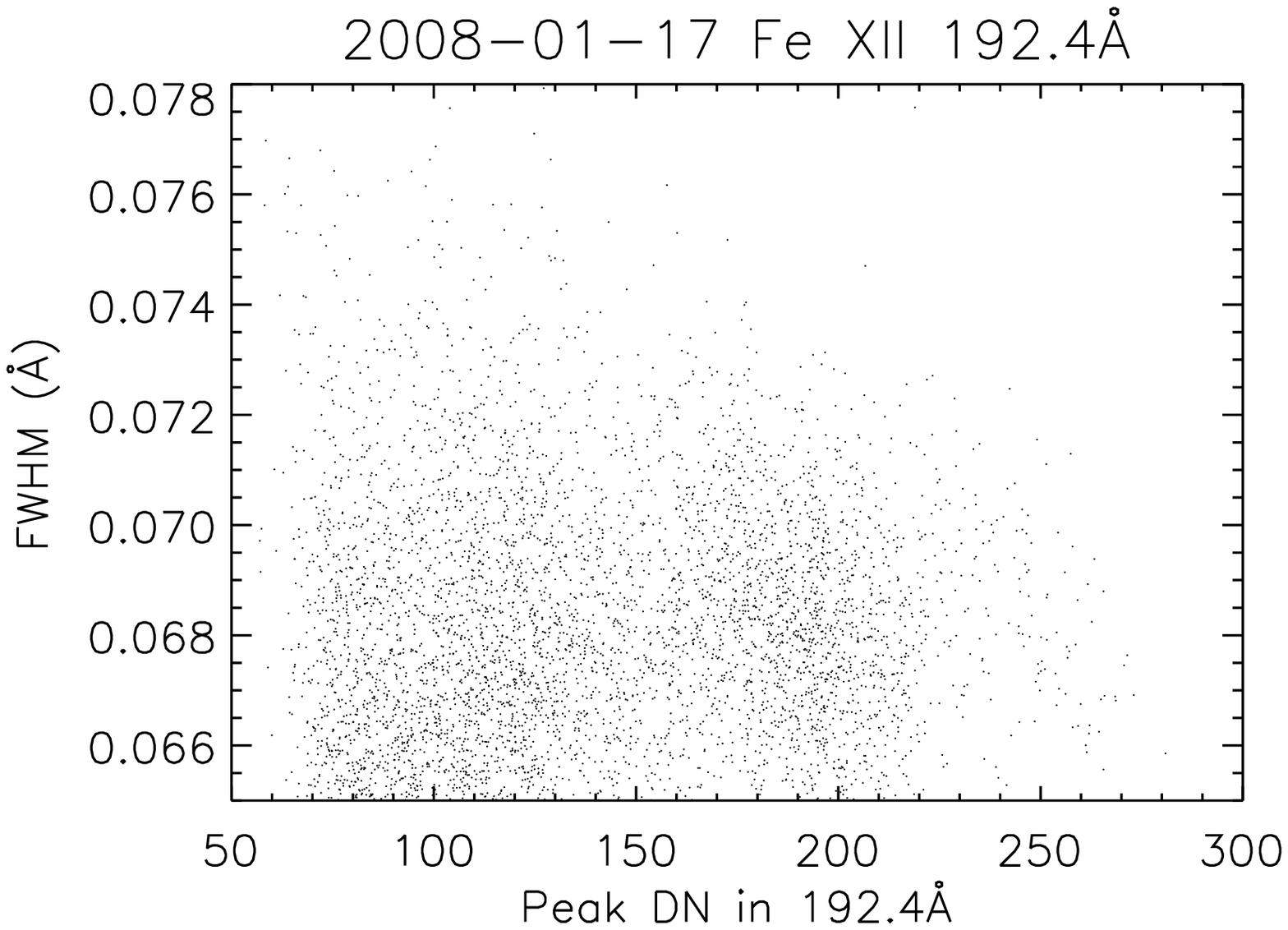}
\includegraphics[width=5.5cm,angle=0]{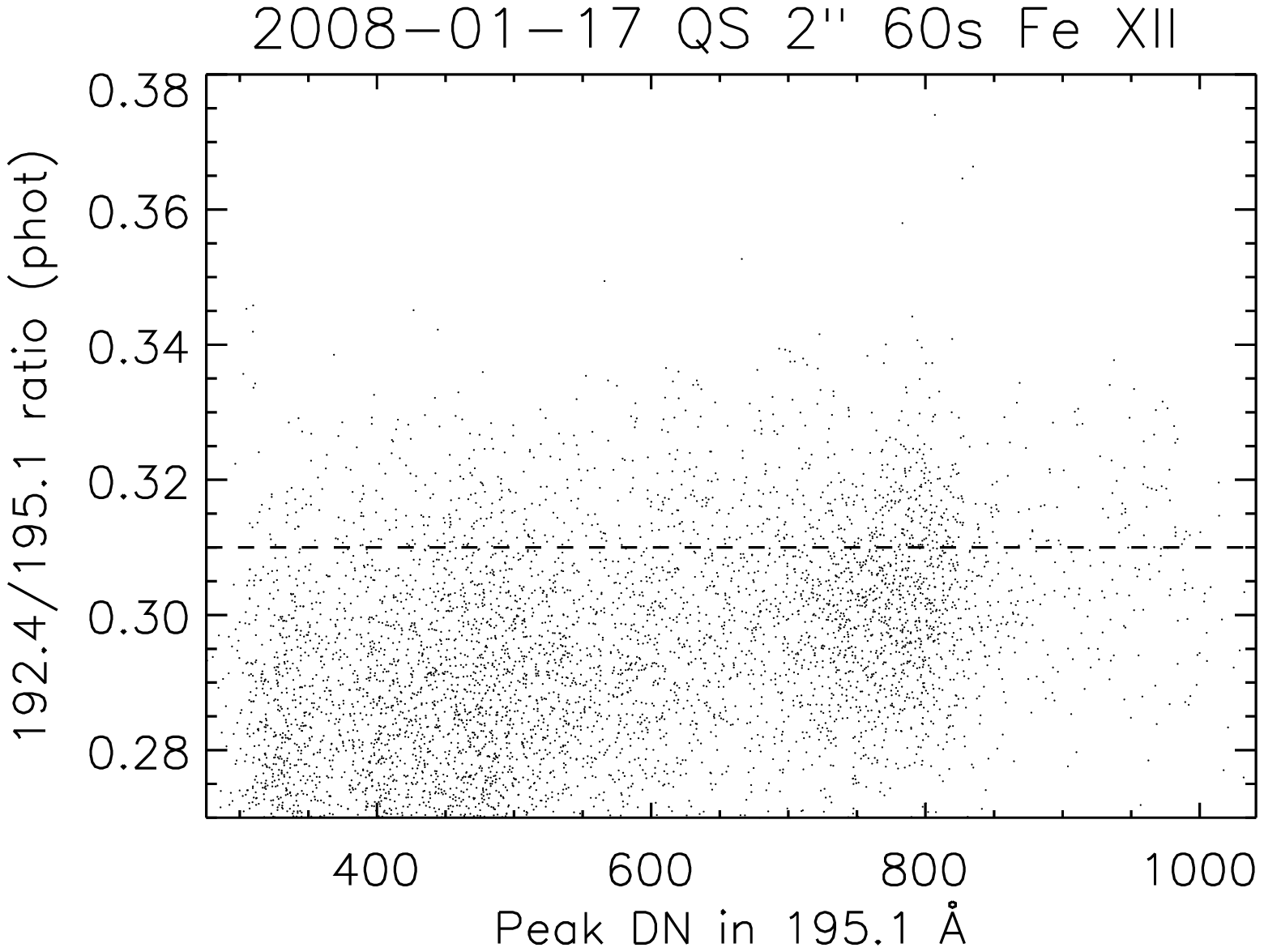}
\includegraphics[width=5.5cm,angle=0]{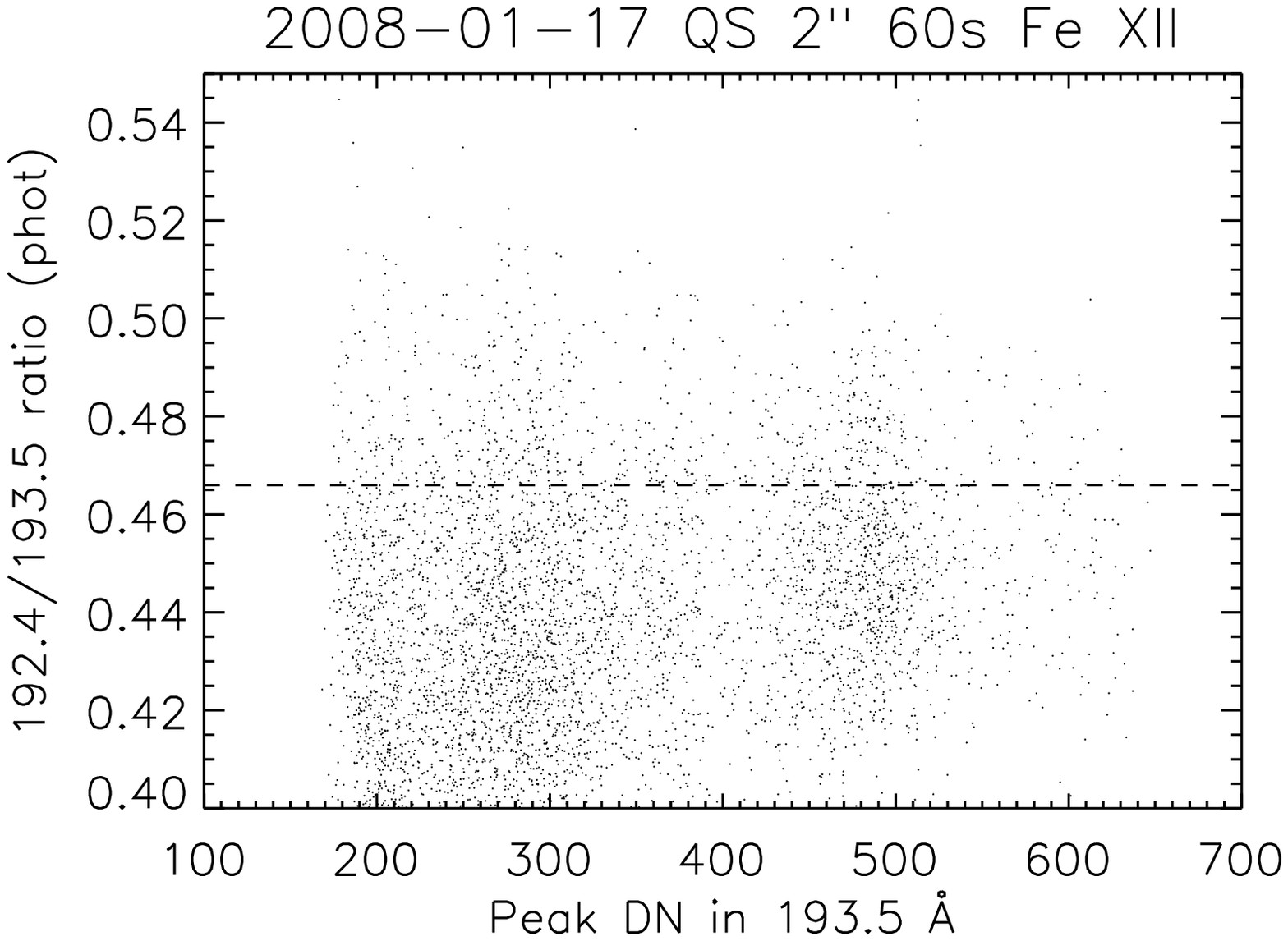}}
\caption{One of the earliest full spectral range quiet Sun on-disk observations with long
exposures and 
with the 2\arcsec\ slit (Atlas\_120x120\_s2\_40s study), on 2008-01-17.
}
\label{fig:2008-01-17}
\end{figure*}

We searched for EIS rasters 
with long exposures,  the full spectral range and the 2\arcsec\ slit
on the quiet Sun and found that one of the first ones was an 
Atlas\_120x120\_s2\_40s with 40s exposures on 2008-01-17.
There is good signal in the \ion{Fe}{xii} lines. 
Fig.~\ref{fig:2008-01-17} shows the main results. 
We see no trends in the widths. The counts in the lines are not very
high.  The  \ion{Fe}{xii}  lines have however  consistently different widths,
as in the 1\arcsec\ slit spectra.
The line ratios are slightly below theory, indicating a 
change in the EIS calibration, in agreement with the results 
discussed in \cite{delzanna:13_eis_calib}.


We searched for even longer exposures with the full spectral range and the 2\arcsec\ slit
and found the first useful one was an Atlas\_120 taken on 2010-10-08. 
In spite of the long exposures, the strongest 195~\AA\ line only had less then 3000
DN at line centre, see Fig.~\ref{fig:2010-10-8}.
There is a small difference (3 m~\AA) between the FWHM in the 192 and 195~\AA\
lines. Above 2500 DN in the peak of the 195~\AA\ line, some departure in the 
line ratio is  visible. There is no obvious E-W variation of the line widths, 
as shown in the bottom of Fig.~\ref{fig:2010-10-8}. 
The 192 vs. 195~\AA\ and 192 vs. 193~\AA\ ratios are  lower than expected
in the dimmest regions, indicating instrumental degradation. 

\begin{figure*}[!htbp]
\centerline{\includegraphics[width=6.0cm,angle=0]{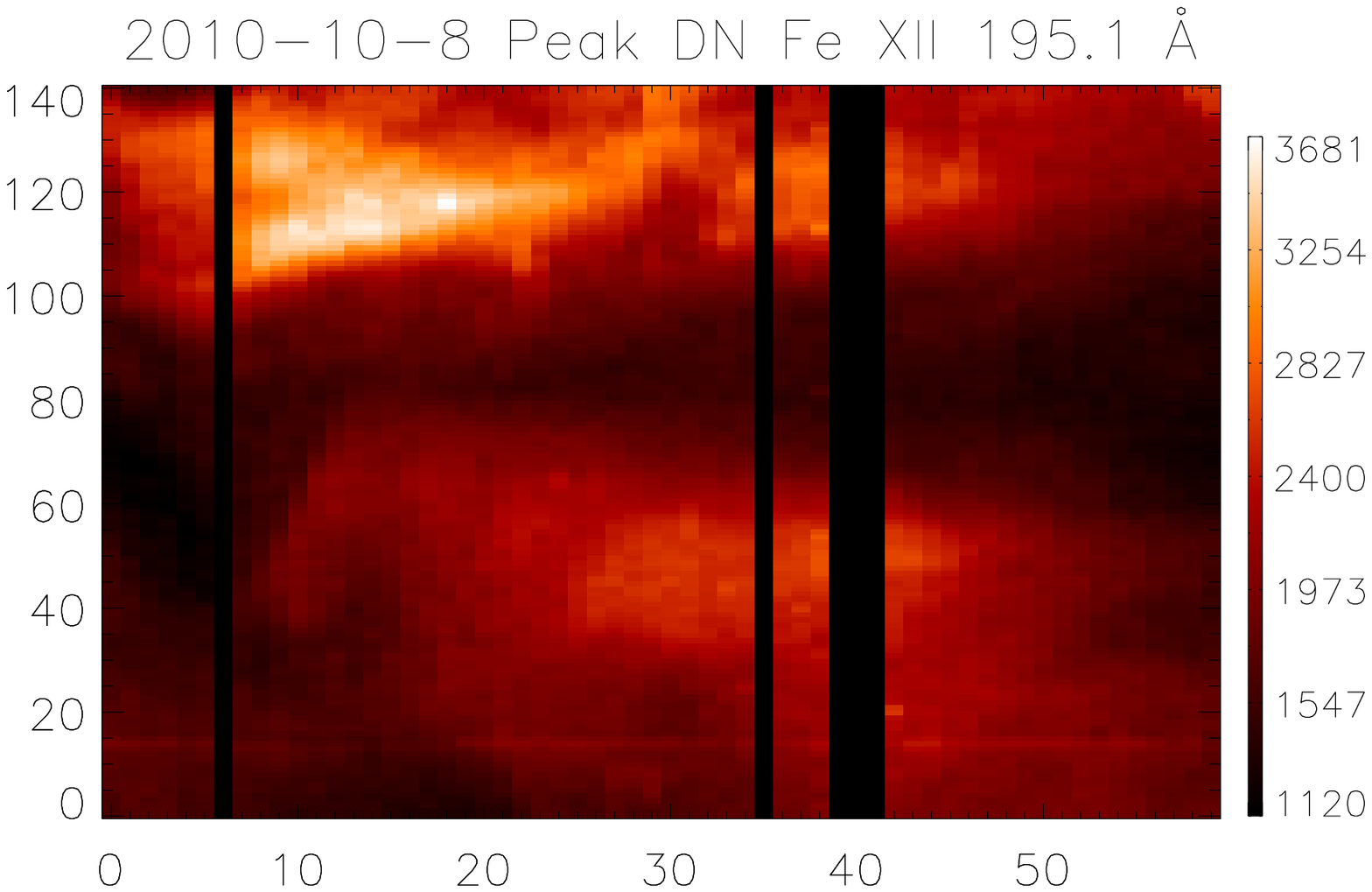}
\includegraphics[width=6.0cm,angle=0]{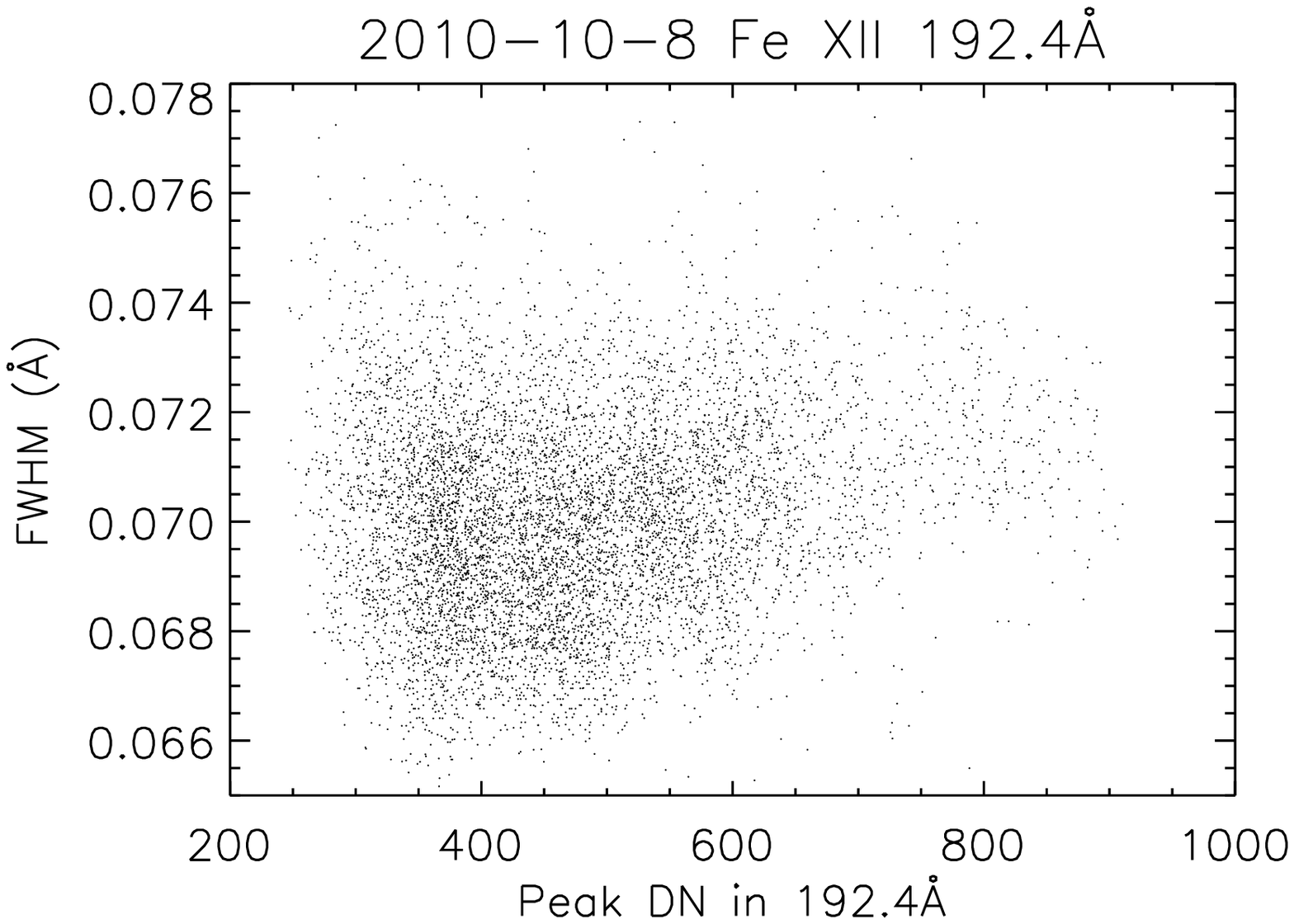}
\includegraphics[width=6.0cm,angle=0]{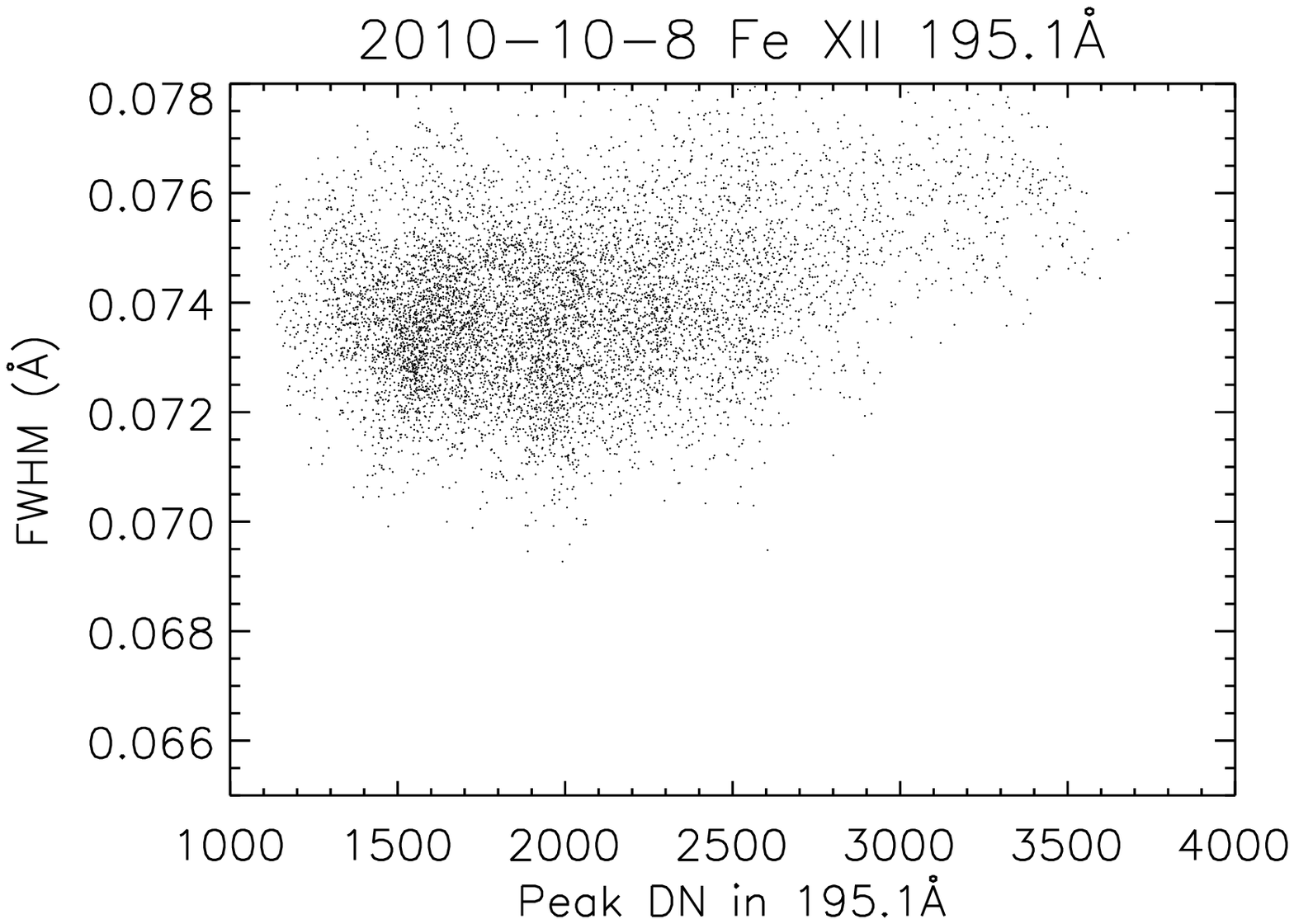}
}
\centerline{\includegraphics[width=6.0cm,angle=0]{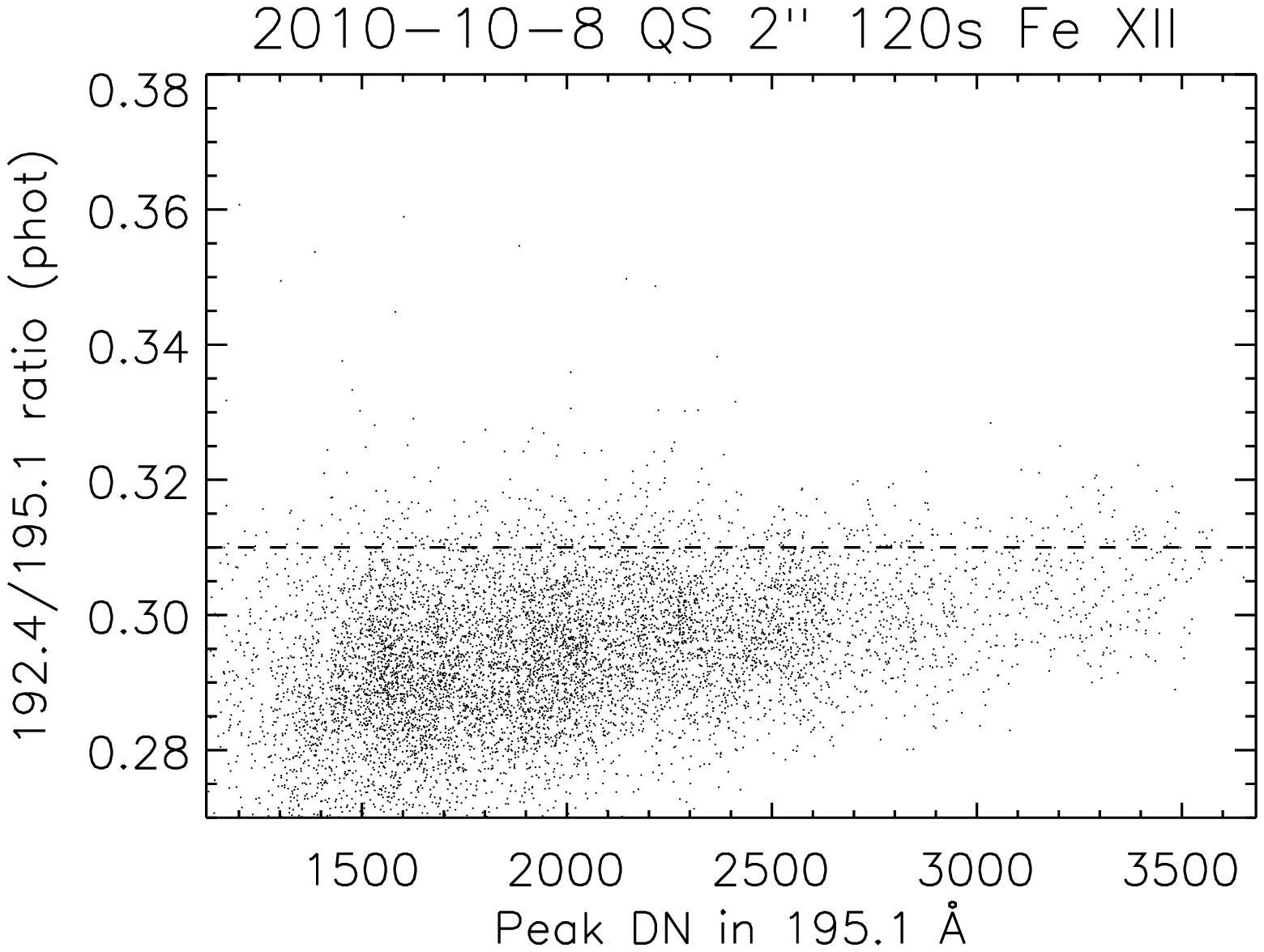}
\includegraphics[width=6.0cm,angle=0]{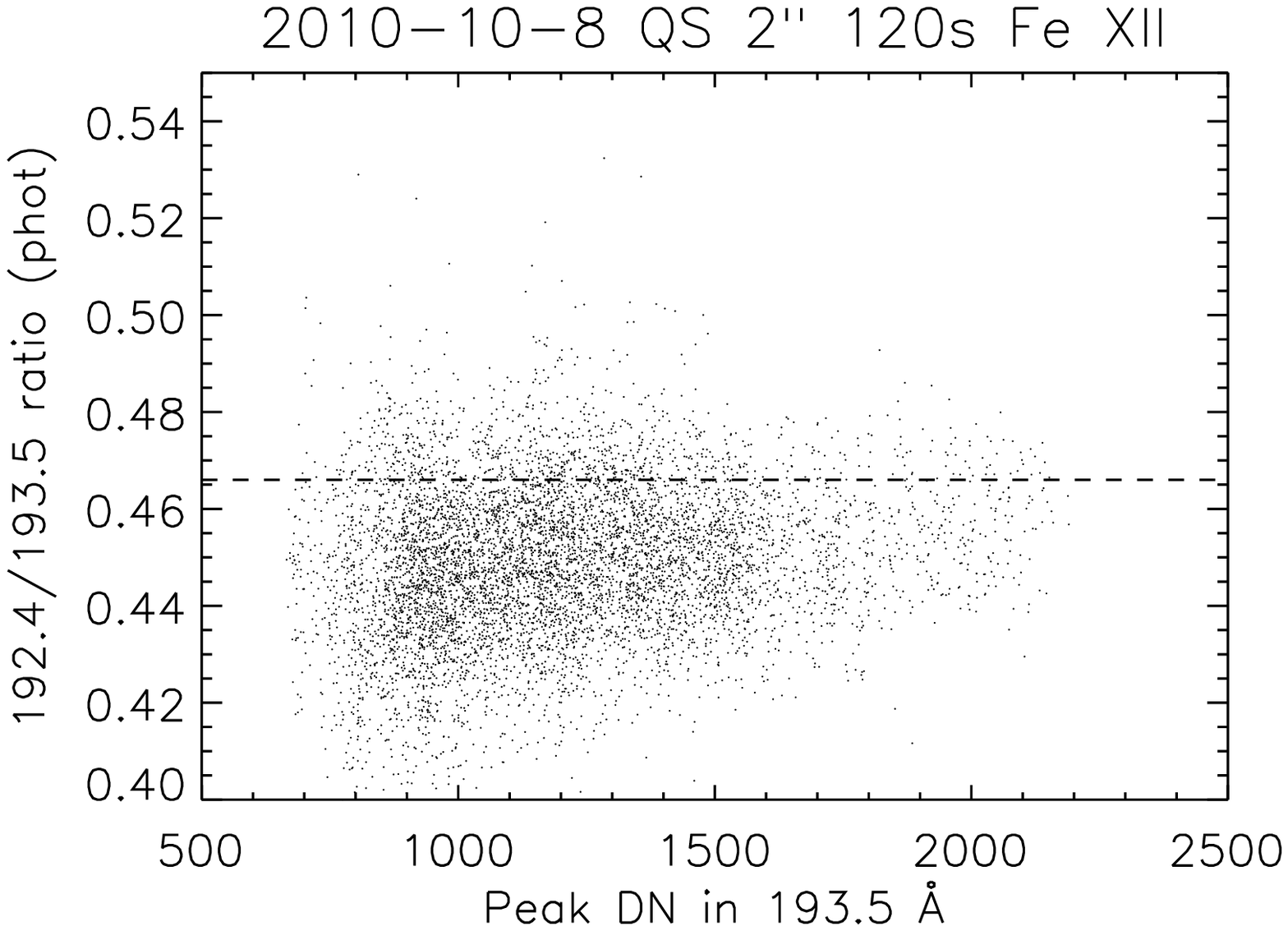}
\includegraphics[width=6.0cm,angle=0]{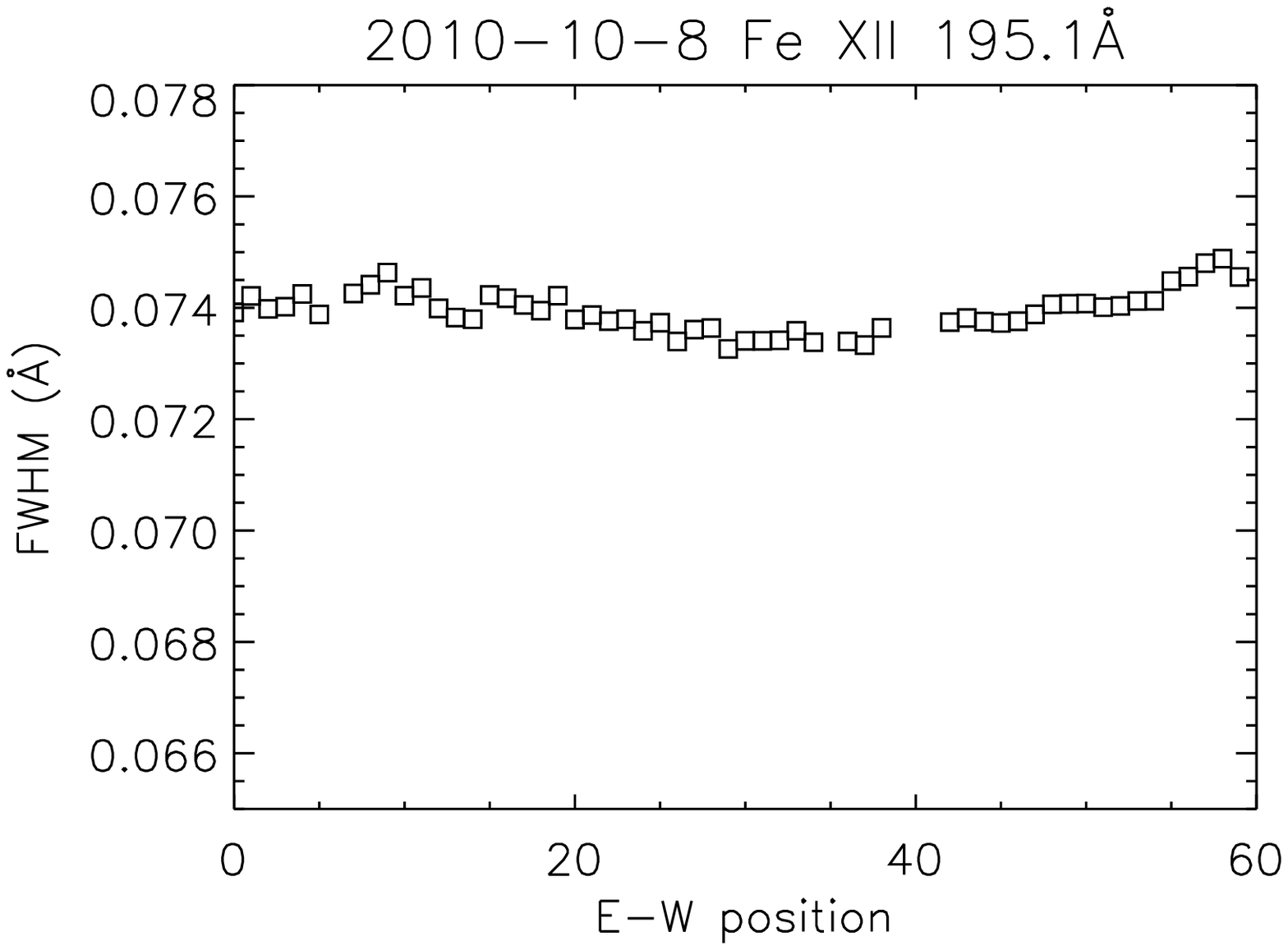}
}
\caption{Quiet Sun on-disk observation with the  2\arcsec\ slit
and the Atlas\_120 study, on 2010-10-08. 
}
\label{fig:2010-10-8}
\end{figure*}

\section{Quiet Sun at the limb}

Observations of the quiet Sun at the limb sometimes  show strong anomalies, sometimes
not.
Fig.~\ref{fig:2007-03-11} shows an observation with the  1\arcsec\ slit  on 2007-03-11.
There is little variation in the widths of the  \ion{Fe}{xii} lines, 
although the two stronger lines always have larger widths than the weaker line.
The intensity ratios do not vary much.

\begin{figure*}[!htbp]
\centerline{\includegraphics[width=5.5cm,angle=0]{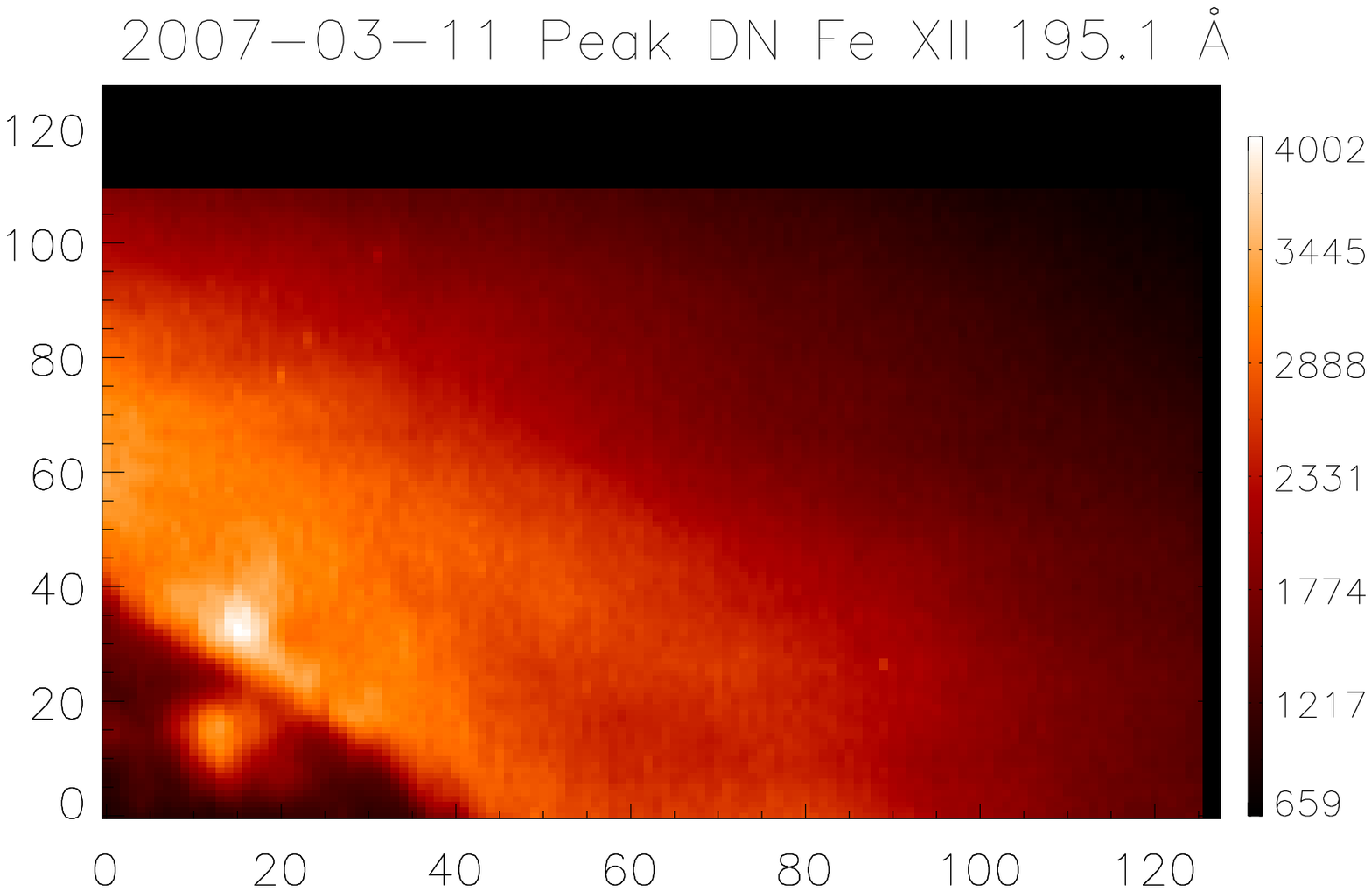}
\includegraphics[width=5.5cm,angle=0]{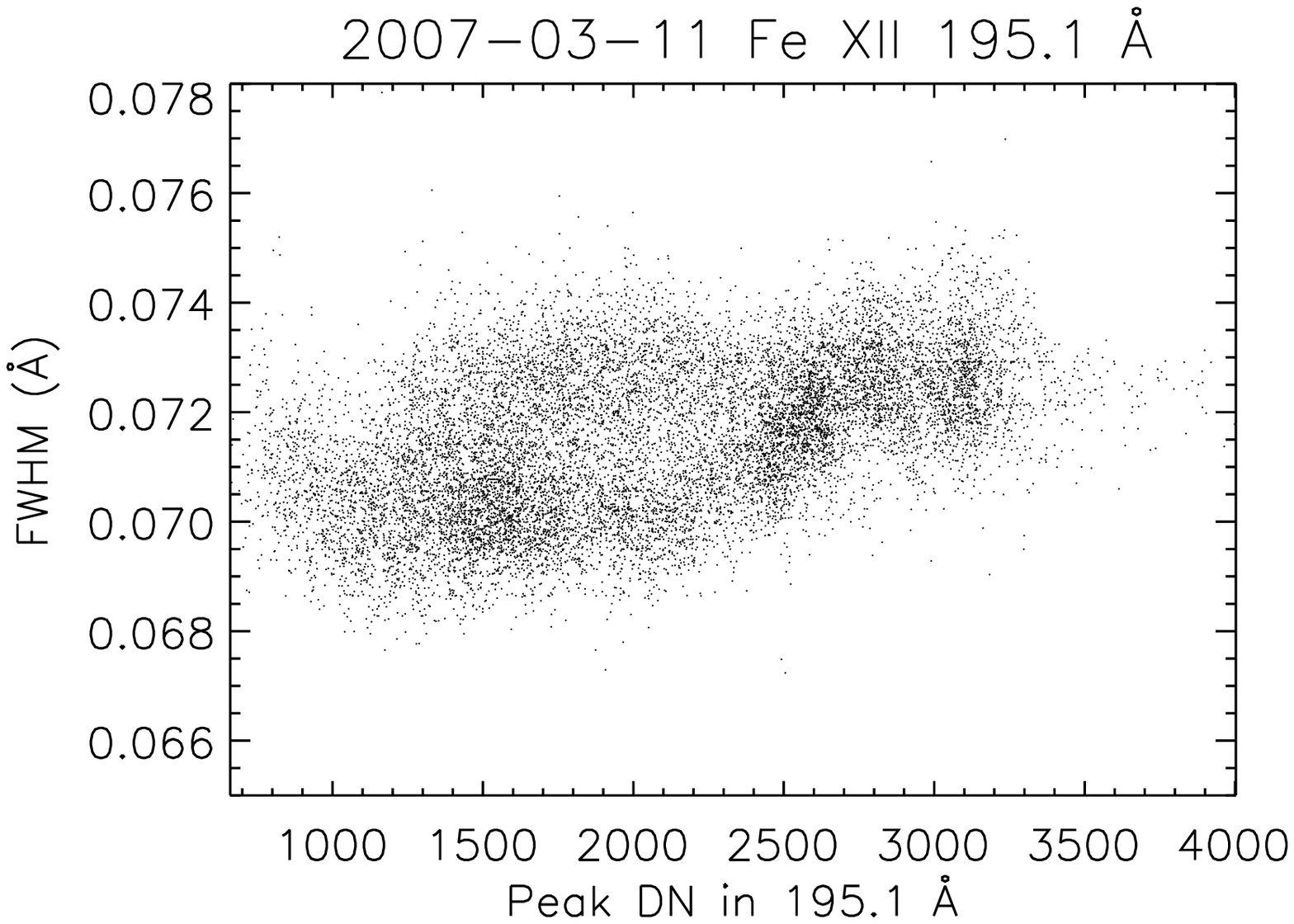}
\includegraphics[width=5.5cm,angle=0]{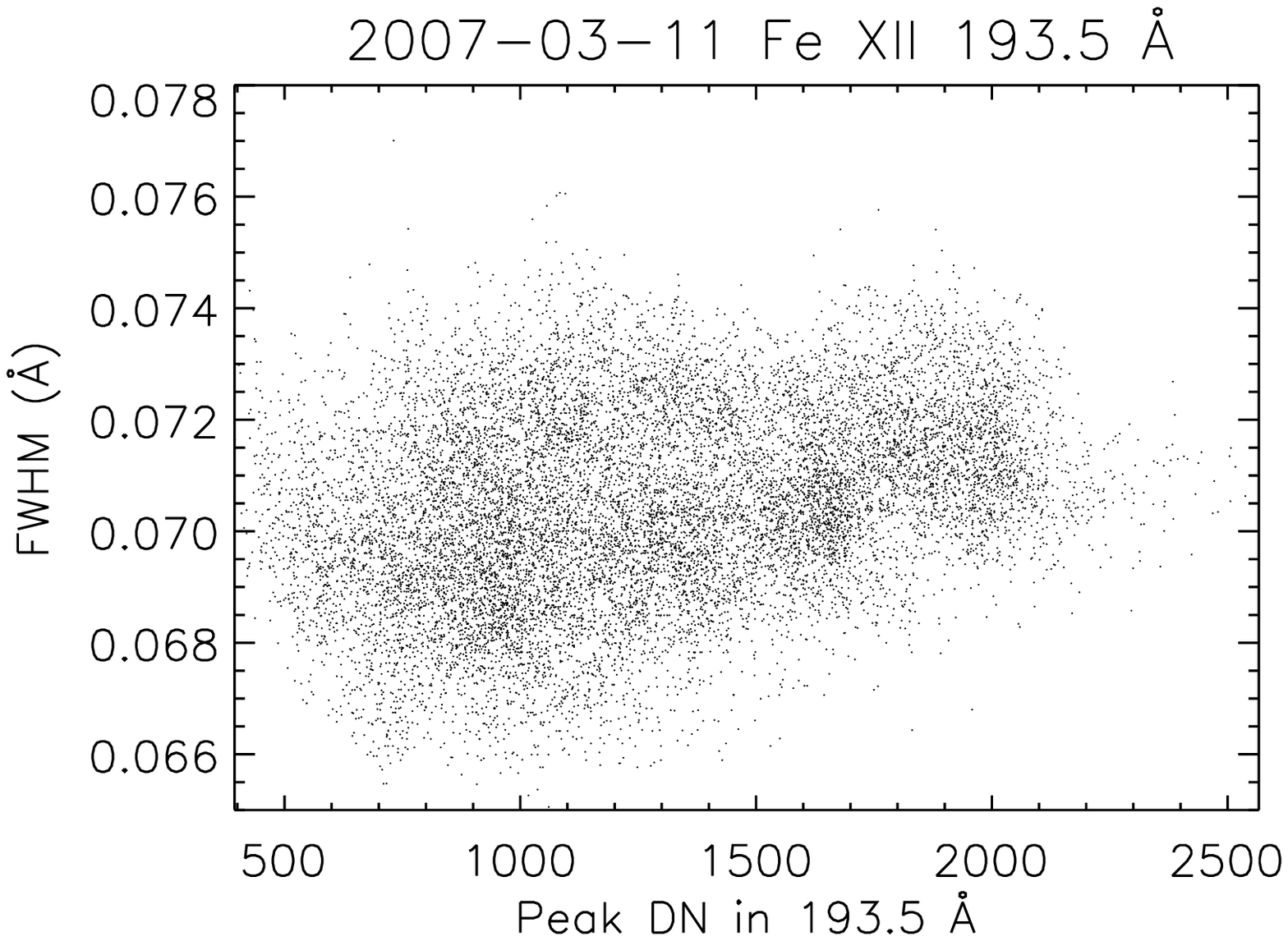}
}
\centerline{\includegraphics[width=5.5cm,angle=0]{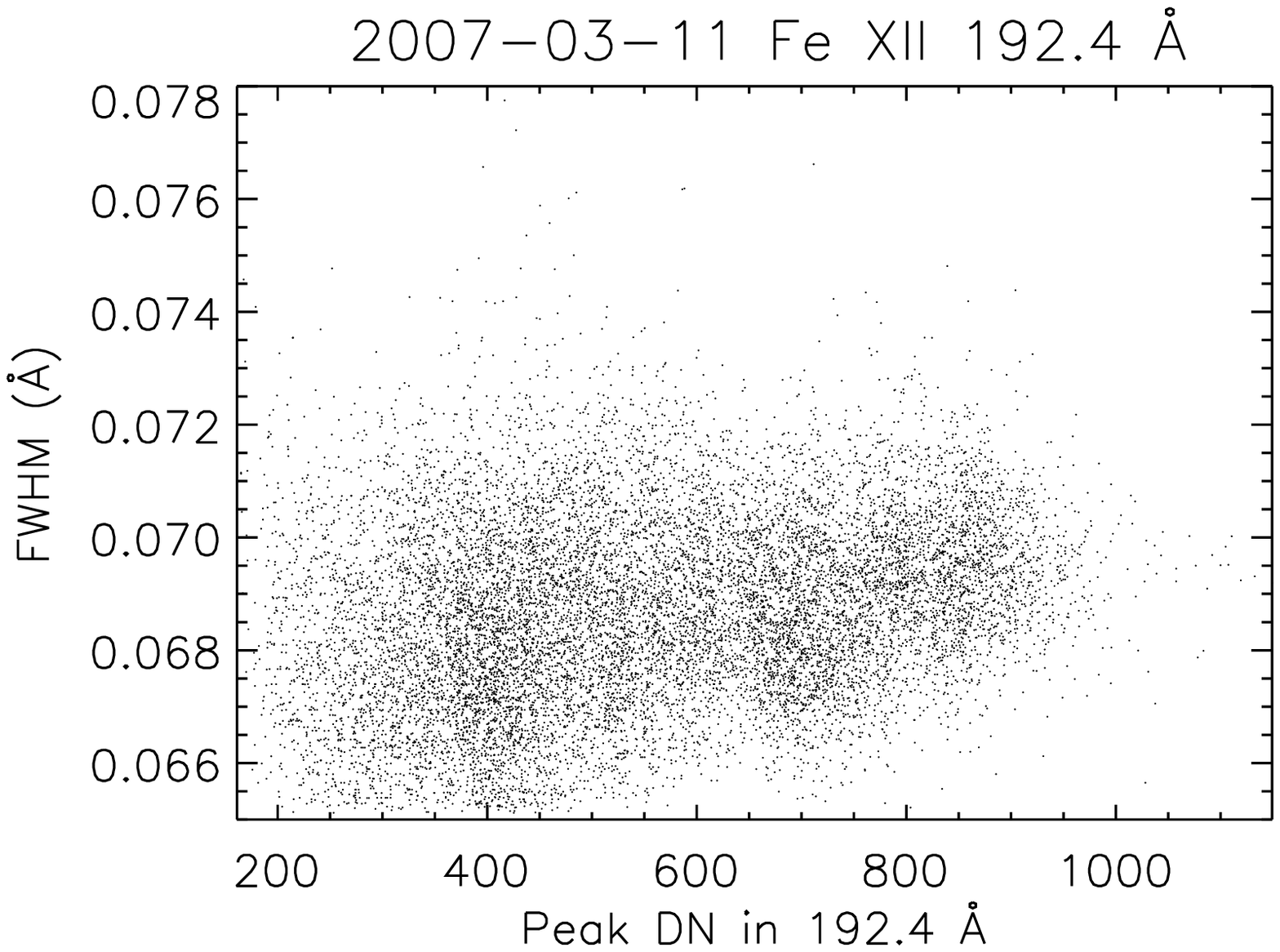}
\includegraphics[width=5.5cm,angle=0]{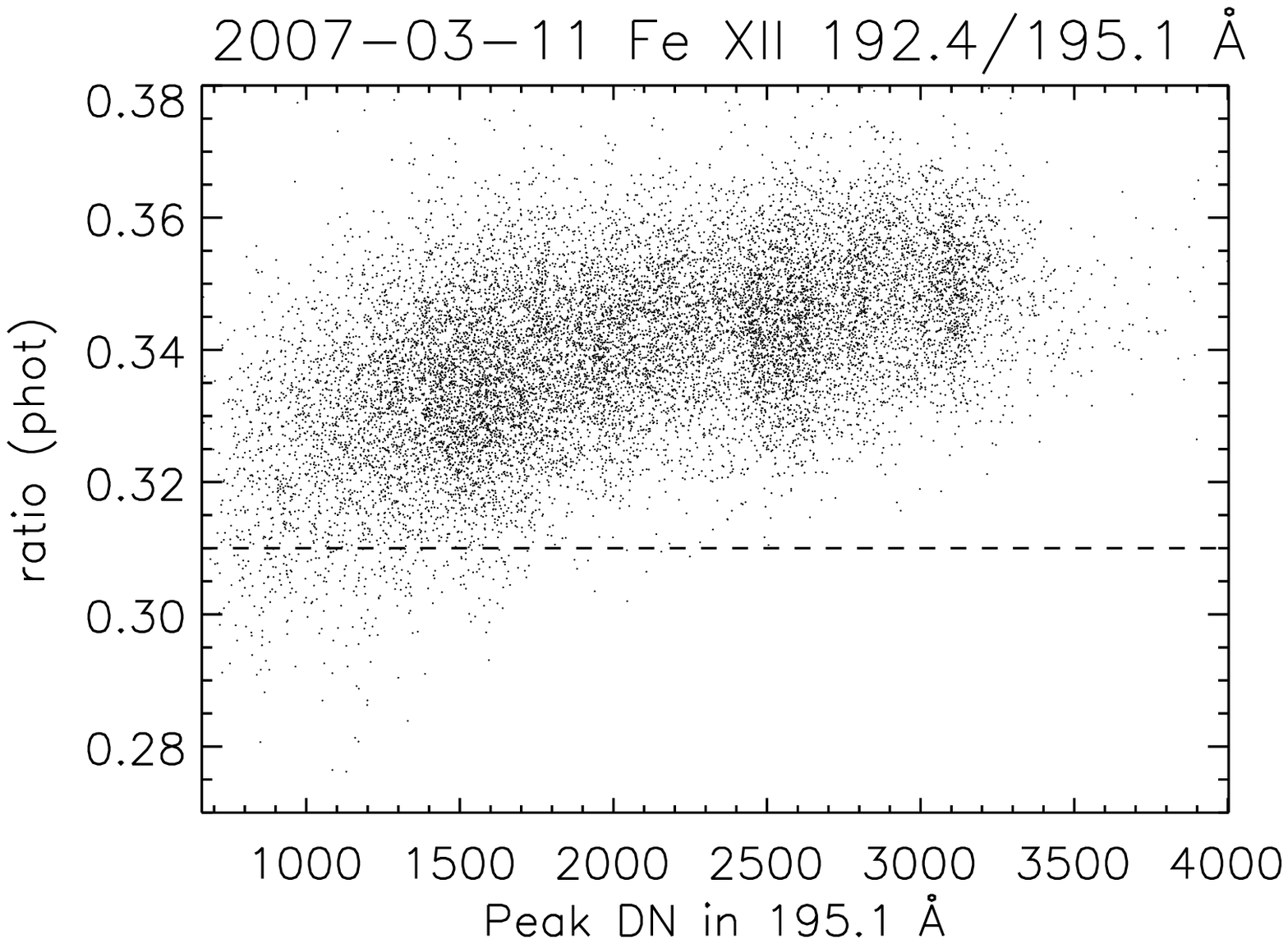}
\includegraphics[width=5.5cm,angle=0]{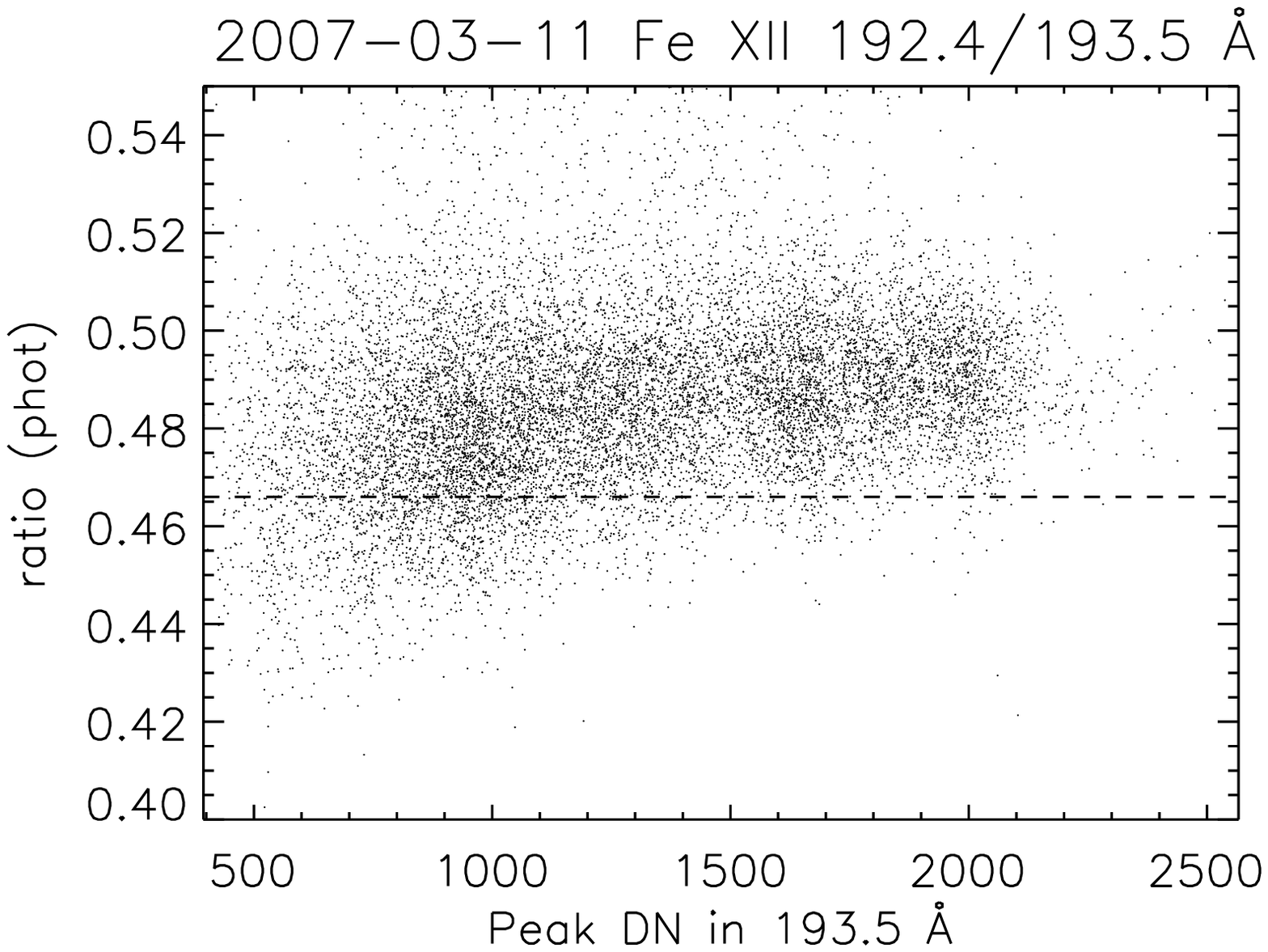}
}
\caption{QS  off-limb observation  with the 1\arcsec\ slit, 90s exposures on 2007-03-11.
}
\label{fig:2007-03-11}
\end{figure*}


We searched for QS observations with long exposures near the equator,
and found that the first useful one was an 
Atlas\_120 taken on 2011-01-04 at 12:59  UT. It observed up to 
about 100\arcsec\ above the limb.
The strongest 195~\AA\ line was nearly saturated near the limb, see
Fig.~\ref{fig:2011-01-4}. The 192~\AA\ and 195~\AA\ lines indicate
a small decrease in the line width off-limb. 
The 192 vs. 195~\AA\ and 192 vs. 193~\AA\ ratios are lower than expected
in the dimmest regions, indicating an instrumental degradation.
The ratios show the anomalous effect in the brightest 
regions, most likely due to opacity.

\begin{figure*}[!htbp]
\centerline{\includegraphics[width=6.0cm,angle=0]{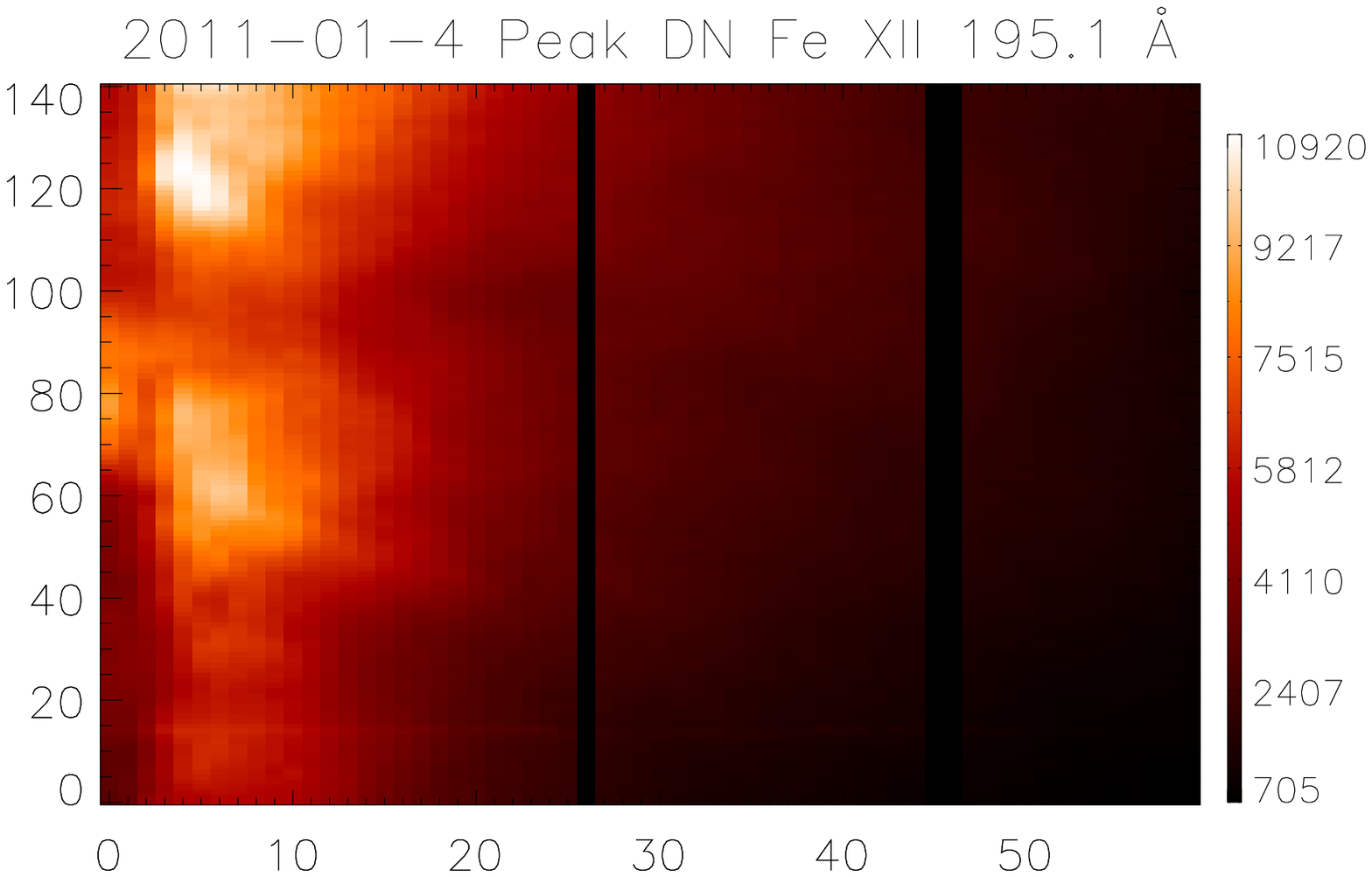}
\includegraphics[width=6.0cm,angle=0]{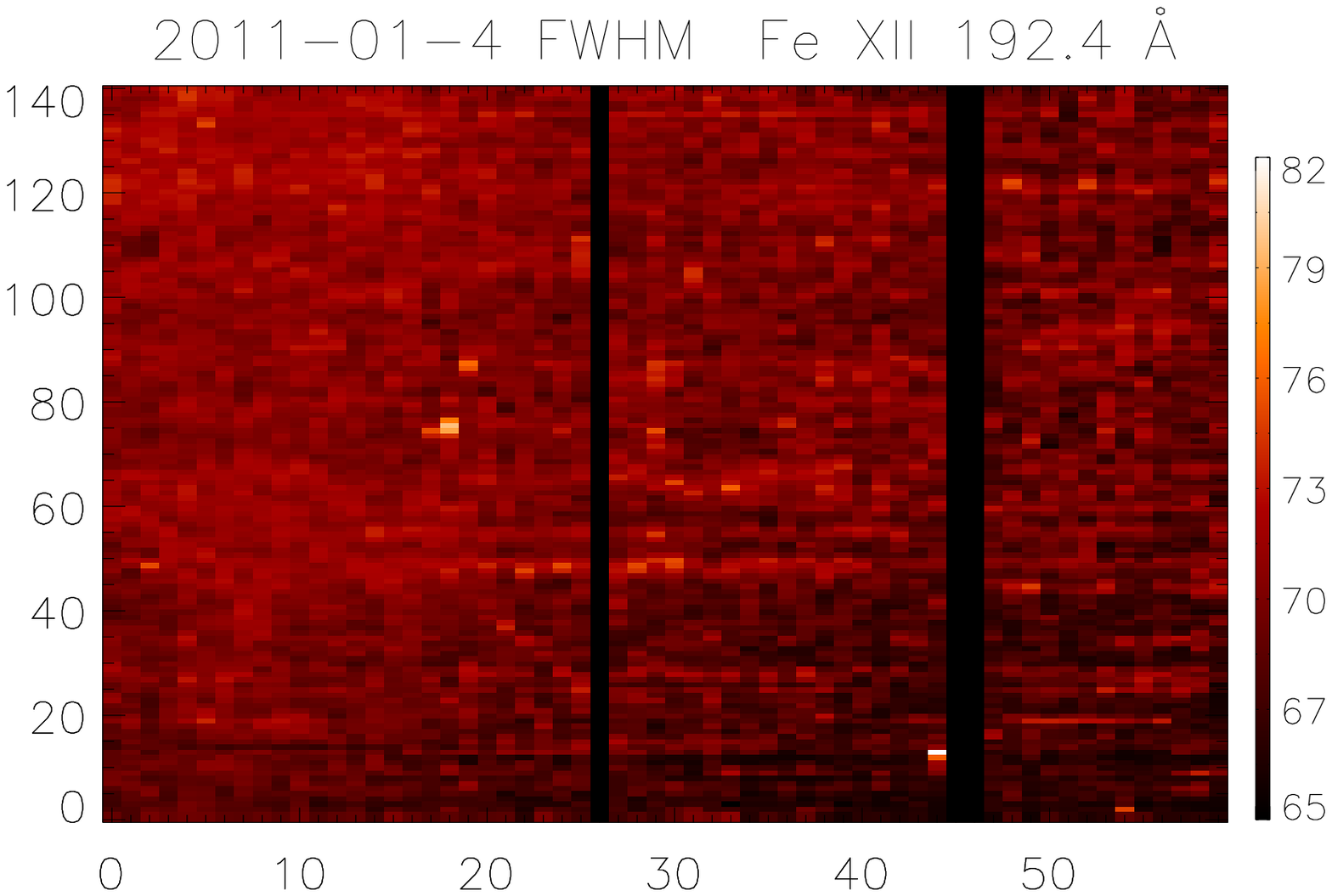}
\includegraphics[width=6.0cm,angle=0]{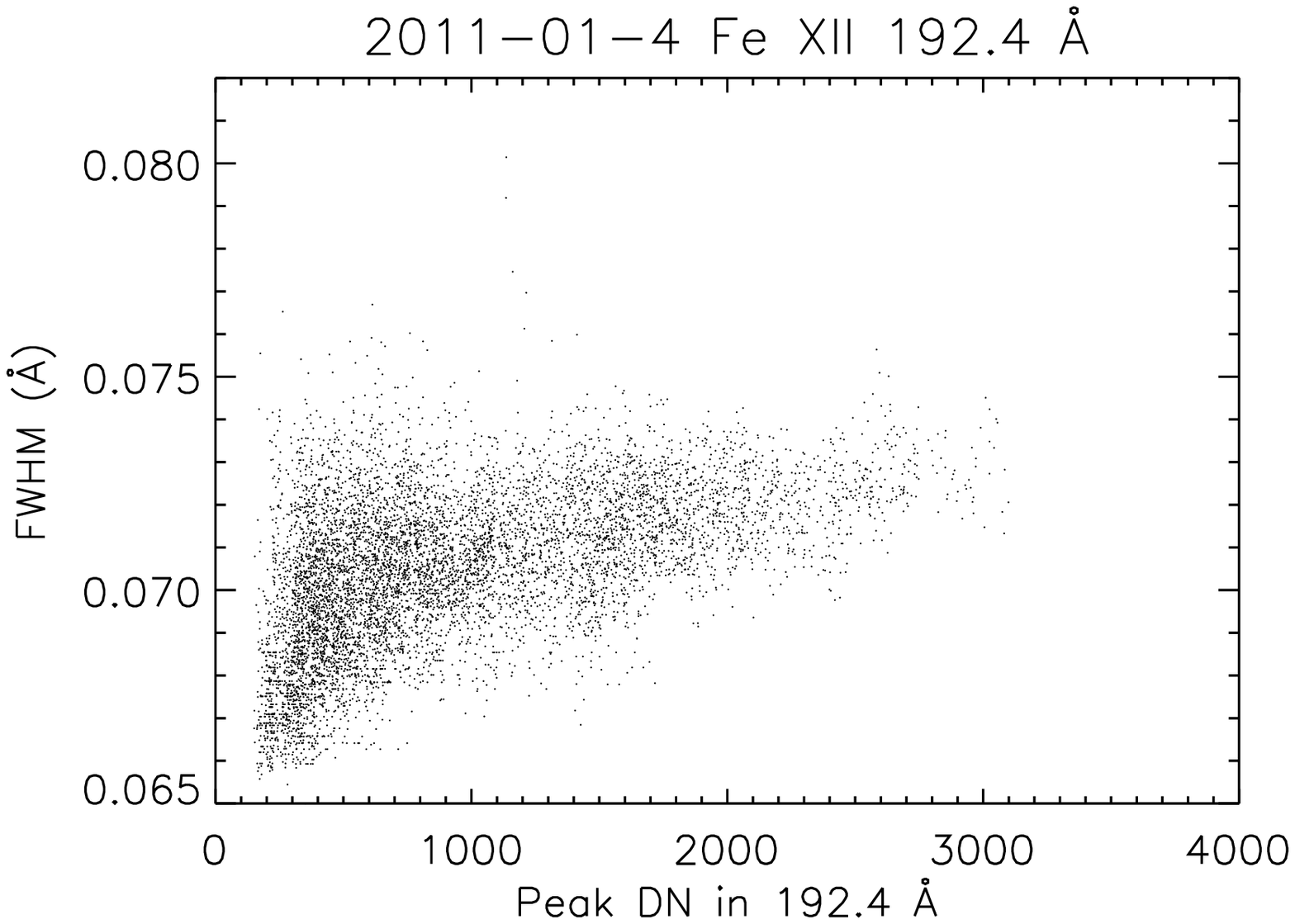}
}
\centerline{\includegraphics[width=6.0cm,angle=0]{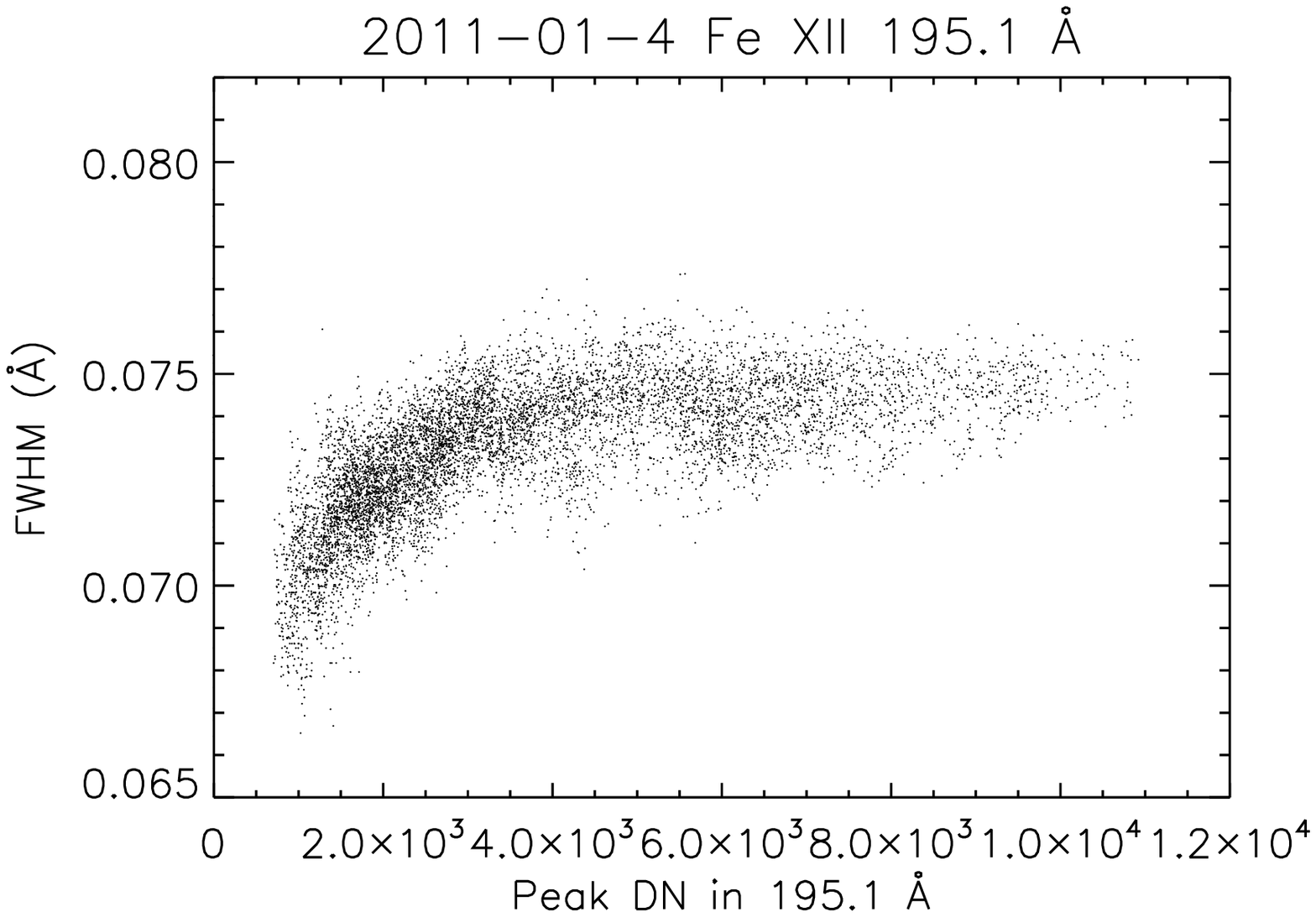}
\includegraphics[width=6.0cm,angle=0]{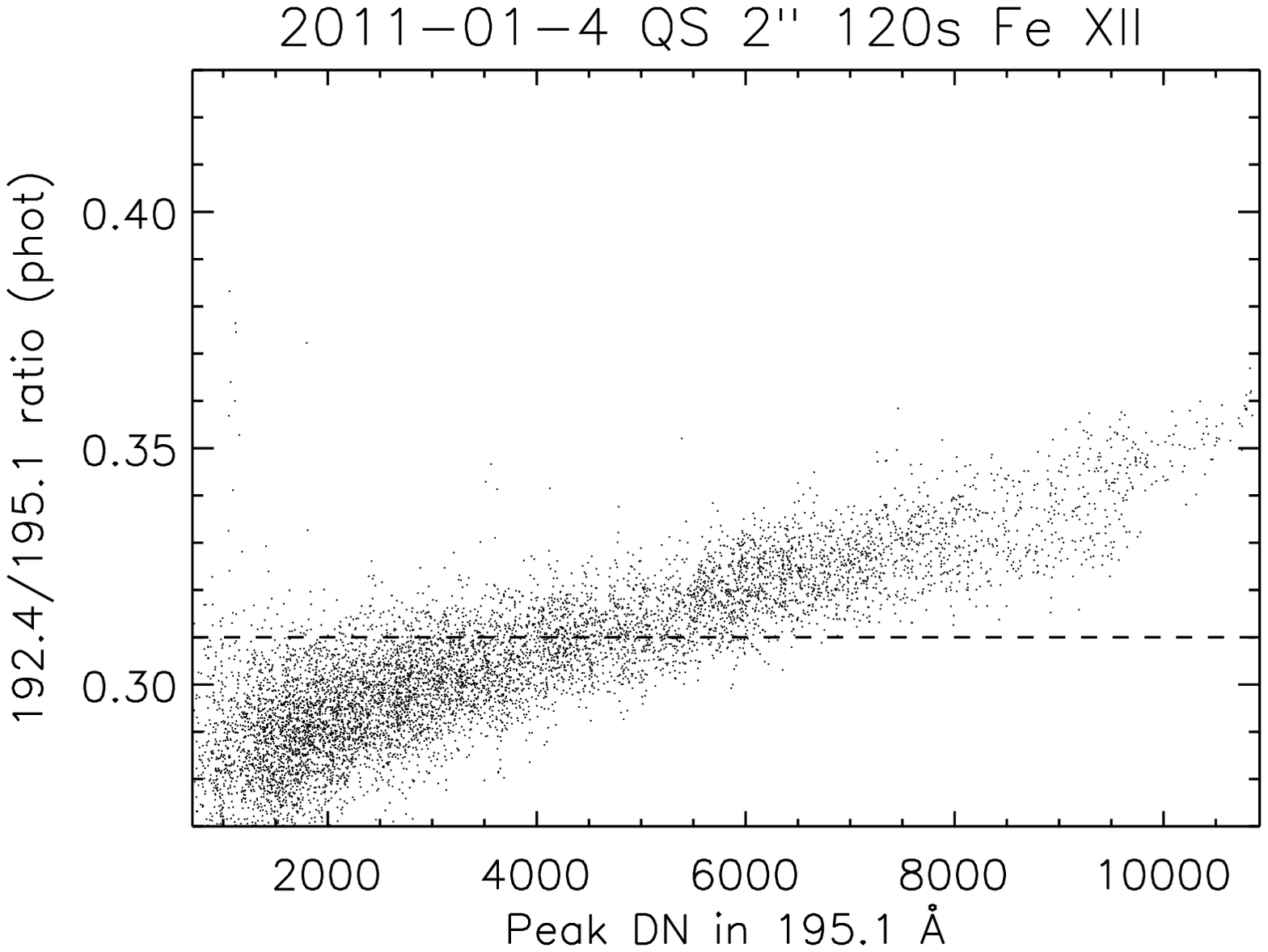}
\includegraphics[width=6.0cm,angle=0]{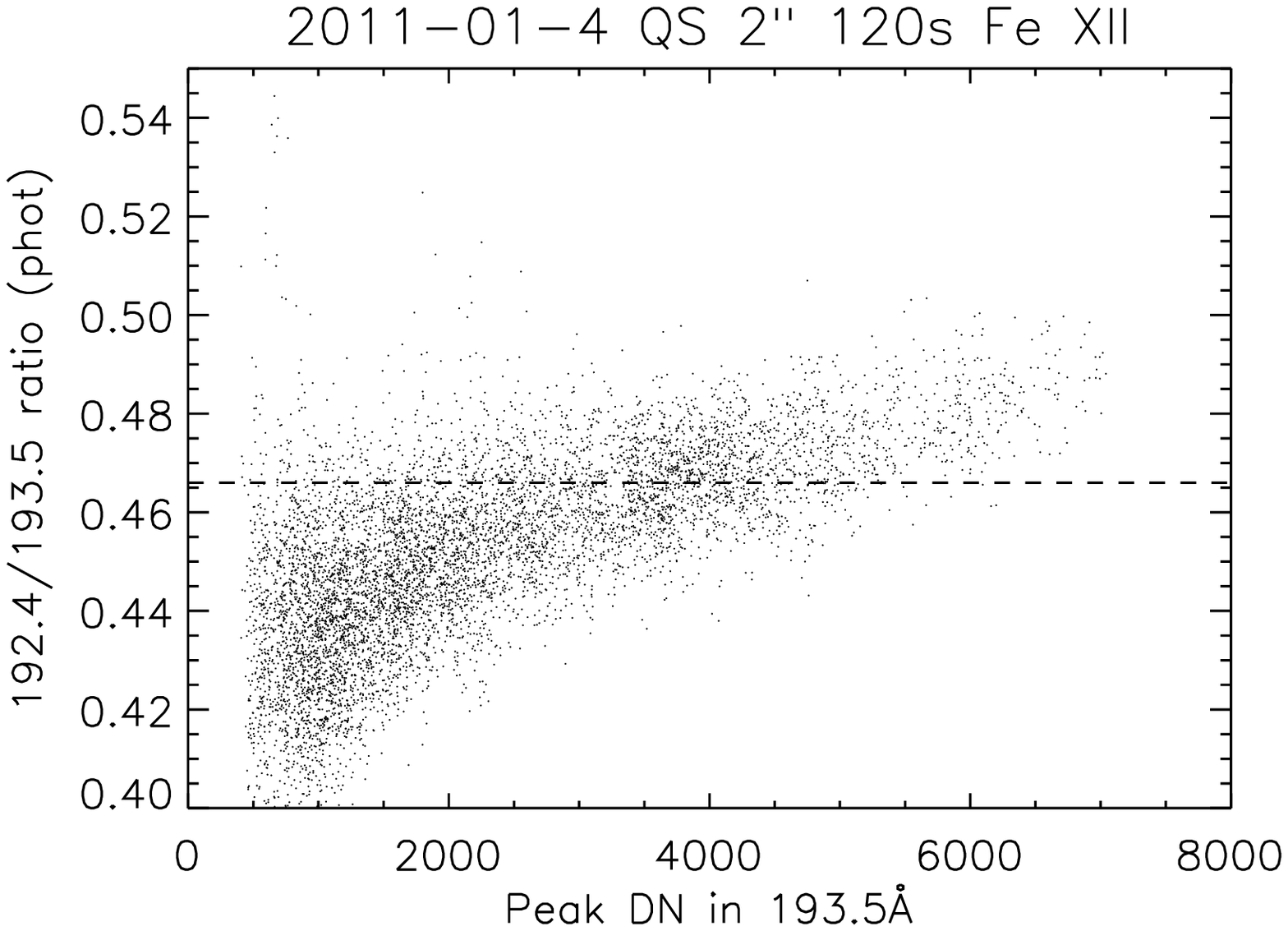}
}
\caption{Quiet Sun observation at the limb with the 
2~\arcsec\ slit and an Atlas\_120 taken on 2011-01-04 at 12:59  UT.
}
\label{fig:2011-01-4}
\end{figure*}

\section{2018 observations - Quiet Sun - bottom of the slit}

We obtained QS rasters with the bottom of the slit, pointed near Sun centre,
to use as a reference, mostly to see if there were any E-W variations
in the line widths in this part of the slit.
The signal was very low, so only the strongest 195~\AA\ line was usable.
Fig.~\ref{fig:20180712} (top) shows the image of the width of the line,
while Fig.~\ref{fig:20180712} (bottom) shows the averaged values along the 
E-W direction, indicating no obvious variations.

\begin{figure}[!htbp]
\centerline{\includegraphics[width=6.0cm,angle=0]{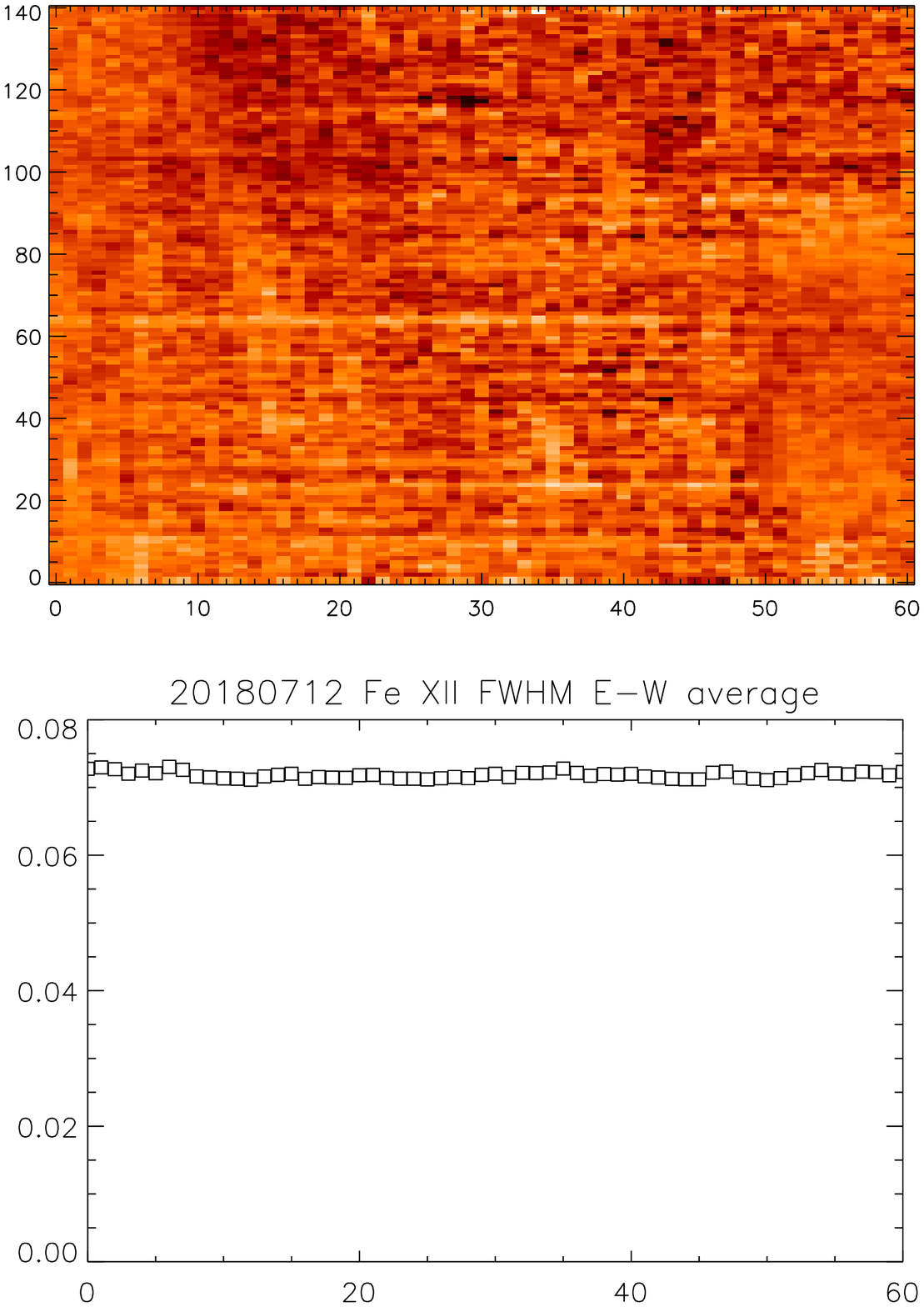}}
\caption{An observation of the FWHM of the strongest 195~\AA\ line, in the
bottom of the EIS 2\arcsec\ slit on the Quiet Sun near Sun centre, 
indicating no obvious E-W variations.
}
\label{fig:20180712}
\end{figure}

\section{The off-limb QS observations out to 1.5~\rsun\ on 2007-05-08}

{
Fig.~\ref{fig:8-may_1} and Fig.~\ref{fig:8-may_2} show a sample 
of results from the analysis of the off-limb observation 
on 2007-05-8.
The presence of the active region at the limb somewhat affected the nearby corona,
which can be seen by a small  increase in the ionisation  temperatures,
compared to the values we measured two days later.  
The observed line widths are also nearly constant, with a small increase
 with radial distance.

The excess widths were estimated from the \ion{Fe}{xiii} line, 
assuming the temperatures from the \ion{Fe}{xii}/\ion{Fe}{x} ratio, 
and adding a (small) uncertainty of 3 m~\AA\ in the observed widths. 
The boxes show the results with the instrumental width as in the 
EIS software. The triangles assume a constant instrumental width of 
64$\pm$2 m~\AA. 

}

\begin{figure}[!htbp]
\centerline{\includegraphics[width=5.0cm,angle=0]{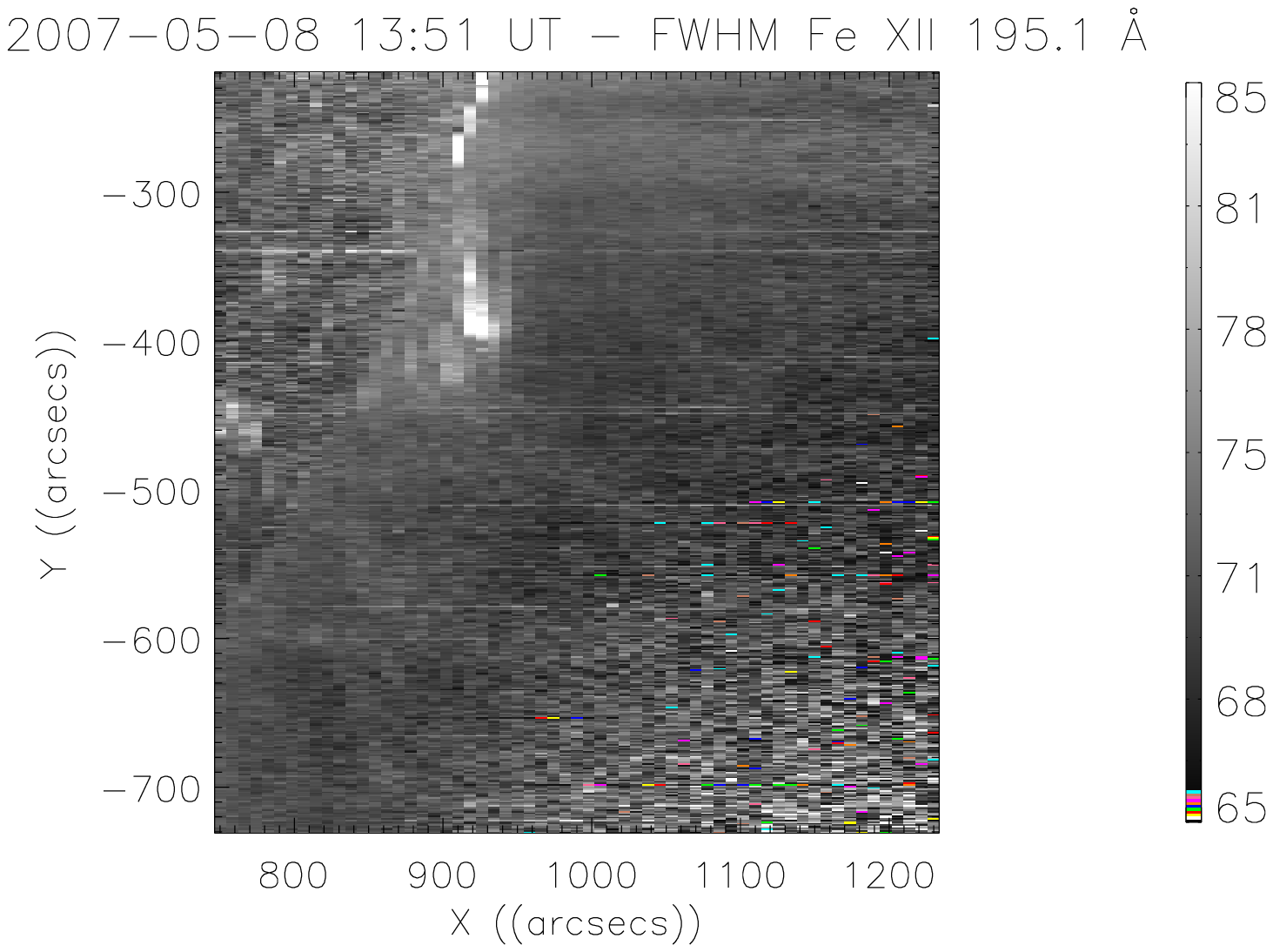}
\includegraphics[width=5.0cm,angle=0]{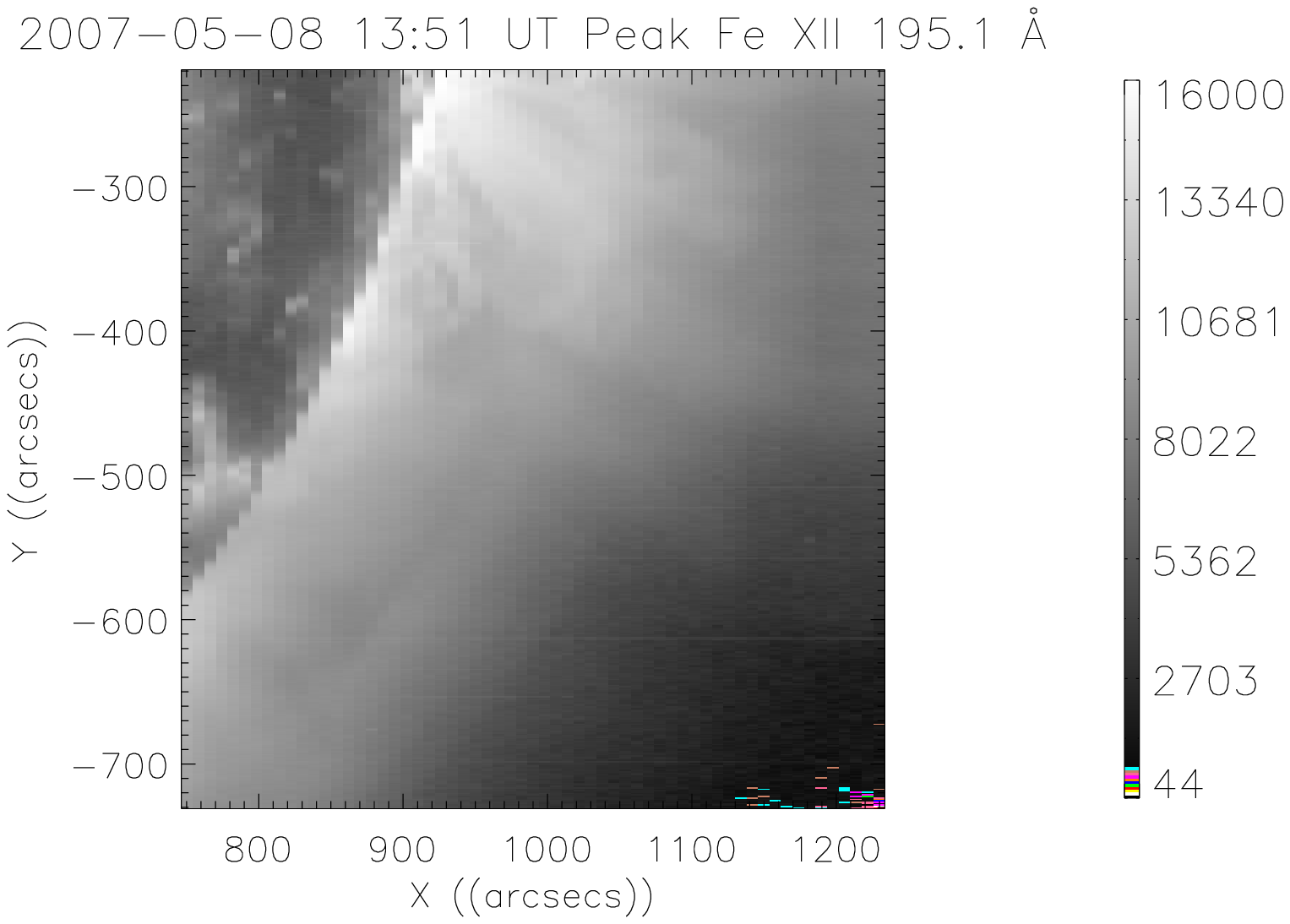}}
\centerline{\includegraphics[width=6.0cm,angle=0]{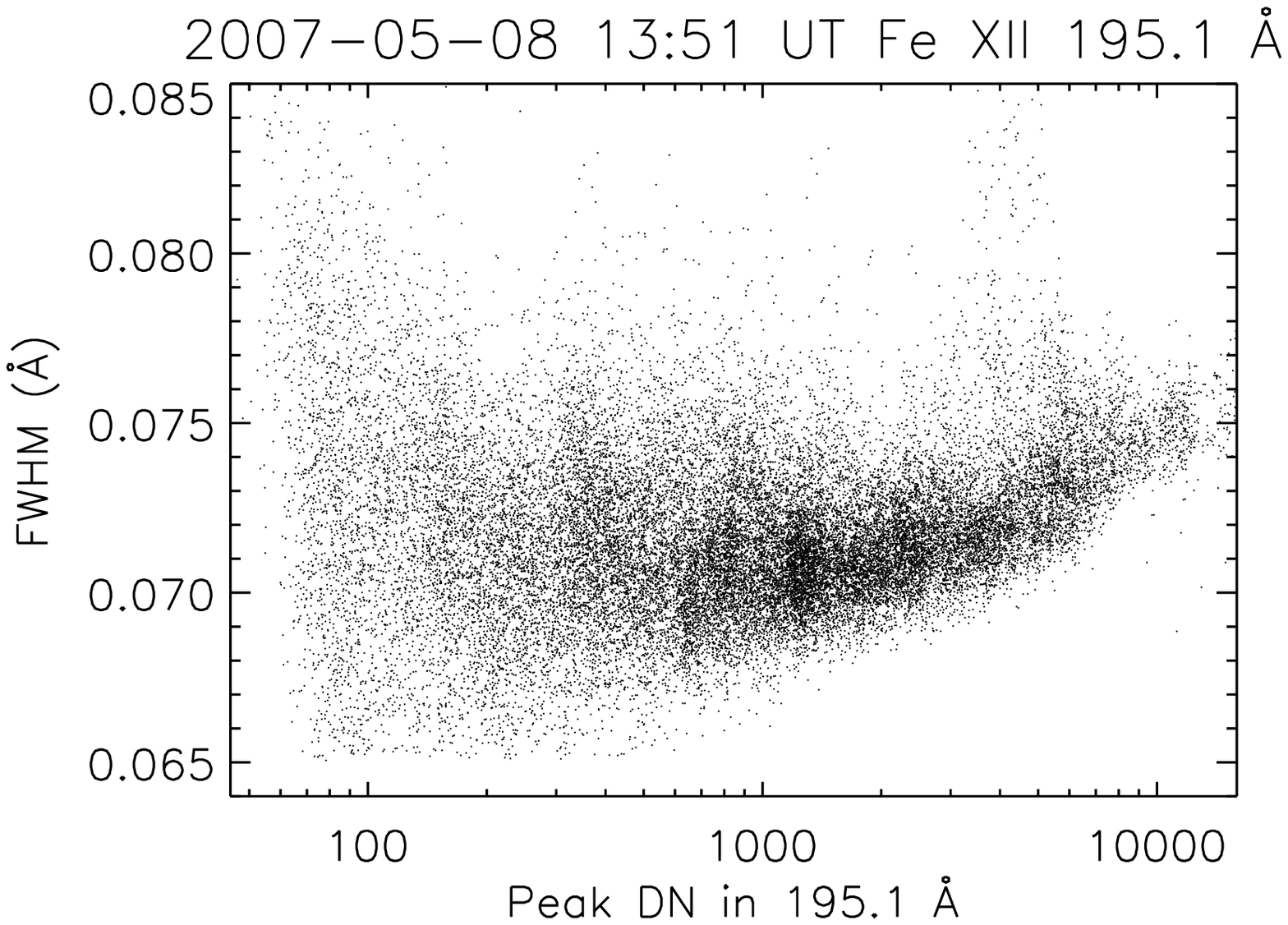}}
\caption{Selected results from the Hinode EIS off-limb observation on 2007-05-8 at 13:51 UT.
From top to bottom: image of the FWHM and peak DN in the \ion{Fe}{xii} 195.1~\AA\ line;
scatter plot of the FWHM as a function of peak intensity.
}
\label{fig:8-may_1}
\end{figure}

\begin{figure}[!htbp]
\centerline{\includegraphics[width=7.0cm,angle=0]{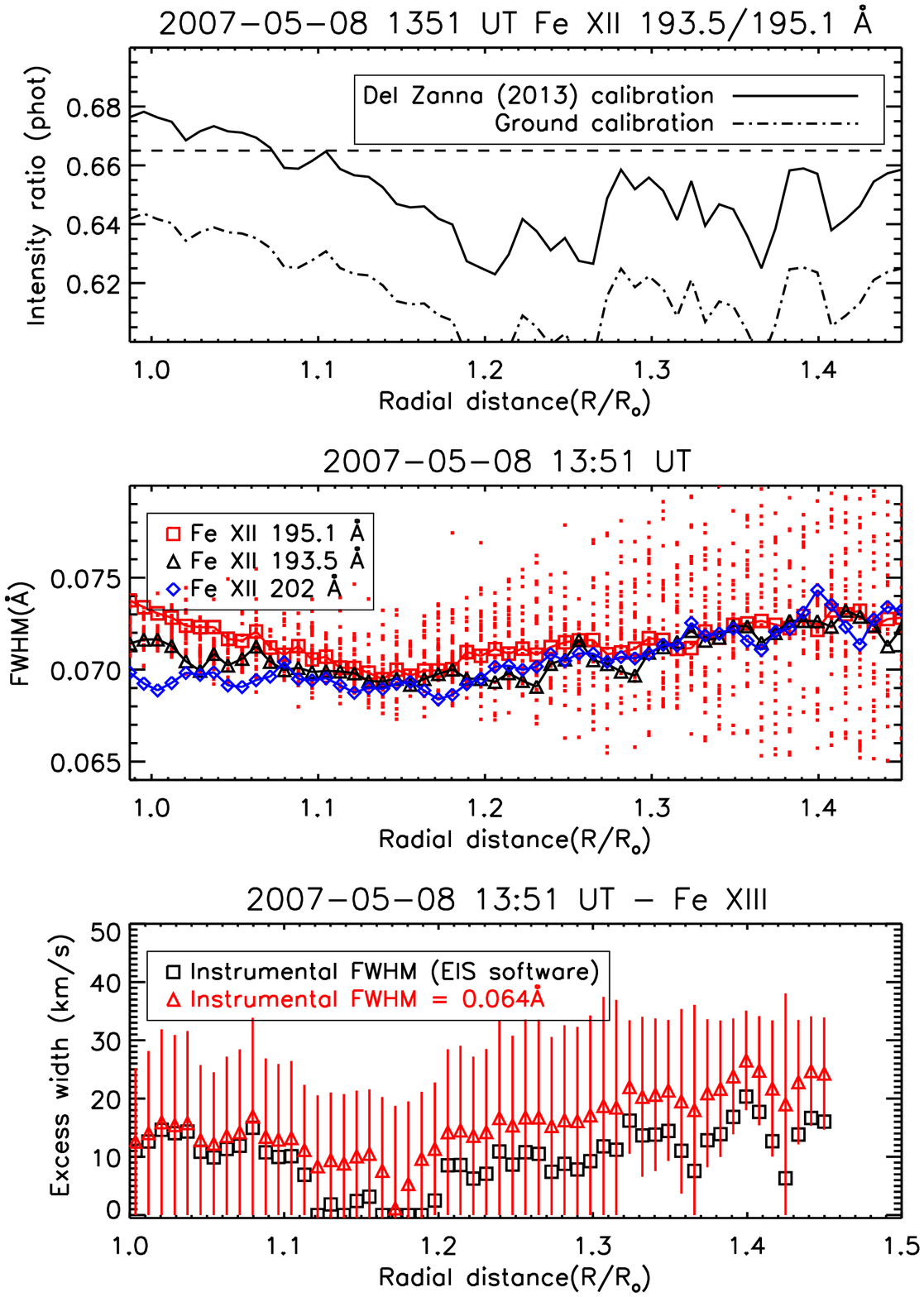}}
\centerline{\includegraphics[width=5.0cm,angle=90]{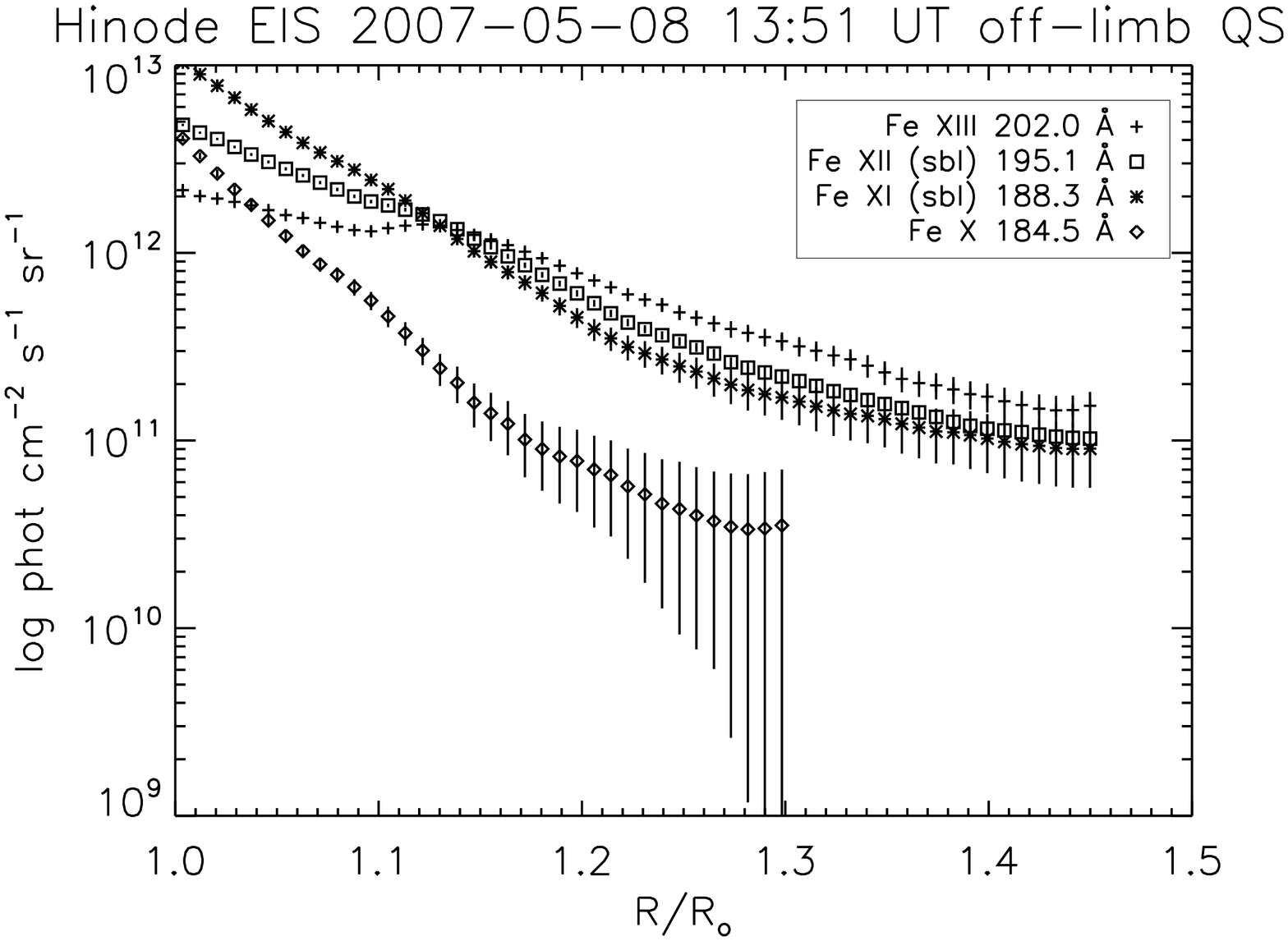}}
\centerline{\includegraphics[width=5.0cm,angle=90]{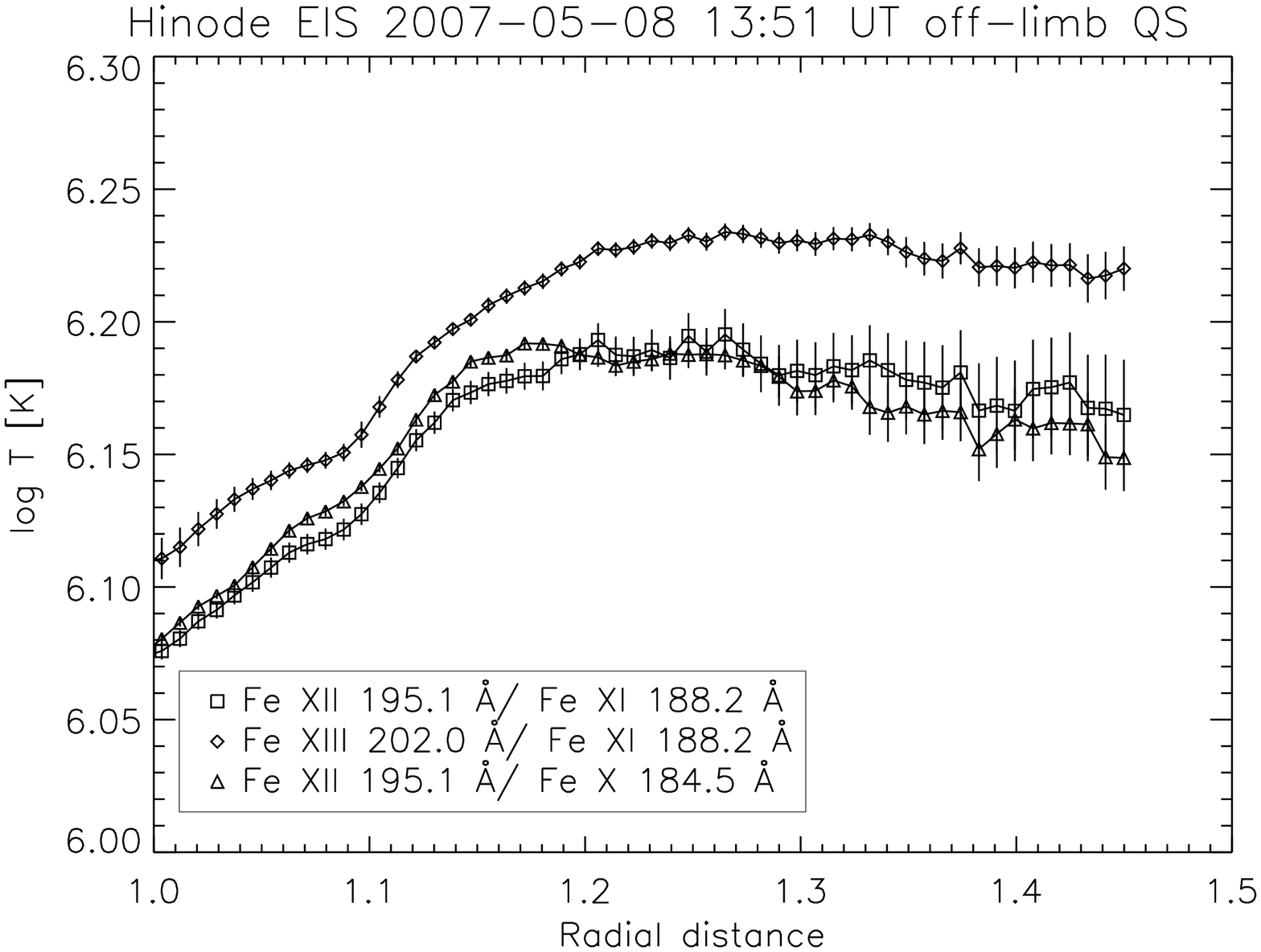}}
\caption{From top to bottom: measured intensity ratio of the 
\ion{Fe}{xii} 193.5 vs. 195.1~\AA\ lines 
on 2007-05-8, as function of radial distance in the QS sector; 
observed FWHM of the main three lines,
from averaged spectra along the radial QS sector, with the pixel-by-pixel FWHM 
of the  195.1~\AA\ line (points); excess width from the 
\ion{Fe}{xiii} line;  
radiances of selected lines along the radial QS sector; 
isothermal temperatures  along the radial QS sector.
}
\label{fig:8-may_2}
\end{figure}

\clearpage

\subsection{2018 July QS off-limb observations out to 1.5\rsun}

As most of the 2007 observations were affected somewhat by the presence 
of the active region, we have recently obtained new observations with the 
special EIS study gdz\_off\_limb1\_60  in July 2018,
when the Sun was very quiet. The pointing was similar to the 2007 observations,
in the south-west.
As the EIS instrument has degraded and the solar signal is lower, these 
observations are of less quality than the previous ones. 
Fig.~\ref{fig:2018-07-06} shows the widths obtained  from the two main lines, obtained 
from a radial sector. As in the 2007 observations, a small decrease around the 
limb is visible. However, the variation is not significant considering the 
various uncertainties we have described above.
The small increase above 1.2~\rsun\ is also most probably not real, as the lines 
were very weak. 

\begin{figure}[!htbp]
\centerline{\includegraphics[width=6.0cm,angle=0]{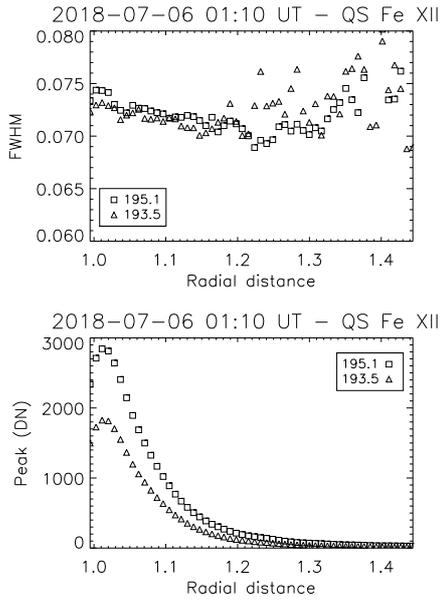}}
\caption{QS off-limb observations out to 1.5\rsun, on 2018-07-06. 
The top plot shows the variation of the FWHM in the two strongest  \ion{Fe}{xii} 
lines, as a function of  distance along a radial sector. The bottom plot 
shows the peak counts in the lines. 
}
\label{fig:2018-07-06}
\end{figure}

\clearpage

\section{AR observations on-disk } 

We searched for observations on active regions with the  full  spectral range and the 1\arcsec\ slit. 
The earliest  useful one was a study 
HPW001\_FULLCCD\_RAST\_128x128\_90S\_SLIT1, i.e. with 90s exposures and 128 
slit positions.
The strongest lines are saturated in the brightest regions. We removed the 
pixels where the lines have more than 120000 DN in their peak 
intensity. The range of peak values considered for the 195~\AA\ line
 goes from 1000 to 120000 DN, as shown in   Fig.~\ref{fig:2006-12-25}. 
The widths of the lines tend to be narrower in the regions with lowest intensities. 
There is a small difference  between the FWHM in the lines.
The 192 vs. 195~\AA\ and 192 vs. 193~\AA\ ratios are slightly lower than theory
in the dimmest regions, and show significant instrumental effects.

\begin{figure*}[!htbp]
\centerline{\includegraphics[width=6.0cm,angle=0]{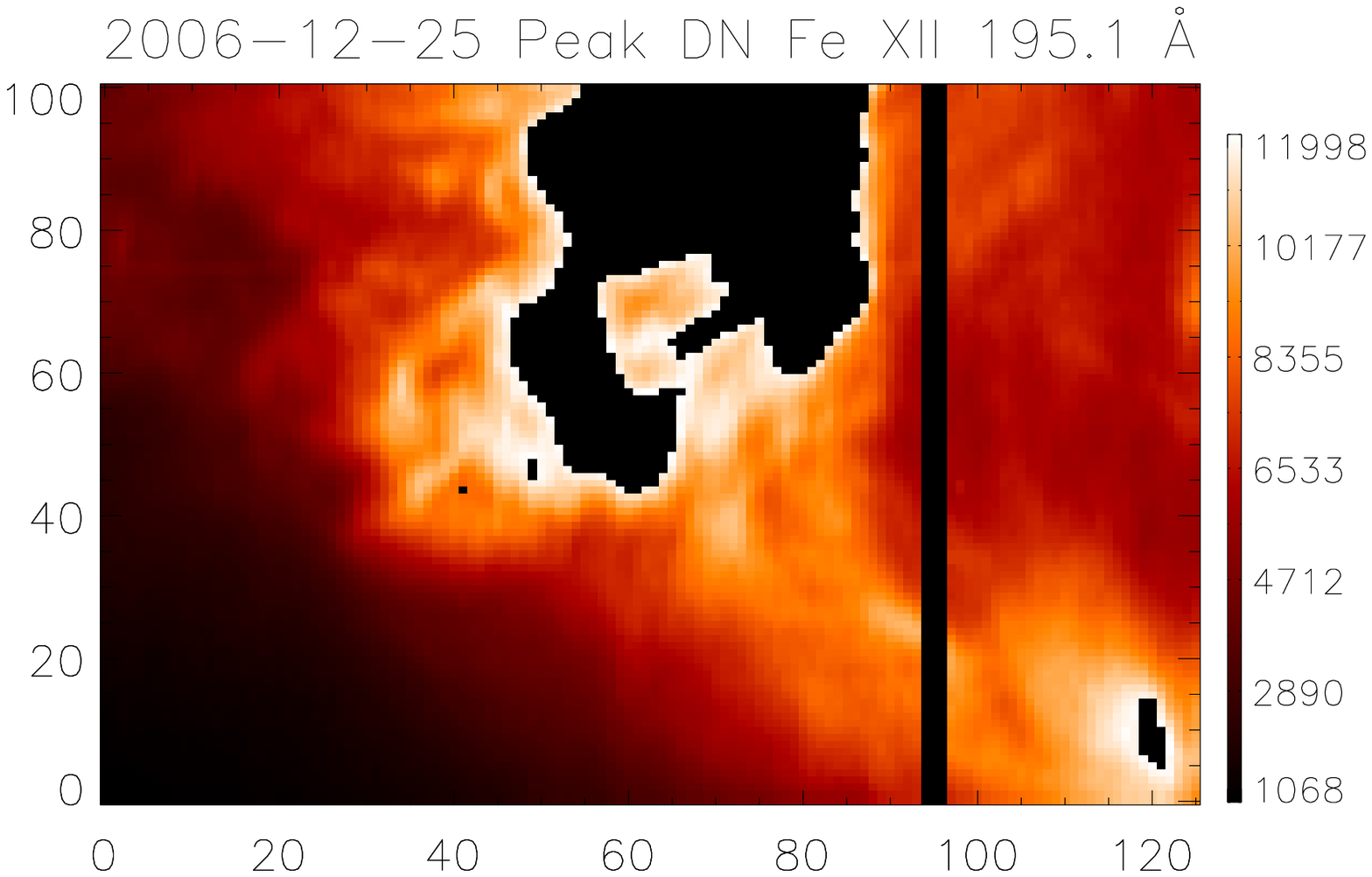}
\includegraphics[width=6.0cm,angle=0]{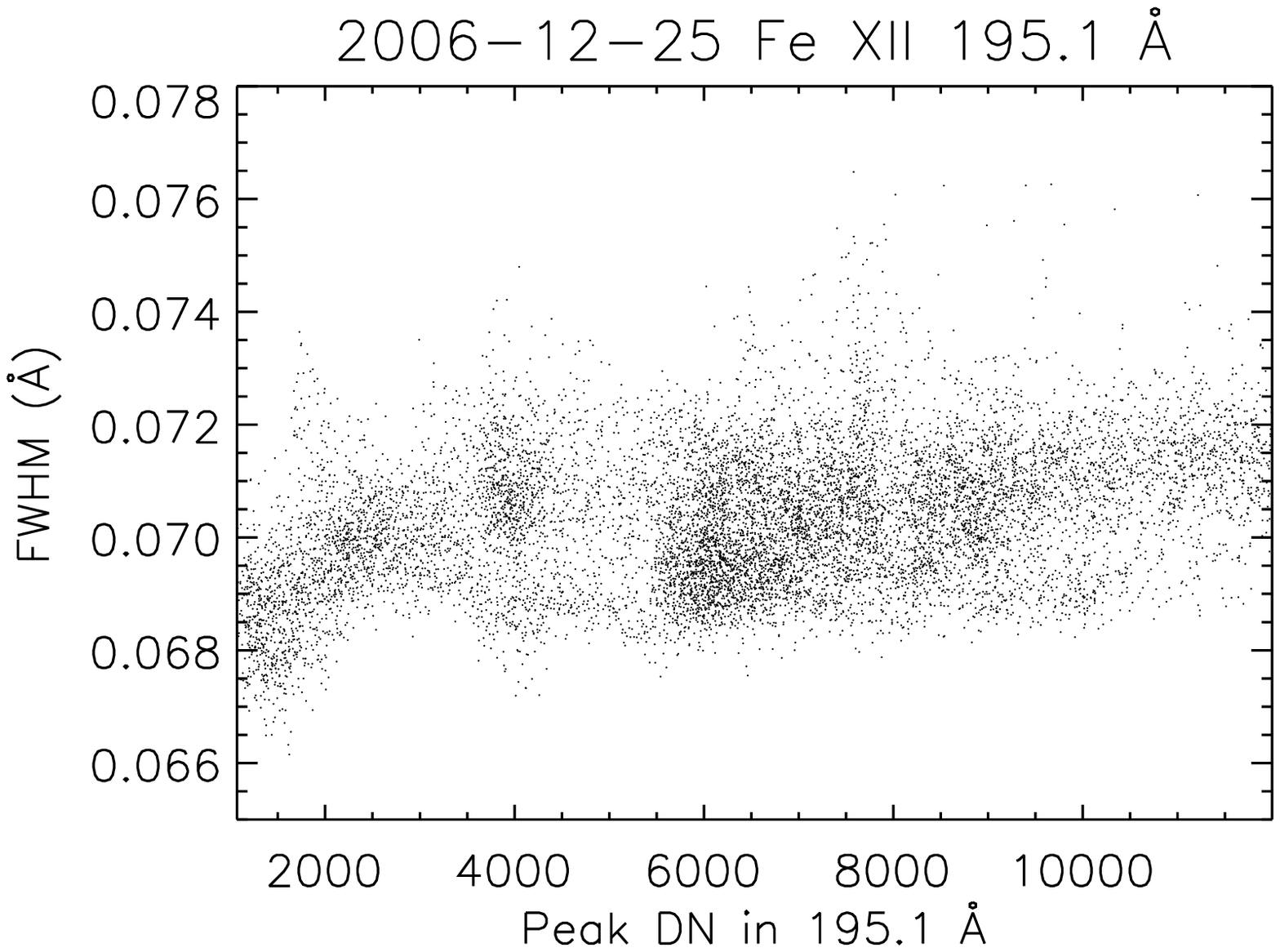}
\includegraphics[width=6.0cm,angle=0]{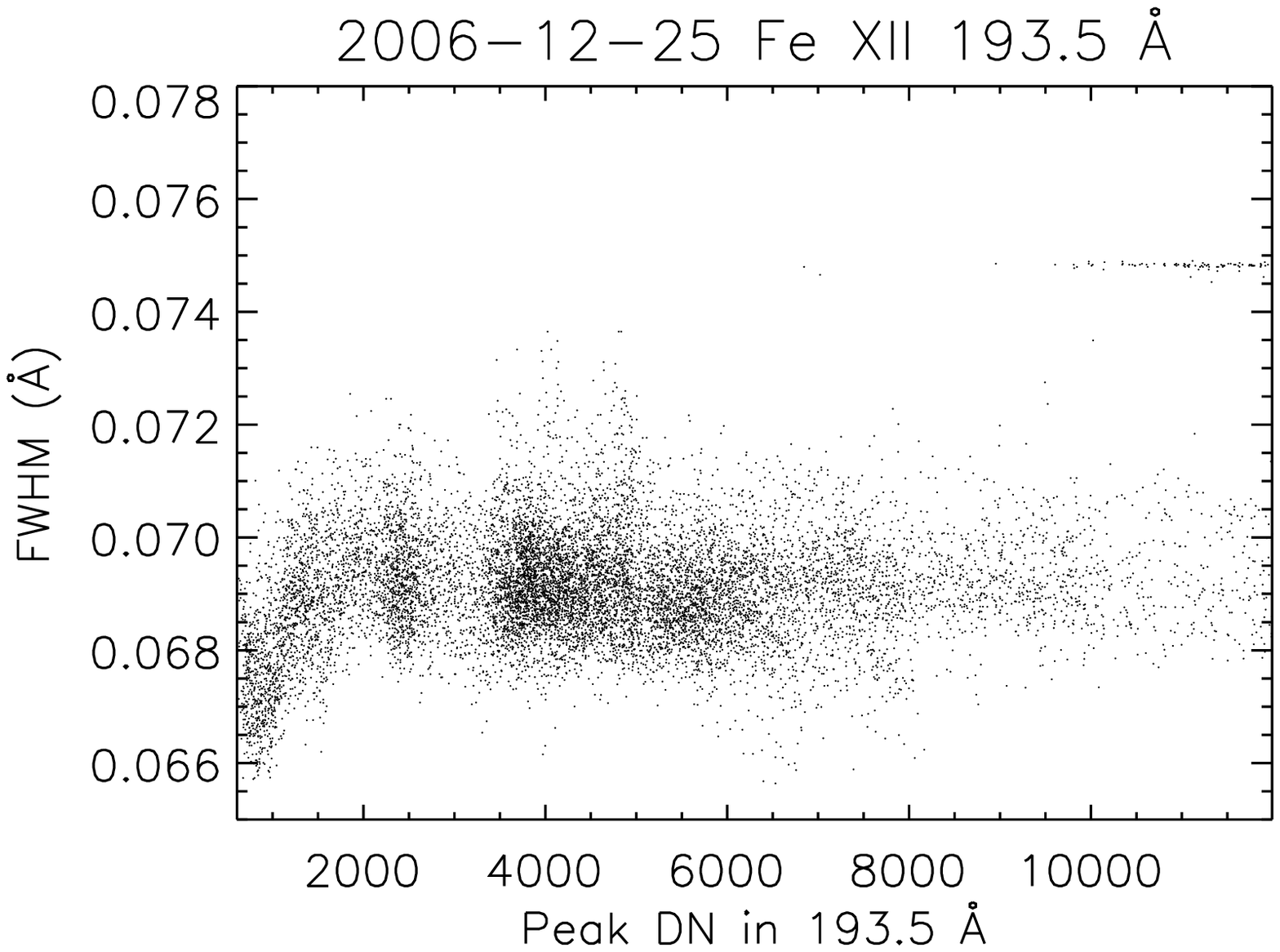}
}
\centerline{\includegraphics[width=6.0cm,angle=0]{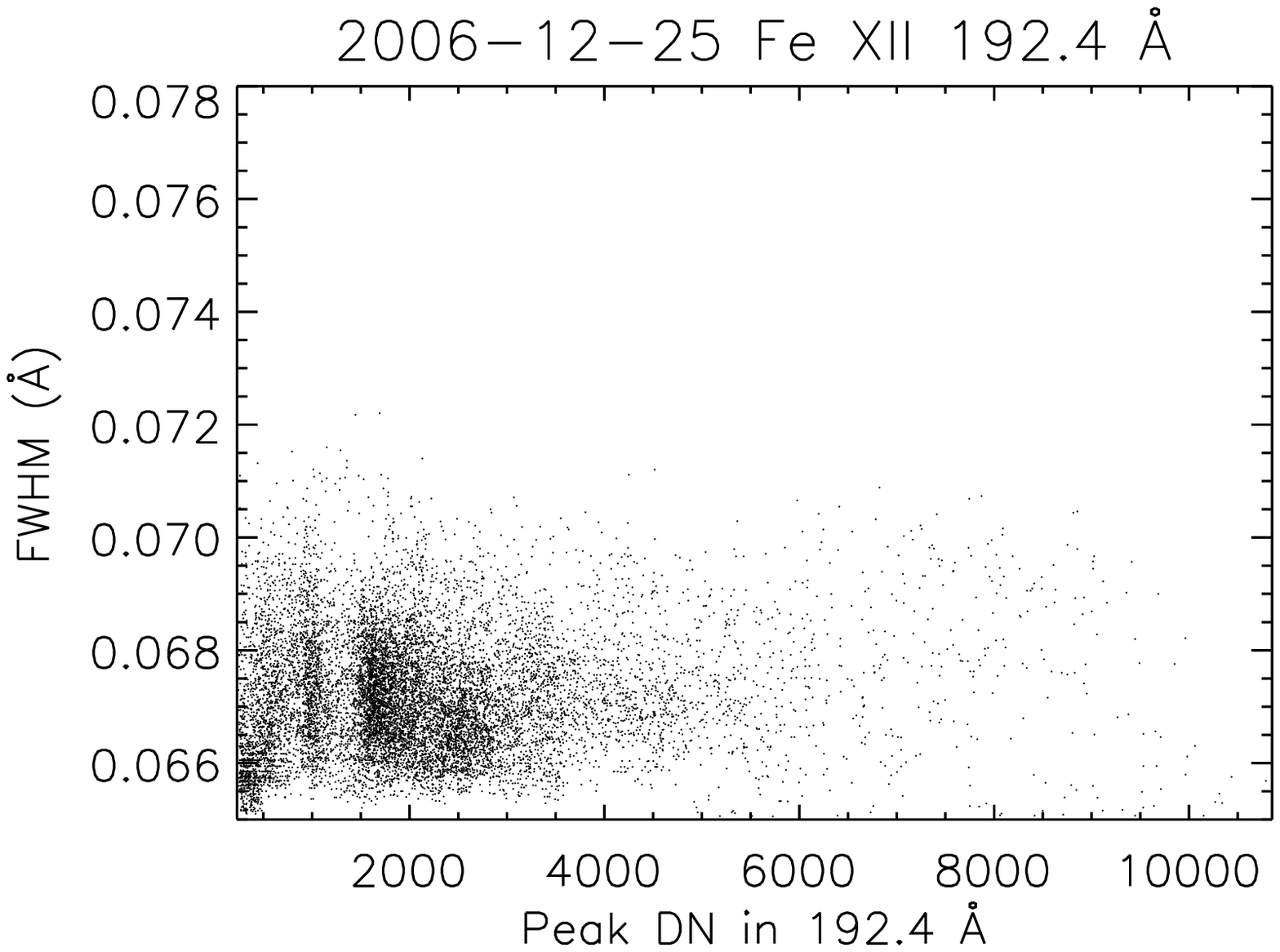}
\includegraphics[width=6.0cm,angle=0]{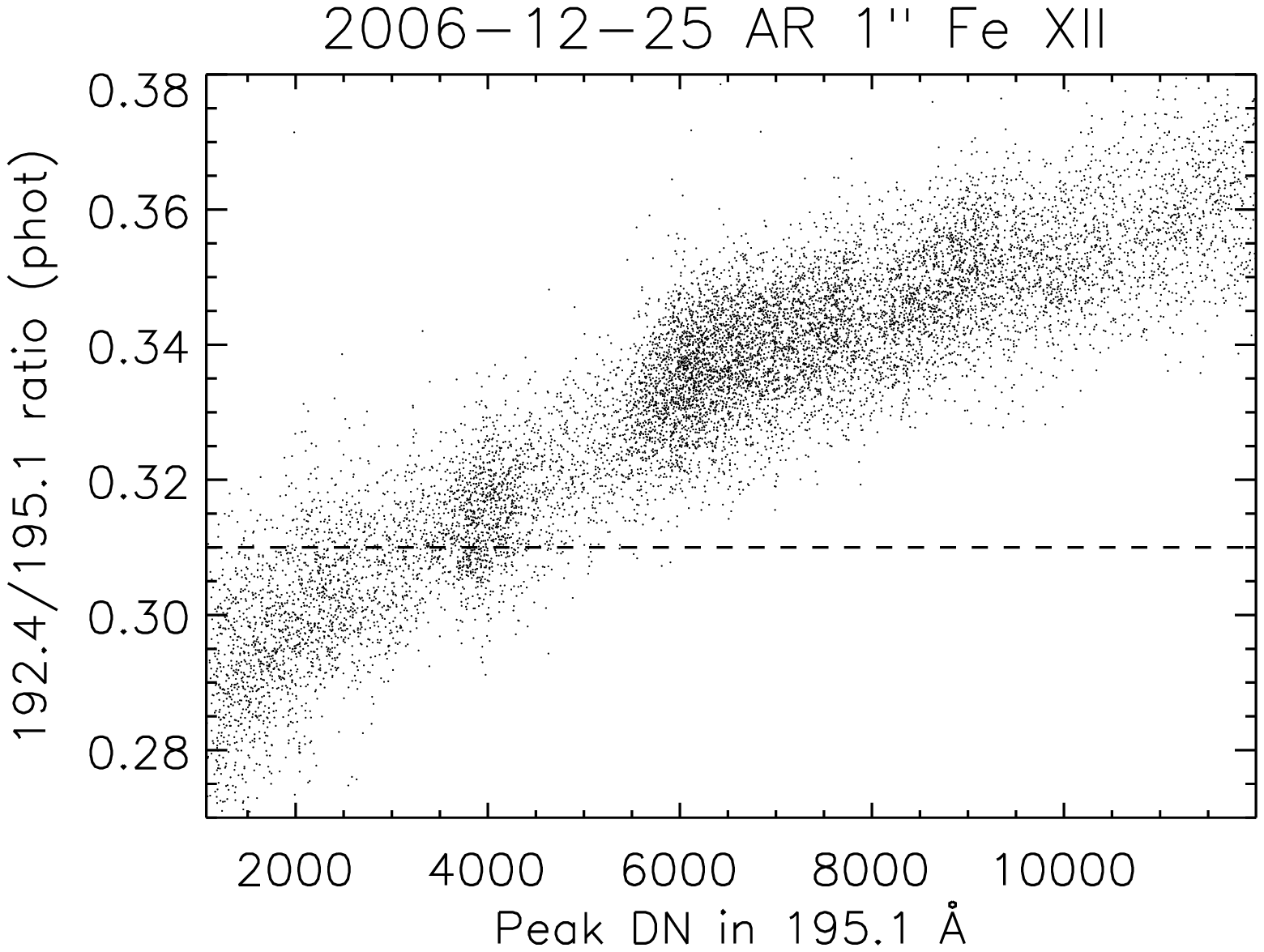}
\includegraphics[width=6.0cm,angle=0]{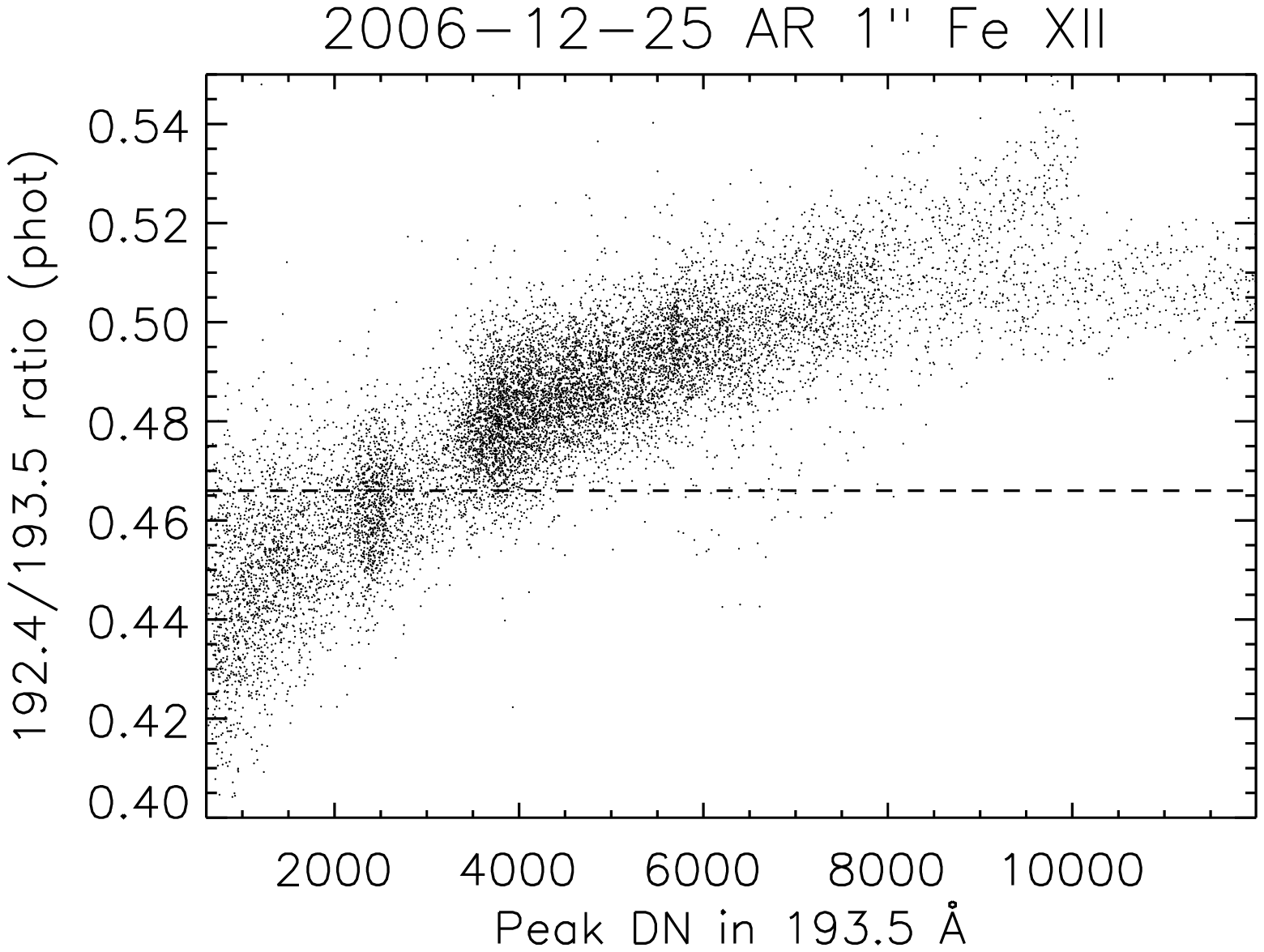}}
\caption{Active region on-disk observation on 2006-12-25 with the 
the 1\arcsec\ slit, with the HPW001\_FULLCCD\_RAST\_128x128\_90S\_SLIT1 study.
}
\label{fig:2006-12-25}
\end{figure*}


We searched for observations on active regions with the 
full spectral range and the 2\arcsec\ slit
and found the first useful one was a study with 40s  exposures, taken on
2007-05-21  at 07:27 UT. This observation was studied by 
\cite{delzanna_ishikawa:09} as it contained a small flare.
The strongest lines are saturated in the brightest regions. We removed the 
pixels where the lines have more than 120000 DN in their peak 
intensity. There is no obvious variation with line width and intensity.
However, the 195~\AA\ line consistently has a larger width.
The 192 vs. 195~\AA\ and 192 vs. 193~\AA\ ratios are close to their expected values
in the dimmest regions, but  show significant opacity  effects in the brightest regions.

\begin{figure*}[!htbp]
\centerline{\includegraphics[width=6.0cm,angle=0]{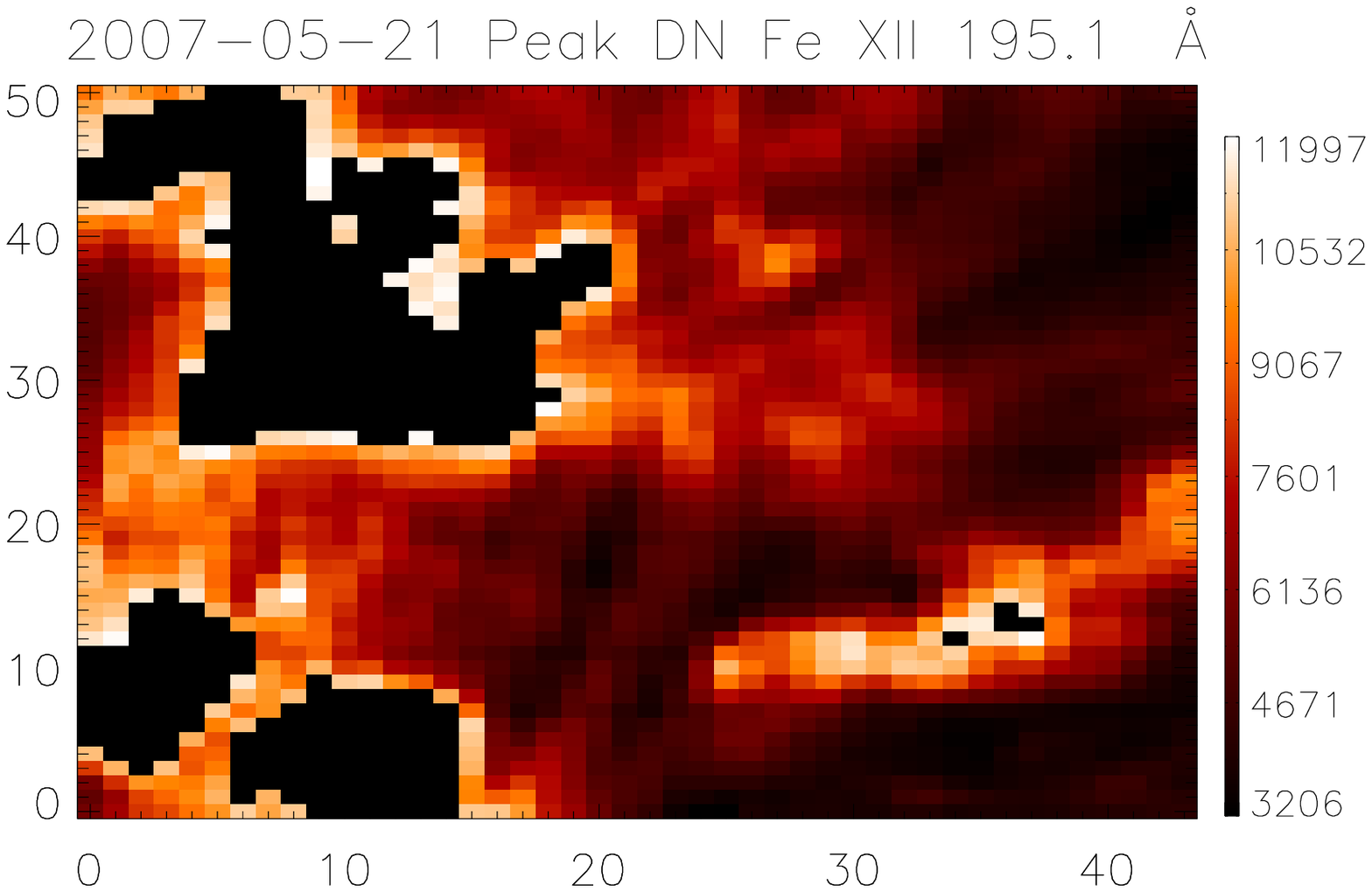}
\includegraphics[width=6.0cm,angle=0]{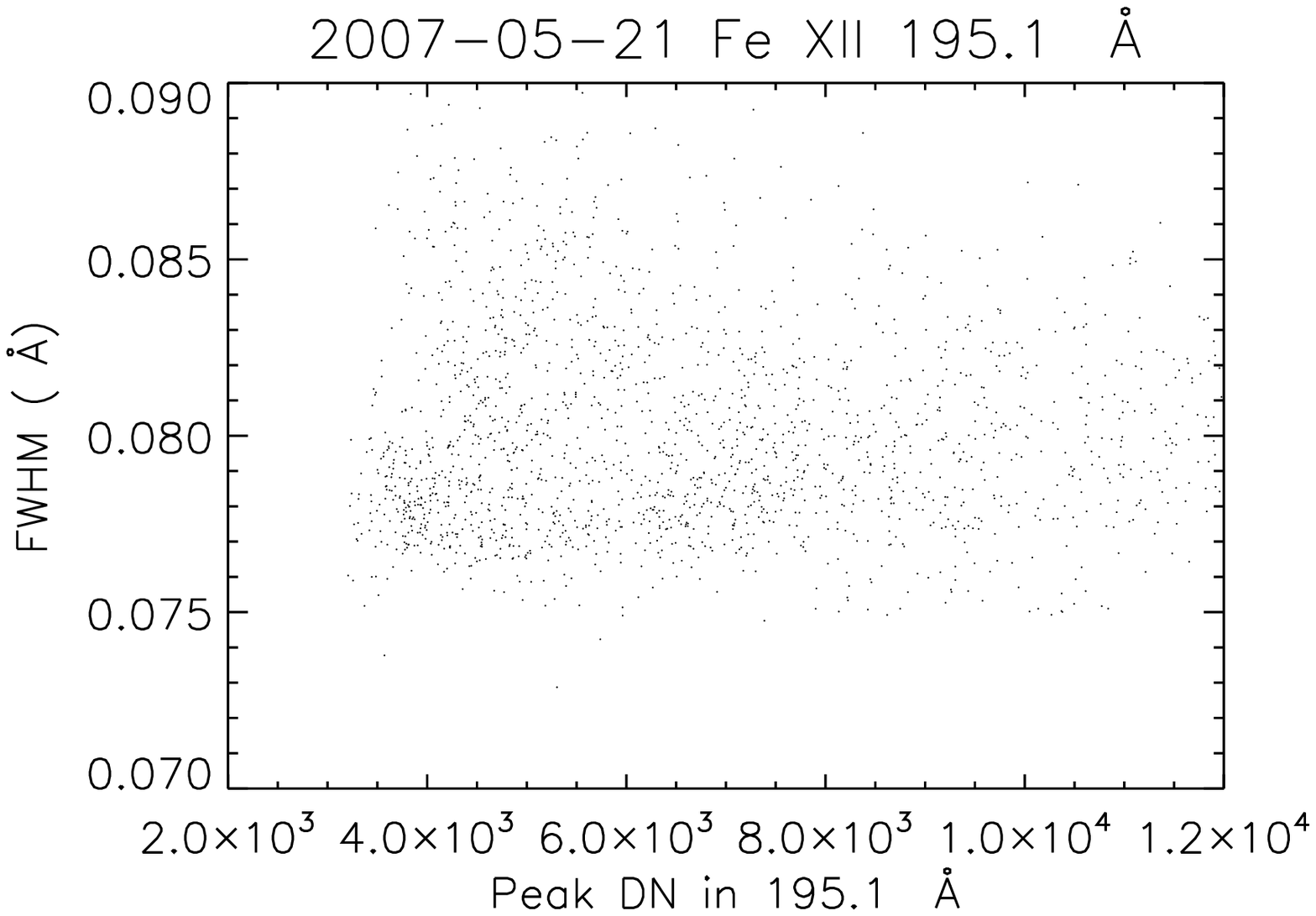}
\includegraphics[width=6.0cm,angle=0]{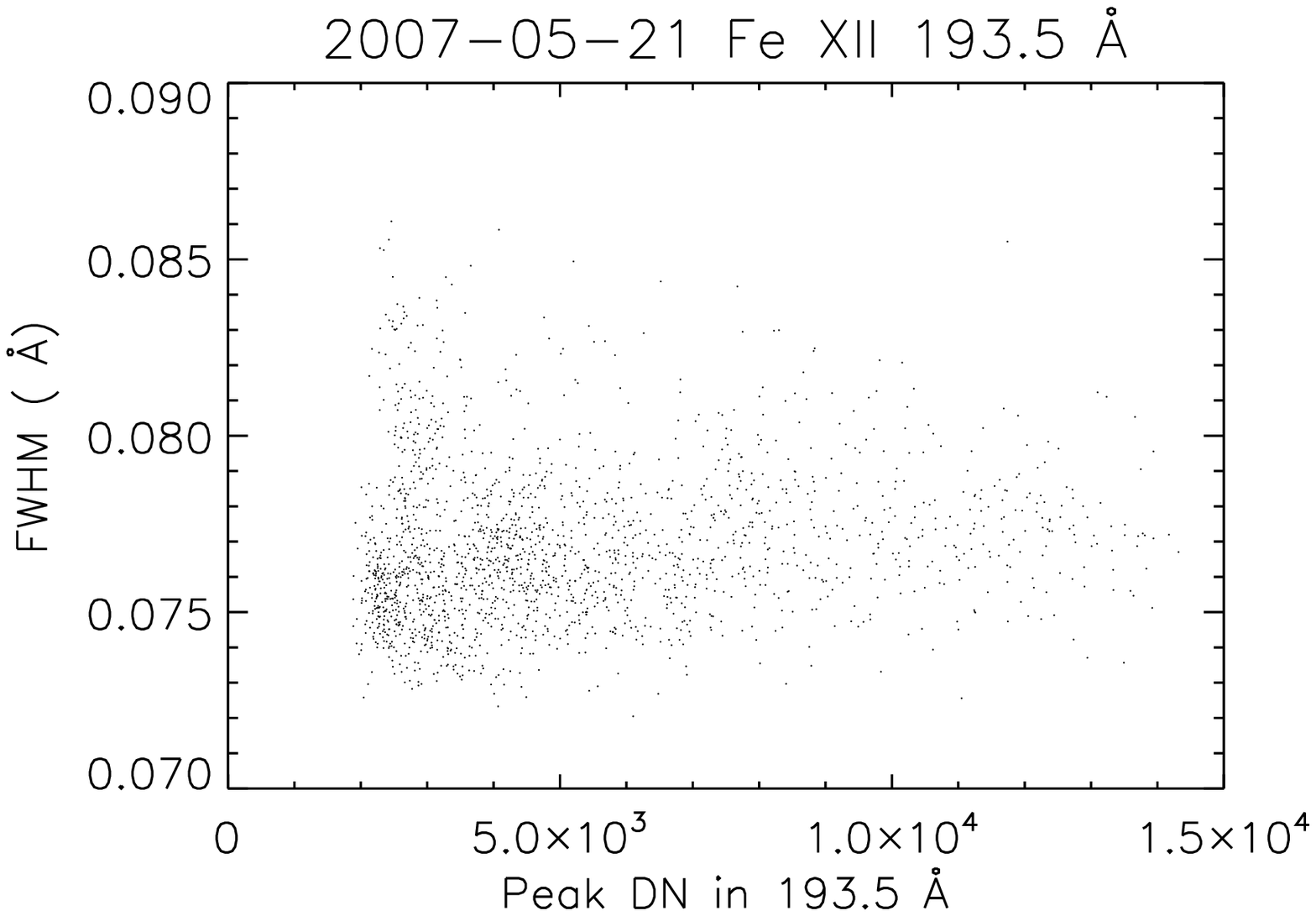}
}
\centerline{\includegraphics[width=6.0cm,angle=0]{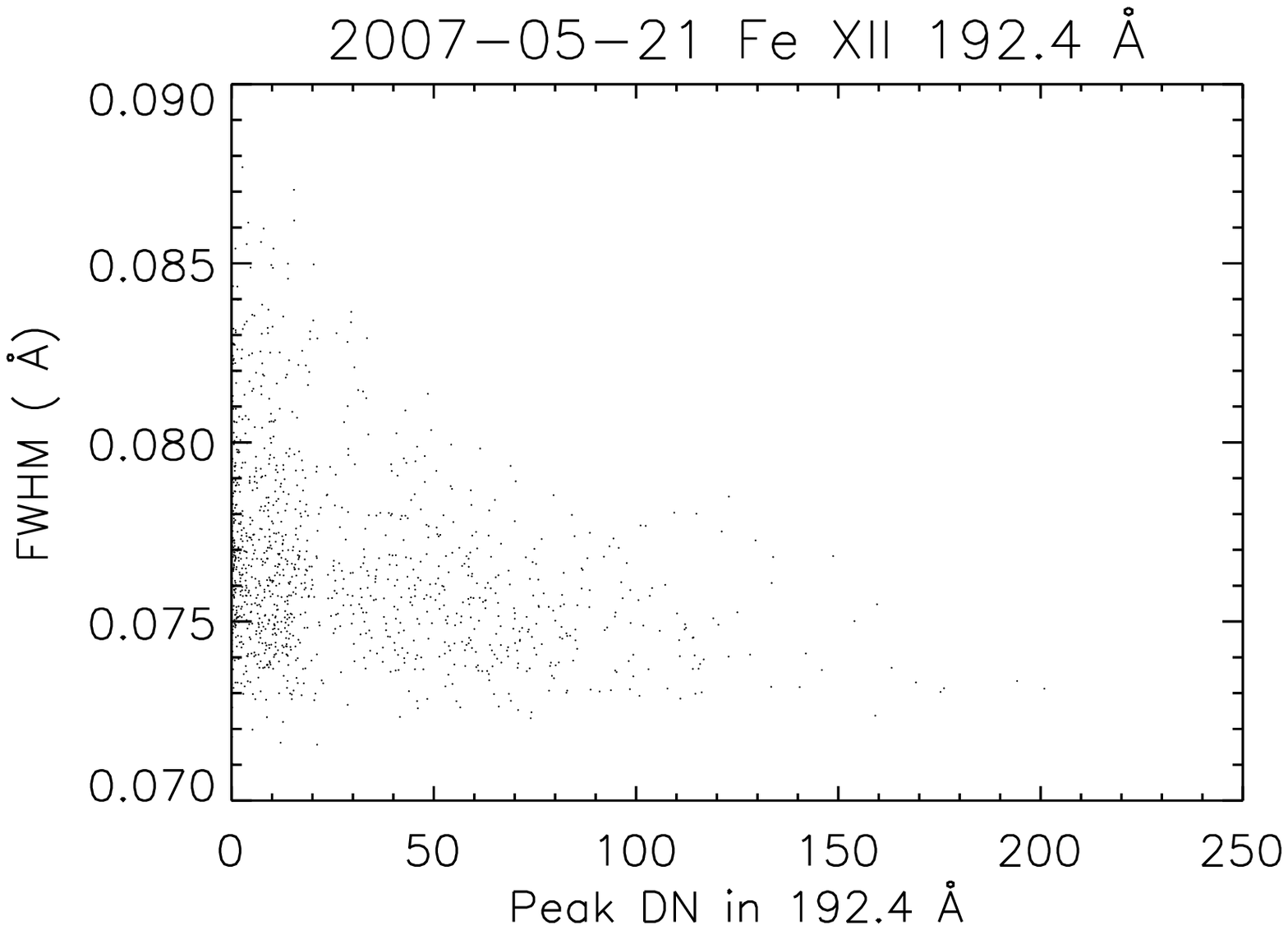}
\includegraphics[width=6.0cm,angle=0]{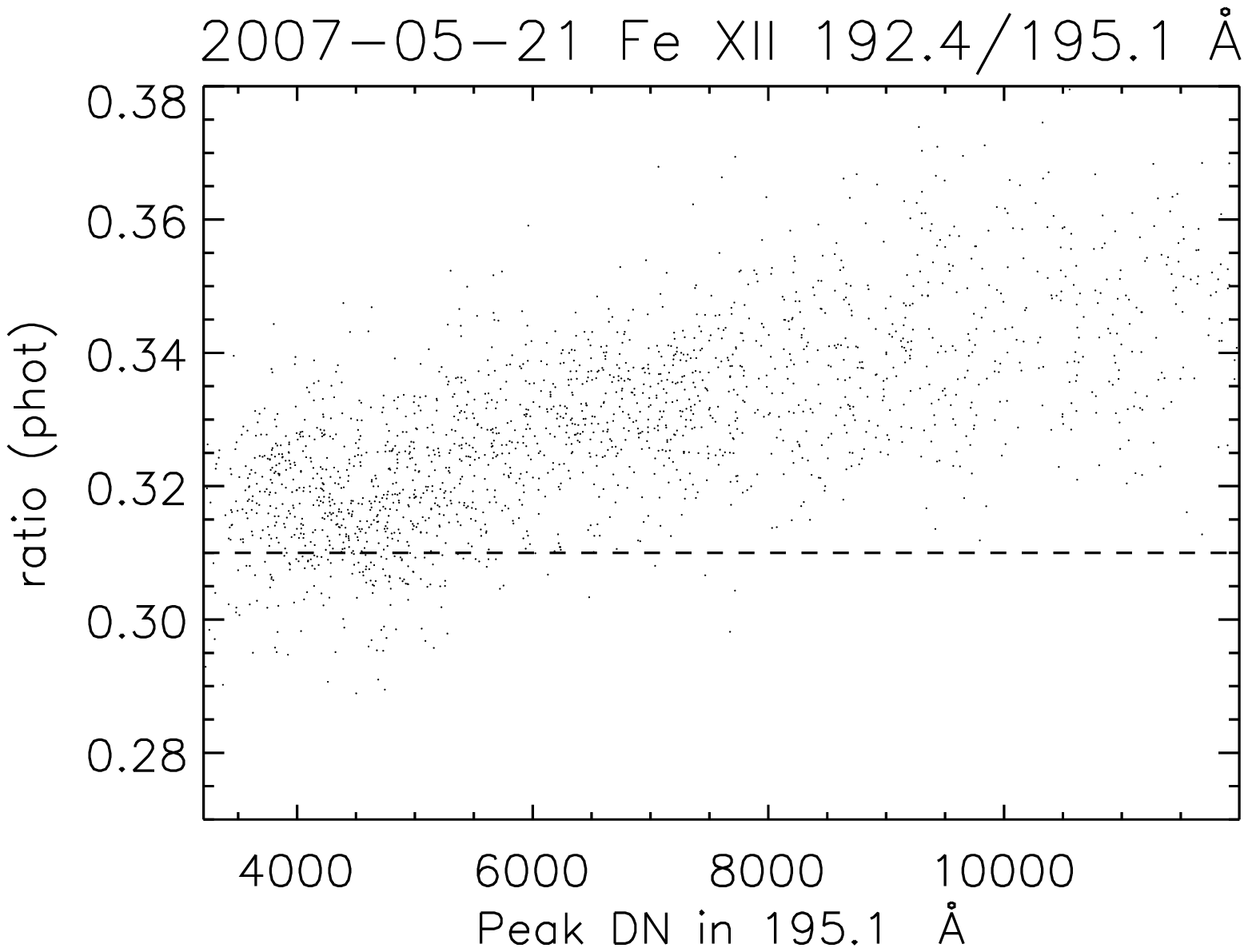}
\includegraphics[width=6.0cm,angle=0]{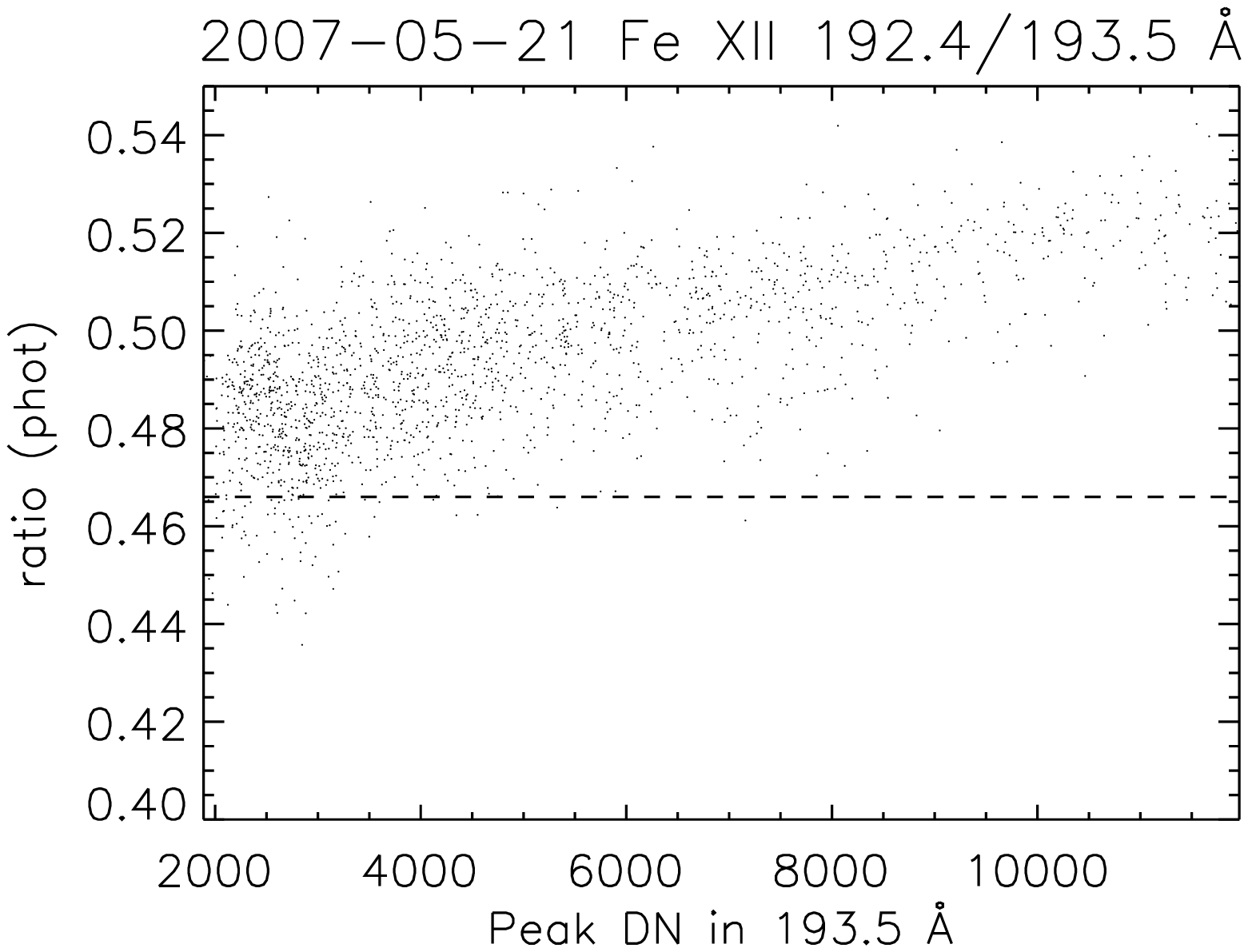}}
\caption{Active region on-disk observation on 2007-05-21 with  
the 2\arcsec\ slit
}
\label{fig:2007-05-21}
\end{figure*}

For active region observations, assuming  a FWHM of 0.03~\AA,
 a typical averaged density of $2\, 10^{9}$ cm$^{-3}$, 
`coronal' abundances (increased by 3.2 over photospheric, 
see \citealt{delzanna_mason:2018}), $N_l / N({\rm Fe XII}) = 0.75$,
an opacity at line centre of one would require a path length
 $\Delta S = 4 \, 10^{4}$ Km, which is a large but not 
 unreasonable path length.

\clearpage

\section{On the  \ion{Fe}{xii} line ratios in Young et al. (2009)}

{

\cite{young_etal:2009} presented a detailed analysis of 
EIS \ion{Fe}{xii} and  \ion{Fe}{xiii} 
line ratios to measure electron densities. Two datasets were considered
and variations along two slit positions were presented. 
One of the 1\arcsec\ slit exposures was across a system of active region 
loops, observed on   2007 May 6. In the Appendix, the 
 192 vs. 195~\AA\ and  193 vs. 195~\AA\ intensity ratios were shown. 
It was noted that the ratios increased in the top part of the 
section of the slit telemetred to the ground (304\arcsec), and it was suggested
that perhaps this was due to a change in the instrument sensitivity along the 
slit. 

We analysed the same EIS raster using our procedure. 
The chosen slit location was not described in their Appendix.
However, a comparison of the intensity images indicates that the chosen 
one is probably the No. 204 (counted from the East over 330 exposures).
Fig.~\ref{fig:young_2009_may6a} (top) shows the logarithmically-scaled
intensity of the strongest lines, as in Fig.12 of \cite{young_etal:2009},
showing reasonable agreement (changing the exposure within a few pixels
provides similar results). Fig.~\ref{fig:young_2009_may6a} (bottom)
shows the 192 vs. 195~\AA\ and  193 vs. 195~\AA\ intensity ratios,
using the EIS ground calibration. The results are virtually the same as
those shown in Fig.~B1 in \cite{young_etal:2009}. 
The drop around pixel 220 in the 193 vs. 195~\AA\ ratio is caused by a 
large piece of dust present on the EIS CCD at that location where the 
 193~\AA\ line is.

\begin{figure}[!htbp]
\centerline{\includegraphics[width=7.0cm,angle=0]{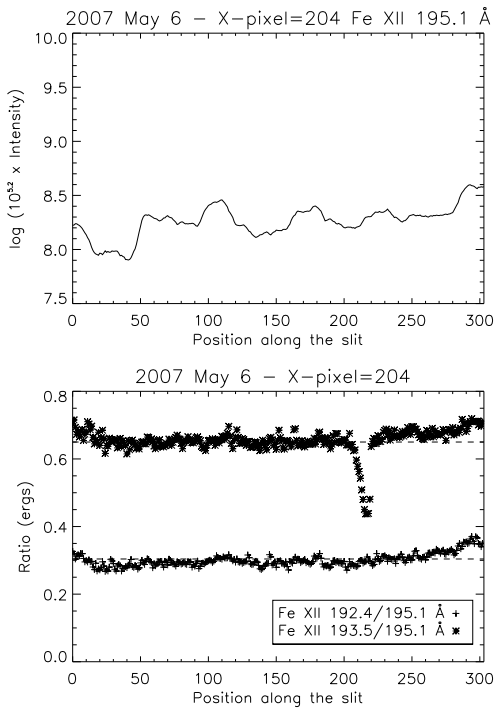}}
\caption{Top: the logarithmically-scaled intensity of the \ion{Fe}{xii}  195~\AA\ line
along  304\arcsec\ of the  1\arcsec\ slit, across AR loops observed on 2007 May 6,
as in Fig.12 of  \cite{young_etal:2009}. Bottom: the intensity 
ratios of the  \ion{Fe}{xii} lines, using the ground calibration.}
\label{fig:young_2009_may6a}
\end{figure}

\begin{figure}[!htbp]
\centerline{\includegraphics[width=7.0cm,angle=0]{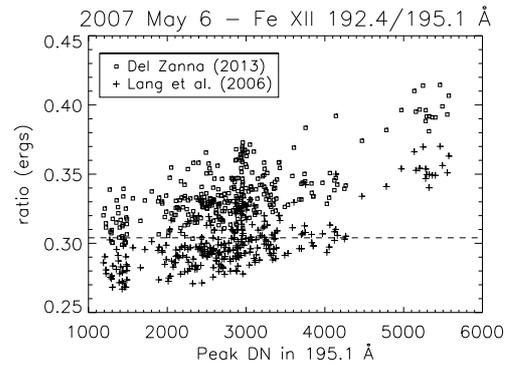}}
\caption{The intensity ratio of the \ion{Fe}{xii} 192 vs. 195~\AA\
lines along the same slit shown in Fig.~\ref{fig:young_2009_may6a}, as a 
function of the peak intensity in the stronger line, and for two 
different radiometric calibrations.
}
\label{fig:young_2009_may6b}
\end{figure}

It turns out that the increased ratio noted by  \cite{young_etal:2009}
was in a bright region in the top region. 
Fig.~\ref{fig:young_2009_may6b} shows the  \ion{Fe}{xii} 192 vs. 195~\AA\
intensity ratio this time plotted as a function of the peak 
intensity of the  195~\AA\ line, showing the clear trend which we have seen 
in most active region observations. The variation is significant, about 30\%.
The ratio has the same behaviour across the whole active region.
The pluses in Fig.~\ref{fig:young_2009_may6b} are obtained with the 
EIS ground calibration, while the boxes are with the 
Del Zanna (2013) calibration.  As we have already mentioned, the 
Del Zanna (2013) calibration was obtained assessing line ratios in quiet 
Sun areas in 2006-2007 in this wavelength region, hence provides a 
very good agreement with the expected values 
in the low-intensity areas in active regions. 

}

\end{document}